\documentclass[12pt,english,american]{article}
\usepackage[T1]{fontenc}
\usepackage[latin9]{inputenc}
\usepackage{geometry}
\usepackage{color}
\usepackage{babel}
\usepackage{amsthm}
\usepackage{amstext}
\usepackage{amssymb}
\usepackage{mathrsfs}
\usepackage{amsmath}
\usepackage{cancel}   
\usepackage{yfonts}
\usepackage{soul}
\usepackage{tikz}
\usepackage[unicode=true,pdfusetitle,
 bookmarks=true,bookmarksnumbered=false,bookmarksopen=false,
 breaklinks=true,pdfborder={0 0 0},backref=false,colorlinks=true]
 {hyperref}
\hypersetup{
 linkcolor=blue,citecolor=blue, urlcolor=blue}

\makeatletter

\def\qed{$\Box$\medskip}

\newcommand{\beq}{\begin{equation}}
\newcommand{\eeq}{\end{equation}}
\newcommand{\beqa}{\begin{eqnarray}}
\newcommand{\eeqa}{\end{eqnarray}}
\newcommand{\ben}{\begin{arabicenumerate}}
\newcommand{\een}{\end{arabicenumerate}}

\def\bel{\begin{lem} } 
\def\eel{\end{lem} }
\def\bet{\begin{thm}}
\def\eet{\end{thm}}
\def\bed{\begin{defn}}
\def\eed{\end{defn} }

\def\bec{\begin{cor}}
\def\eec{\end{cor}}
\def\ber{\begin{rem}}
\def\eer{\end{rem}}

\usepackage{enumitem}		
\theoremstyle{plain}
\newtheorem{thm}{\protect\theoremname}[section]
\theoremstyle{definition}
\newtheorem{defn}[thm]{\protect\definitionname}
\theoremstyle{plain}

\theoremstyle{plain}
\theoremstyle{remark}
\newtheorem{rem}[thm]{\protect\remarkname}
\theoremstyle{plain}
\newtheorem{lem}[thm]{\protect\lemmaname}
\theoremstyle{plain}
\newtheorem{cor}[thm]{\protect\corollaryname}

\usepackage{txfonts,refstyle,xcolor}

\newref{con}{name = Conjecture\ }
\newref{prop}{name = Proposition\ }
\newref{def}{name = Definition\ }
\newref{sec}{name = Section\ }
\newref{sub}{name = Section\ }
\newref{thm}{name = Theorem\ }
\newref{lem}{name = Lemma\ }
\newref{cor}{name = Corollary\ }
\newref{fig}{name = Figure\ }
\newref{rem}{name =  Remark\ }

\usepackage{bbm}
\newcommand{\charf}{\mathbbm{1}}

\usepackage[all]{xy}

\newcommand{\xyR}[1]{%
     \makeatletter
     \xydef@\xymatrixrowsep@{#1}
     \makeatother
}

\newcommand{\xyC}[1]{%
     \makeatletter
     \xydef@\xymatrixcolsep@{#1}
     \makeatother
}

\newcommand{\ncol}[1]{\color{normalcolor}}

\makeatother

\addto\captionsamerican{\renewcommand{\corollaryname}{Corollary}}
\addto\captionsamerican{\renewcommand{\definitionname}{Definition}}
\addto\captionsamerican{\renewcommand{\lemmaname}{Lemma}}
\addto\captionsamerican{\renewcommand{\propositionname}{Proposition}}
\addto\captionsamerican{\renewcommand{\remarkname}{Remark}}
\addto\captionsamerican{\renewcommand{\theoremname}{Theorem}}
\addto\captionsenglish{\renewcommand{\corollaryname}{Corollary}}
\addto\captionsenglish{\renewcommand{\definitionname}{Definition}}
\addto\captionsenglish{\renewcommand{\lemmaname}{Lemma}}
\addto\captionsenglish{\renewcommand{\propositionname}{Proposition}}
\addto\captionsenglish{\renewcommand{\remarkname}{Remark}}
\addto\captionsenglish{\renewcommand{\theoremname}{Theorem}}
\providecommand{\corollaryname}{Corollary}
\providecommand{\definitionname}{Definition}
\providecommand{\lemmaname}{Lemma}
\providecommand{\propositionname}{Proposition}
\providecommand{\remarkname}{Remark}
\providecommand{\theoremname}{Theorem}

\addto\captionsamerican{\renewcommand{\corollaryname}{Corollary}}
\addto\captionsamerican{\renewcommand{\definitionname}{Definition}}
\addto\captionsamerican{\renewcommand{\lemmaname}{Lemma}}
\addto\captionsamerican{\renewcommand{\propositionname}{Proposition}}
\addto\captionsamerican{\renewcommand{\remarkname}{Remark}}
\addto\captionsamerican{\renewcommand{\theoremname}{Theorem}}
\addto\captionsenglish{\renewcommand{\corollaryname}{Corollary}}
\addto\captionsenglish{\renewcommand{\definitionname}{Definition}}
\addto\captionsenglish{\renewcommand{\lemmaname}{Lemma}}
\addto\captionsenglish{\renewcommand{\propositionname}{Proposition}}
\addto\captionsenglish{\renewcommand{\remarkname}{Remark}}
\addto\captionsenglish{\renewcommand{\theoremname}{Theorem}}
\providecommand{\corollaryname}{Corollary}
\providecommand{\definitionname}{Definition}
\providecommand{\lemmaname}{Lemma}
\providecommand{\propositionname}{Proposition}
\providecommand{\remarkname}{Remark}
\providecommand{\theoremname}{Theorem}

\begin{document}
\title{\emph{Local} iterative block-diagonalization of gapped Hamiltonians: a new tool in singular perturbation theory  }
  \author{S. Del Vecchio\footnote{Institut f\"ur Theoretische Physik, ITP - Universit\"at Leipzig, Germany
 / email:simone.del\textunderscore vecchio@physik.uni-lepzig.de }\,,  J. Fr\"ohlich\footnote{Institut f\"ur Theoretiche Physik, ETH-Z\"urich , Switzerland / email: juerg@phys.ethz.ch}\,, A. Pizzo \footnote{Dipartimento di Matematica, Universit\`a di Roma ``Tor Vergata", Italy
/ email: pizzo@mat.uniroma2.it}\,, S. Rossi\footnote{Dipartimento di Matematica, Universit\`a degli studi di Bari Aldo Moro, Italy
/ email: stefano.rossi@uniba.it}}

\date{02/11/2020}

\maketitle

\abstract{In this paper the \emph{local} iterative Lie-Schwinger block-diagonalization method, introduced 
in \cite{FP}, \cite{DFPR1}, and \cite{DFPR2} for quantum chains, is extended to higher-dimensional quantum lattice 
systems with Hamiltonians that can be written as the sum of an unperturbed gapped operator, consisting of a sum of 
on-site terms, and a perturbation consisting of bounded interaction potentials of short range mutltiplied by a real 
coupling constant $t$. Our goal is to prove that the spectral gap above the ground-state energy of such Hamiltonians 
persists for sufficiently small values of $\vert t \vert$, \textit{independently} of the size of the lattice.

\noindent
New ideas and concepts are necessary to extend our method to systems in dimension $d>1$: 
As in our earlier work, a sequence of 
\emph{local} block-diagonalization steps based on judiciously chosen unitary conjugations of the original Hamiltonian is 
introduced. The supports of effective interaction potentials generated in the course of these block-diagonalization steps can 
be identified with what we call \textit{minimal rectangles} contained in the lattice, a concept that serves to tackle 
combinatorial problems that arise in the course of iterating the block-diagonalization steps. For a given minimal 
rectangle, control of the effective interaction potentials generated in each block-diagonalization step with support 
in the given rectangle is achieved by exploiting a variety of rather subtle mechanisms which
include, for example, the use of \emph{weighted sums of paths} consisting of overlapping rectangles and of 
\emph{large denominators}, expressed in terms of sums of orthogonal projections, that serve to control analogous sums of projections in the numerators resulting from 
 the unitary conjugations of the interaction potential terms involved in the local
block-diagonalization step.}
\\
\section{Models of gapped quantum lattice systems, and survey of results}
\setcounter{equation}{0}

In this paper we introduce and study a family of quantum lattice systems describing insulating materials 
in two or more dimensions. We are interested in analyzing the low-energy spectrum of the Hamiltonians 
of these systems and, in particular, in showing that the ground-state energies of these Hamiltonians are 
separated from the rest of their energy spectrum by a strictly positive gap. Our analysis is based on 
a novel method consisting in iteratively  block-diagonalizing the Hamiltonians with respect to the 
ground-state subspace. The block-diagonalization is accomplished by a sequence of unitary 
conjugations of the Hamiltonians. Our analysis is motivated in part by recent interest in characterizing 
\textit{``topological phases''}, see e.g. \cite{BN, BH, BHM}; more specifically by studying Hamiltonians of \textit{``topological insulators''} 
whose ground-state energy is separated from the higher-lying spectrum by a strictly positive energy gap.
But the scope of our techniques is actually more general. 

To be concrete we consider tight-binding models of 
electrons hopping on a lattice $\mathbb{Z}^{d}, d\geq 2$, with Hamiltonians that
are given as the sum of an \textit{unperturbed operator}, $K_0$, and a 
\textit{perturbation}, $K_{I}$, consisting of a sum of bounded \textit{interaction potentials}. The operator $K_0$ 
can be written as a sum of terms, $H_{\mathbf{i}}$, only depending on the degrees of freedom located at 
single sites $\mathbf{i} \in \mathbb{Z}^{d}$, while the interaction potentials contributing to $K_{I}$ only couple degrees 
of freedom located on subsets of the lattice of strictly bounded diameter. We focus our attention on unperturbed operators 
$K_0$ with a unique ground-state, $\Omega$, and a positive energy gap above their ground-state energy; (but our methods 
can be extended to families of unperturbed operators with degenerate ground-state energies). 
Our aim is to iteratively construct an 
anti-self-adjoint operator $S\equiv S(t)= -S(t)^{*}, t\in \mathbb{R},$ such that the ground-state of the operator 
$e^{S}\big(K_0 + t\cdot K_{I} \big)e^{-S}$ is given again 
by $\Omega$, and the spectrum of the restriction of this operator to the subspace orthogonal to $\Omega$ lies strictly 
above the ground-state energy, provided the absolute value of the coupling constant $t$ is small enough.
Our method to construct the operator $S=S(t)$ is inspired by a novel technique introduced in \cite{FP}, 
which, in its original form, has been limited to chains, i.e., to one-dimensional systems. This technique represents
an interesting example of multi-scale, iterative perturbation theory: it consists in successively block-diagonalizing the 
Hamiltonians associated with sequences of bounded, connected subsets of the lattice. In one dimension, 
such subsets are intervals. 
But, for $d>1$, the number of connected subsets of a given cardinality, $R$, containing a fixed point 
of the lattice grows exponentially in $R$, and this causes certain difficulties that make it necessary to refine 
the methods in \cite{FP} in a rather subtle way; see \mbox{Sect. \ref{method}.}

\noindent
We remark that the procedure described here is amenable to be extended to analogous lattice systems but with unbounded interactions \cite{DFPR4}.

It is appropriate to comment on earlier work addressing problems closely related to the ones treated in our paper. 
Actually, it is primarily the \textit{mathematical methods} used in our analysis that are novel. Our main results are very similar to ones that can be found in the literature. In \cite{Y} and \cite{KT}, results reminiscent of ours have been obtained by using cluster expansions based on operator methods; in \cite{DS} fermionic path integral methods have been used for the same purpose, and in \cite{NSY}, \cite{H}, \cite{MZ} quasi-adiabatic flows have been constructed to establish results related to ours.  Ideas sharing some similarities with the ones presented in our paper have been used in \cite{DRS} for purposes analogous to ours, and in \cite{I1, I2} for a partial analysis of many-body localization in one dimension. 

\subsection{A family of quantum lattice systems}

We consider a finite,  $d$-dimensional lattice, $\Lambda_{N}^{d}\subset \mathbb{Z}^{d}$, with sides consisting of 
$N$ vertices, where $N<\infty$ is arbitrary (but fixed). Each vertex in $\Lambda_{N}^{d}$ is labelled by a multi-index $\bold{i}:=(i_1, \dots, i_d)$, with 
$i_j\in (1, \dots, N)$, $j=1,\dots,d$.  The Hilbert space of pure state vectors of the quantum lattice systems 
studied in this paper is given by
\begin{equation}\label{Hilbert}
\mathcal{H}^{(N)}:=\bigotimes_{\bold{i}\in \Lambda_{N}^{d}}  \,\mathcal{H}_{\bold{i}}\,,\quad \text{with }\quad 
\mathcal{H}_{\bold{i}}\simeq \mathbb{C}^M, \, \forall \, \bold{i}\in \Lambda_N^d\,,
\end{equation}
where $M$ is an arbitrary, but finite, $N$-independent integer. Let $H$ be a non-negative $M\times M$ matrix with the properties that $0$ 
is an eigenvalue of $H$ corresponding to an eigenvector $\Omega \in \mathbb{C}^M$, and 
$$H \upharpoonright_{\lbrace \mathbb{C} \Omega \rbrace^{\perp}} \geq \charf \,,$$
where $\charf $ is the identity matrix.

\noindent
We define
\begin{equation}\label{H_i}
H_{\bold{i}}:= ( \bigotimes_{\Lambda_{N}^{d} \ni\bold{j} \neq \bold{i}} \charf_{\bold{j}})\otimes \underset{\underset{\bold{i}^{th} \text{slot}}{\uparrow}}{H} \,
\end{equation}
where $\charf_{\bold{j}}$ is the identity matrix on $\mathcal{H}_{\bold{j}}$.
By $P_{\Omega_{\bold{i}}}$ we denote the orthogonal projection onto the subspace
\begin{equation}\label{vacuum_i}
 ( \bigotimes_{\Lambda_{N}^{d} \ni \bold{j} \neq \bold{i}}\mathcal{H}_{\bold{j}})\otimes  \underset{\underset{\bold{i}^{th} \text{slot}}{\uparrow}}{\{\mathbb{C}\Omega\}} 
\subset \mathcal{H}^{(N)}\,, \quad \text{  and}\quad   P_{\Omega_{\bold{i}}}^{\perp} := \charf - P_{\Omega_{\bold{i}}}\,.
\end{equation}
Then 
\begin{equation*} 
H_{\bold{i}}=P^{\perp}_{\Omega_{\bold{i}}}\,H_\bold{i}\, P^{\perp}_{\Omega_{\bold{i}}}+P_{\Omega_{\bold{i}}}\, H_\bold{i} \, P_{\Omega_{\bold{i}}}\,,
\end{equation*}
with
\begin{equation}\label{gaps}
P_{\Omega_{i}}\,H_\bold{i}\, P_{\Omega_{\bold{i}}}=0\,,\quad\text{and} \quad P^{\perp}_{\Omega_{\bold{i}}}\, H_{\bold{i}}\, P^{\perp}_{\Omega_{\bold{i}}}\geq P^{\perp}_{\Omega_{\bold{i}}}\,.
\end{equation}

We study quantum systems on the lattice $\Lambda_N^d$ with Hamiltonians of the form
\begin{equation}\label{Hamiltonian}
K_{N}\equiv K_{N}(t):= \underbrace{\sum_{\bold{i}\in \Lambda^{d}_N}H_{\bold{i}}}_{K_0}\,\,+\,\, t \cdot \underbrace{\sum_{J_{\bold{k},\bold{q}} \subset \Lambda_{N}^{d}\,,\, k \leq \bar{k}} V_{J_{\bold{k},\bold{q}}}}_{K_{I}}\,, 
\end{equation}
where:

\begin{enumerate}
\item[i)]
$J_{\bold{k},\bold{q}}\equiv J_{k_1, \dots, k_d\,;\,q_1,\dots,q_d}$ denotes the rectangle in $\Lambda_N^{d}$ with sides of lengths $k_1, k_2, \dots, k_d$, respectively, whose $2^d$  corners are the sites given by $(q_1 + \varepsilon_1 k_1, \dots, q_d + \varepsilon_d k_d)$,\,\, $\varepsilon_j = 0 \text{ or } 1,$ for $j=1,\dots, d$.
(Notice that $\Lambda_{N}^{d} \equiv J_{\bold{N}-\bold{1},\bold{1}}$, where $\bold{N}-\bold{1}=(N-1, \dots, N-1)$ and $\bold{1}=(1, \dots, 1)$\,).
\item[ii)]
$k\equiv |\bold{k}|$ denotes the circumference ($=$ sum of the side lengths) of a rectangle $J_{\bold{k},\bold{q}}$, i.e.,
\begin{equation}
 k\equiv |\bold{k}|:=\sum_{i=1}^{d}k_i\,.
\end{equation}
\item[iii)] The range of the interaction potentials, namely the integer $\bar{k} < \infty$ with the property that $\vert \mathbf{k}\vert \leq \bar{k}\,, \,\forall$ rectangles  $J_{\mathbf{k}, \mathbf{q}}$ appearing in (\ref{Hamiltonian}), is arbitrary, but fixed, and \textit{$N-$independent}.
\item[iv)]   $V_{J_{\bold{k},\bold{q}}}$ is a symmetric matrix on $\mathcal{H}^{(N)}$ with the property that 
\begin{equation}\label{potential}
V_{J_{\bold{k},\bold{q}}} \,\, \text{  acts as the identity on  }\,\, \bigotimes_{\bold{j}\in \Lambda_{N}^{d}\, , \, \bold{j}\notin J_{\bold{k},\bold{q}}}  \,\mathcal{H}_{\bold{j}}\,,\quad \text{and}\quad \|V_{J_{\bold{k},\bold{q}}}\|\leq 1\,, \end{equation}
for all \,$\mathbf{k}, \mathbf{q}$\,, with $\vert \mathbf{k} \vert \leq \bar{k}< \infty$, as in iii), (and $V_{J_{\bold{k},\bold{q}}} =0$ whenever $\vert \mathbf{k} \vert > \bar{k}$). The rectangle $J_{\bold{k},\bold{q}}$ is called the ``support'' of $V_{J_{\bold{k},\bold{q}}}$.
\item[v)] $t \in \mathbb{R}$ is a coupling constant independent of $N$.
\end{enumerate}

\subsection{Main result}\label{main}
Our main result is the following theorem proven in Section \ref{proofs} (see Theorem \ref{main-res}).\\

\noindent
{\bf{Theorem.}}
\textit{Under the assumption that (\ref{gaps}) and (\ref{potential}) hold, for an arbitrary, but fixed finite range $\bar{k}<\infty$, 
the Hamiltonian $K_{N}(t)$ defined in (\ref{Hamiltonian}) has the following properties:\\
There exists some $t_d  > 0$ independent of $N$ such that, for any coupling constant $t\in \mathbb{R}$ with $\vert t \vert < t_d$, and for all $N < \infty$,
\begin{enumerate}
\item[(i)]{ $K_{N}(t)$ has a unique ground-state; and}
\item[(ii)]{ the energy spectrum of $K_{N}(t)$ has a strictly positive gap, $\Delta_{N}(t) \geq \frac{1}{2}$, above the ground-state energy.}\\
\end{enumerate}
}
\noindent Results similar to this theorem have appeared in the literature; see, e.g., \cite{DS}. The main novelty of our paper is the method of proof. \\

We define
\begin{equation}\label{vacuum-proj}
P_{vac}:=\bigotimes_{\bold{i}\in \Lambda^{d}_N} P_{\Omega_{\bold{i}}}\,,
\end{equation}
which is the orthogonal projection onto the ground-state subspace of the unperturbed operator \mbox{$K_{0,N}\equiv K_{N}(t=0)= \sum_{\bold{i}\in \Lambda^{d}_N}H_{\bold{i}}\,.$}
We will construct an anti-symmetric matrix $S_{N}(t)=-S_{N}(t)^{*}$ acting on $\mathcal{H}^{(N)}$ (so that 
exp$\big[\pm S_{N}(t)\big]$ are unitary matrices), with the property that, after conjugation, the operator
\begin{equation}\label{conjug}
e^{S_{N}(t)}K_{N}(t)e^{-S_{N}(t)}=: \widetilde{K}_{N}(t)
\end{equation}
is \textit{``block-diagonal''} with respect to the pair \big($P_{vac}$, $P_{vac}^{\perp}:= \charf - P_{vac}\big)$ of projections, in the sense that $P_{vac}$ projects onto the ground-state of $\widetilde{K}_{N}(t)$,
\begin{equation}\label{block-diag}
\widetilde{K}_{N}(t)= P_{vac} \widetilde{K}_{N}(t) P_{vac} + P_{vac}^{\perp} \widetilde{K}_{N}(t) P_{vac}^{\perp}\,,
\end{equation}
and 
\begin{equation}\label{gapss}
\text{infspec}\left(P_{vac}^{\perp}\widetilde{K}_{N}(t) P_{vac}^{\perp} \upharpoonright_{P_{vac}^{\perp} \mathcal{H}^{(N)}}\right)
\geq \text{infspec} \left(P_{vac} \widetilde{K}_{N}(t) P_{vac} \upharpoonright_{P_{vac}\mathcal{H}^{(N)}}\right) + \Delta_{N}(t)\,,
\end{equation}
with $\Delta_{N}(t) \geq \frac{1}{2}$, for $\vert t \vert < t_d$, \textit{uniformly} in $N$.
\\

The Hamiltonian we will study in the following has the special form
\begin{equation}\label{initial-ham}
K_{N}(t):=\sum_{\bold{i}\in \Lambda^{(d)}_N}H_{\bold{i}}+t\sum_{j=1}^{d}\sum_{q_1=1}^{N}\dots \sum_{q_j=1}^{N-1} \dots \sum_{q_d=1}^{N}\,V_{J_{\bold{1}_j,\bold{q}}}
\end{equation}
where 
\begin{equation}\label{def-1_j}
(\bold{1}_j,\bold{q}):=(0,\dots, k_j=1, \dots,0\,;\,q_1,\dots,q_d)\,,
\end{equation}
i.e., the range of the interaction potentials is $\bar{k}=1$.
We could study potentials with an arbitrary finite range. But, in order to keep our exposition as transparent as possible, we restrict our attention to nearest neighbor ``hopping terms''. For simplicity, we also assume that the coupling constant is positive, i.e., $t>0$. 
\\

\noindent
{\bf{Organization of the paper.}} 
In Sect.  \ref{method}, we explain the formal aspects of our construction. In Sect. \ref{minrec}, we introduce the notion of \textit{``minimal rectangles''} that will play an important role in our analysis. In Sect. \ref{trasf-ham}, we describe the \emph{local} (so-called Lie-Schwinger) conjugations of the Hamiltonian associated with minimal rectangles. Next, in Sect. \ref{algo}, we introduce an algorithm that describes the flow of effective interactions determined by the iterative conjugations of the Hamiltonian used to  block-diagonalize it. Moreover, we outline the new features and the complications of our strategy arising in dimensions $d\geq 2$, as compared to the one used in \cite{FP} for chains. 

\noindent
In Sect. \ref{tree} we describe a scheme of re-expansions of collections of effective interaction potentials and a method to derive estimates on the norms of these operators that involve keeping track of paths of connected rectangles.

\noindent
In Sect. \ref{gap-section} we recall how to provide a lower bound on the spectral gap $\Delta_{N}(t)$, for sufficiently small values of the coupling constant $t$, following the same procedure as in \cite{FP}.

\noindent
In Sect. \ref{proofs} the proof of convergence of our construction of the operator $S_{N}(t)$ is presented, with a few 
technicalities deferred to Appendix \ref{appendix}. Theorem \ref{th-norms} is the core result in our proof of convergence, 
enabling us to control the norms of the effective interactions by using a composite  strategy combining different mechanisms, 
depending on the regime of the growth processes of rectangles; see Sect. \ref{algo}. 
>From Theorem \ref{th-norms}, the final result of this paper, Theorem \ref{main-res}, follows.
\\

\noindent
{\bf{Notation}}\\

\noindent
1) For chains, i.e., $d=1$, the rectangles $J_{\bold{k},\bold{q}}$ coincide with the connected one-dimensional graphs, 
$I_{k,q}$, $k \in \mathbb{N}$, used in \cite{FP},  with $k$ edges connecting  the $k+1$ vertices $q,1+q,\dots, k+q$, that can also be seen as  ``intervals'' of length $k$ whose left end-point coincides with $q$. 
\\

\noindent
2) We use the same symbol for the operator $O_{\bold{j}}$ acting on $\mathcal{H}_{\bold{j}}$ and the corresponding operator $$ O_{\bold{j}} \otimes \charf_{J_{\bold{k},\bold{q}}\setminus \{\bold{j}\}}$$ acting on $ \bigotimes_{\bold{i}\in J_{\bold{k},\bold{q}} }\mathcal{H}_{\bold{i}}$, for any $ \bold{j}\in J_{\bold{k},\bold{q}}$. Similarly, with a slight abuse of notation, we do not make a distinction between an operator
$O_{J_{\bold{l}, \bold{i}}}$ acting on $\mathcal{H}_{J_{\bold{l}, \bold{i}}}:=\bigotimes_{\bold{j}\in J_{\bold{l}, \bold{i}}} \mathcal{H}_{\bold{j}}$ and the corresponding operator acting on the whole Hilbert space $\mathcal{H}^{(N)}$ which is obtained out of $O_{J_{\bold{l}, \bold{i}}}$ by tensoring by the identity matrix operator on all the remaining sites.
\\

\noindent
3) With the symbol ``$\subset$" we denote strict inclusion, otherwise we use the symbol  ``$\subseteq$". 
\\

\noindent
4) The multiplicative constant implicit in the symbol $\mathcal{O}(\cdot)$ can depend on the spatial dimension $d$.
\\


{\bf{Acknowledgements.}}
A.P.  thanks  the Pauli Center, Z\"urich, for hospitality in Spring 2017 when this project got started, and also acknowledges the MIUR Excellence Department Project awarded to the Department of Mathematics, University of Rome Tor Vergata, CUP E83C18000100006.
S.D.V. is supported by the Deutsche Forschungsgemeinschaft (DFG) within the Emmy Noether grant CA1850/1-1.

\section{Outline of the proof strategy }\label{method}
The conjugations used to block-diagonalize the Hamiltonian in (\ref{Hamiltonian}) determine a flow of effective
Hamiltonians. These operators are expressed in terms of effective interaction potentials with supports that can be represented 
as connected unions of the rectangles $J_{\bold{k},\bold{i}}$ labelling interaction terms in formula (\ref{Hamiltonian}). 
Whereas for chains, $d= 1$, when starting from a family of intervals (i.e., $I_{k,q}\equiv J_{\bold{k},\bold{q}}$ with 
$\bold{k}=k$ and $1\leq q \leq N-k$), the connected sets associated with the new interaction potentials are again intervals, 
the situation is much more complicated in higher dimensions, $d>1$, because connected sets of arbitrary shape arise in 
the flow. The control of growth processes giving rise to each fixed shape that can appear in our construction is crucial in 
order to accomplish the block-diagonalization of the Hamiltonian. For an arbitrary connected set of a fixed shape, 
the number of growth processes scales factorially in the number of edges of the set. This crude estimate is, however,
not good enough to control the norms of the interaction potentials associated with a given shape, since the 
expected prefactor, $t^{n}$, in the norm of the interaction potential labelled by a connected set of cardinality 
$n$ with a fixed shape arising from all possible growth processes terminating in the given shape cannot compensate the number, $\mathcal{O}(n!)$, of such growth processes when $n$ tends to $\infty$; (here $t$ is the coupling constant). Hence in our estimates we cannot  simply count all growth processes giving rise to each fixed shape since some of them are in fact forbidden by the ordering encoded in the block-diagonalization procedure.
 In this paper we circumvent this problem with a strategy outlined in Sect. \ref{algo}, which involves the notion of \textit{``minimal rectangles''} introduced in the next subsection.
\\

\subsection{Minimal rectangles}\label{minrec}

We recall that the symbol $J_{\bold{k},\bold{q}}\equiv J_{k_1, \dots, k_d\,;\,q_1,\dots,q_d}$ denotes a rectangle in 
$\Lambda_N^{d}$ whose sides have lengths $k_1, k_2, \dots, k_d$, and that $|\bold{k}|$ denotes the sum of these lengths, i.e., $|\bold{k}|:=\sum_{i=1}^{d}k_i$.
The coordinates of the $2^d$ corners of $J_{\bold{k},\bold{q}}$ are $d$-tuples of integers given by either $q_j$ or $ q_j+k_j$ at the $j$-th position, for all $1\leq j \leq d$, with $q_j\leq N-k_j$.

The rectangles $J_{\bold{k},\bold{q}}$ play the role of the intervals $I_{k,q}$ in the one-dimensional case considered in 
\cite{FP}. Similarly to the one-dimensional case, the pairs $(\bold{k},\bold{q})$ label the block-diagonalization steps, 
which are ordered according to the ordering relation \, ``$\succ$'' \, defined as follows.\footnote{For example,  in dimension $d=2$, in order to determine the successor of $(\bold{k},\bold{q})=(k_1, k_2;q_1, q_2)$ we observe that: 

\noindent
a) The elements $(k_1, k_2;q_1+1, q_2)$ and $(k_1, k_2;q_1, q_2+1)$ are both successors of $(k_1, k_2;q_1, q_2)$ but $(k_1, k_2;q_1, q_2+1) \succ (k_1, k_2;q_1+1, q_2)$;  
b) for the elements $(k'_1, k'_2;q'_1, q'_2)$, $(k''_1, k''_2;q''_1, q''_2)$ such that $k'_1+ k'_2=k''_1+ k''_2=k_1+ k_2$, 
if $k_1'>k_1''$ then  $(k''_1, k''_2;q''_1, q''_2)\succ (k'_1, k'_2;q'_1, q'_2)$.}
\begin{equation}\label{ordering}
(\bold{k}',\bold{q}') \succ (\bold{k},\bold{q}) \qquad \text{iff }
\end{equation} 
\begin{itemize}
\item{ $\sum_{j=1}^{d} k'_{j}> \sum_{j=1}^{d}k_j$\,;} 
\item{or, if $\sum_{j=1}^{d} k'_{j}=\sum_{j=1}^{d}k_j$, $k^{'}_j<k_j$, for some $1\leq j\leq d$\,, with $k^{'}_l=k_l$, $\forall  l < j$\,;}
\item{or, if $k'_l=k_l$, for all $l$, \,$q'_j>q_j$\,, for some $1 \leq j \leq d$,\, with $q'_l=q_l$, $\forall l > j$\,.}
\end{itemize}
As will become clear from our description of the block-diagonalization flow in the next section, the ordering amongst rectangles must ensure that rectangles with larger circumference $|\bold{k}|$ succeed those of smaller circumference. With this requirement fulfilled, the  ordering chosen here is convenient;  but it is definitely not the only possible ordering.

With the symbols $(\bold{k},\bold{q})_{+j}$ and $(\bold{k},\bold{q})_{-j}$ we denote the $j$-th successor  and the $j$-th predecessor of $(\bold{k},\bold{q})$, respectively, in the ordering introduced above.  The initial step is $(\bold{0},\bold{N})$, because the ``potentials" associated with the degenerate rectangles consisting of a single point are the on-site terms, $H_{\bold{i}}$, which are already block-diagonal with respect to the pair of projections defined in 
(\ref{pro-minus-multi})-(\ref{pro-plus-multi})\,, below. The final step is $(\bold{N-1},\bold{1})$, where $\bold{N-1}=(N-1,\dots,N-1)$ and $\bold{1}=(1,\dots,1)$.

\begin{defn}
Given an arbitrary rectangle $J_{\bold{k},\bold{q}}$ of sites in $\Lambda_N ^{d}$, we define
\begin{equation}
\mathcal{H}_{J_{\bold{k},\bold{q}}}:= \bigotimes_{\bold{i}\in J_{\bold{k},\bold{q}}}\mathcal{H}_{\bold{i}}\,.
\end{equation}
\end{defn}
\begin{defn}
Consider two rectangles, $J_{\bold{k},\bold{q}}$ and $J_{\bold{k}',\bold{q}'}$, with nonempty intersection. The  \emph{minimal rectangle} associated with $J_{\bold{k},\bold{q}}\cup J_{\bold{k}',\bold{q}'}$ is defined to be the \textit{smallest} rectangle
containing  $J_{\bold{k},\bold{q}}$ and $J_{\bold{k}',\bold{q}'}$. Note that its corners  are the $2^d$ numbers with either
\begin{equation}
 \min\{q_j,q'_j\}\,, \quad\text{or}\quad \max\{q_j+k_j,q'_j+k'_j\}\,
\end{equation}
at the $j$-th position. The minimal rectangle associated with $J_{\bold{k},\bold{q}}$ and $J_{\bold{k}',\bold{q}'}$ is denoted 
by 
\begin{equation}
[J_{\bold{k},\bold{q}}\cup J_{\bold{k}',\bold{q}'}]\,.
\end{equation}
\end{defn}
\begin{defn}
Let $J_{\bold{k},\bold{q}}\subset J_{\bold{l},\bold{i}}$. We define a family, 
$\mathcal{G}^{(\bold{k},\bold{q})}_{J_{\bold{l},\bold{i}}}$, of rectangles by
\begin{equation}\label{def-Gcall}
\mathcal{G}^{(\bold{k},\bold{q})}_{J_{\bold{l},\bold{i}}}:=\Big\{ \, J_{\bold{k}',\bold{q}'}\,\vert \,J_{\bold{k}',\bold{q}'}\neq J_{\bold{l},\bold{i}}\quad \text{and} \quad [J_{\bold{k},\bold{q}}\cup J_{\bold{k}',\bold{q}'}]=J_{\bold{l},\bold{i}}\,\,\Big\}\,.
\end{equation}
\end{defn}

\subsection{Effective Hamiltonians}\label{trasf-ham}
Each conjugation step in the block-diagonalization of the original Hamiltonian is labelled by a rectangle $J_{\mathbf{k}, \mathbf{q}}$ and, consequently, by a pair $(\bold{k},\bold{q})$.  In the effective Hamiltonian arising from a conjugation step, a potential term, $V^{(\bold{k},\bold{q})}_{J_{\bold{l},\bold{i}}}$,  is associated with each rectangle $J_{\bold{l},\bold{i}}$. More precisely, after the conjugation step $(\bold{k},\bold{q})$, the effective Hamiltonian reads
\begin{eqnarray}\label{K-tranf-2}
K_{\Lambda_N^{d}}^{(\bold{k},\bold{q})}
& =&
\sum_{\bold{i}\in \Lambda^{(d)}_N}H_{\bold{i}}+t\sum_{\bold{k}_{(1)}'\,,\,\bold{q}'}V^{(\bold{k},\bold{q})}_{J_{\bold{k}_{(1)}',\bold{q}'}}+t\sum_{\bold{k}_{(2)}'\,,\,\bold{q}'}V^{(\bold{k},\bold{q})}_{J_{\bold{k}_{(2)}',\bold{q}'}}+\dots+t\sum_{\bold{k}_{(|\bold{k}|)}'\,,\,\bold{q}'}V^{(\bold{k},\bold{q})}_{J_{\bold{k}'_{(|\bold{k}|)},\bold{q}'}}\quad\quad\quad \label{K-tranf-1}\nonumber \\
& &
+t\sum_{\bold{k}_{(|\bold{k}|+1)}'\,,\,\bold{q}'}
V^{(\bold{k},\bold{q})}_{J_{\bold{k}'_{(|\bold{k}|+1)},\bold{q}'}}+
\dots+tV^{(\bold{k},\bold{q})}_{J_{\bold{N}-\bold{1},\bold{1}}}\end{eqnarray}
where:
\begin{enumerate}
\item The pairs $(\bold{k}_{(j)}', \bold{q}')$ are used to index all rectangles $J_{\bold{k}',\bold{q}'}$ with 
$|\bold{k}'|=j$;
\item
For a fixed rectangle $J_{\bold{l},\bold{i}}$,  the corresponding potential term may change in each conjugation step of the block-diagonalization procedure, until the step $(\bold{k},\bold{q})= (\bold{l},\bold{i})$ is reached; 
hence $V^{(\bold{k},\bold{q})}_{J_{\bold{l},\bold{i}}}$ is the potential term associated with $J_{\bold{l},\bold{i}}$ 
arising in step $(\bold{k},\bold{q})$ of the block-diagonalization, the superscript $(\bold{k},\bold{q})$ 
keeping track of the changes in the potential term arising in step 
$(\bold{k},\bold{q})$. The operator $V^{(\bold{k},\bold{q})}_{J_{\bold{l},\bold{i}}}$ depends on the coupling constant $t$, but this is not made explicit in our notation; it acts as the identity on the spaces 
$\mathcal{H}_{\bold{j}}$ for $\bold{j} \notin J_{\bold{l},\bold{i}}$. A more precise description of 
how these operators arise in our procedure as well as an outline of the strategy to control their norms are deferred to Section \ref{algo};
\item
For all rectangles $J_{\bold{l},\bold{i}}$ with $(\bold{k},\bold{q})\succ (\bold{l},\bold{i})$, and for the rectangle $J_{\bold{l},\bold{i}} = J_{\bold{k},\bold{q}} $, the associated effective potential$V^{(\bold{k},\bold{q})}_{J_{\bold{l},\bold{i}}}$ is block-diagonal w.r.t. the decomposition of the identity
acting on $\mathcal{H}^{( N)}$ into the sum of projections
\begin{equation}\label{pro-minus-multi}
P^{(-)}_{J_{\bold{l},\bold{i}}}:=\bigotimes_{\bold{j}\in J_{\bold{l},\bold{i}}}  P_{\Omega_{\bold{j}}}\,,
\end{equation}
\begin{equation}\label{pro-plus-multi}
P^{(+)}_{J_{\bold{l},\bold{i}}}:= \Big(\bigotimes_{\bold{j}\in J_{\bold{l},\bold{i}}} P_{\Omega_{\bold{j}}}\Big)^{\perp}\,.
\end{equation}
\end{enumerate}

\noindent
The effective Hamiltonian $K_{\Lambda_N^{d}}^{(\bold{k},\bold{q})}$ of (\ref{K-tranf-2}) is obtained after the conjugation step labeled by $(\bold{k},\bold{q})$. Starting from
\begin{eqnarray}
K_{\Lambda_N^{d}}^{(\bold{k},\bold{q})_{-1}} &=&\sum_{\bold{i}\in \Lambda^{(d)}_N}H_{\bold{i}}+t\sum_{\bold{k}_{(1)}'\,,\,\bold{q}'}V^{(\bold{k},\bold{q})_{-1}}_{J_{\bold{k}_{(1)}',\bold{q}'}}+t\sum_{\bold{k}_{(2)}'\,,\,\bold{q}'}V^{(\bold{k},\bold{q})_{-1}}_{J_{\bold{k}_{(2)}',\bold{q}'}}+\dots+t\sum_{\bold{k}_{(|\bold{k}|)}'\,,\,\bold{q}'}V^{(\bold{k},\bold{q})_{-1}}_{J_{\bold{k}'_{(|\bold{k}|)},\bold{q}'}}\label{K-1}\\
& &+t\sum_{\bold{k}_{(|\bold{k}|+1)}'\,,\,\bold{q}'}
V^{(\bold{k},\bold{q})_{-1}}_{J_{\bold{k}'_{(|\bold{k}|+1)},\bold{q}'}}+\dots+
tV^{(\bold{k},\bold{q})_{-1}}_{J_{\bold{N}-\bold{1},\bold{1}}}\label{K-2}
\end{eqnarray}
the conjugation step labelled by $(\mathbf{k}, \mathbf{q})$ is given by
\begin{equation}\label{def-K-0}
e^{S_{J_{\bold{k},\bold{q}}}}\,  K_{\Lambda_N^{d}}^{(\bold{k},\bold{q})_{-1}}\,e^{-S_{J_{\bold{k},\bold{q}}}}=:K_{\Lambda_N^{d}}^{(\bold{k},\bold{q})}\,,
\end{equation}
where the anti-symmetric matrix $S_{J_{\bold{k},\bold{q}}}$ is chosen in such a way that the interaction potential $V^{(\bold{k},\bold{q}}_{J_{\bold{k}, \bold{q}}}$ is block-diagonal; see Section \ref{gap-section}.
More precisely, following the Lie-Schwinger procedure, $S_{J_{\bold{k},\bold{q}}}$ is built so as to block-diagonalize  the \emph{local} operator given by the sum of all terms in $K_{\Lambda_N^{d}}^{(\bold{k},\bold{q})_{-1}}$ whose support is contained in $J_{\bold{k}, \bold{q}}$. In other words, $S_{J_{\bold{k},\bold{q}}}$ is chosen in such a way that the conjugation in
(\ref{def-K-0}) renders the operator
\begin{equation}
G_{J_{\bold{k},\bold{q}}}+V^{(\bold{k},\bold{q})_{-1}}_{J_{\bold{k}, \bold{q}}}\,,
\end{equation}
block-diagonal, where
\begin{equation}\label{def-G}
G_{J_{\bold{k},\bold{q}}}:=\sum_{\bold{i}\subset J_{\bold{k},\bold{q}} }H_{\bold{i}}+t\sum_{J_{\bold{k}_{(1)}',\bold{q}'}\subset J_{\bold{k},\bold{q}}} V^{(\bold{k},\bold{q})_{-1}}_{J_{\bold{k}_{(1)}',\bold{q}'}}+\dots+t\sum_{J_{\bold{k}'_{(|\bold{k}|-1)},\bold{q}' }\subset J_{\bold{k},\bold{q}}}V^{(\bold{k},\bold{q})_{-1}}_{J_{\bold{k}'_{(|\bold{k}|-1)},\bold{q}'}}\,.
\end{equation}
Here ``block-diagonalization'' refers to the projections $P^{(-)}_{J_{\bold{k},\bold{q}}}$ and $P^{(+)}_{J_{\bold{k},\bold{q}}}$ corresponding to the decomposition of the Hilbert space $\bigotimes_{\bold{i}\in J_{\bold{k},\bold{q}}}\mathcal{H}_{\bold{i}}$  into vacuum subspace and its orthogonal complement, respectively. The operator $G_{J_{\bold{k},\bold{q}}}$
plays the role of the "unperturbed" operator, since it is already block-diagonal w.r.t. the decomposition of  the identity 
$$ \charf = P^{(+)}_{J_{\bold{k},\bold{q}}} + P^{(-)}_{J_{\bold{k},\bold{q}}}\,,$$
i.e., 
\begin{equation}
G_{J_{\bold{k},\bold{q}}}=P^{(+)}_{J_{\bold{k},\bold{q}}}G_{J_{\bold{k},\bold{q}}}P^{(+)}_{J_{\bold{k},\bold{q}}}+P^{(-)}_{J_{\bold{k},\bold{q}}}G_{J_{\bold{k},\bold{q}}}P^{(-)}_{J_{\bold{k},\bold{q}}}\,.
\end{equation}
The construction outlined here works, because one can show inductively that the energy gap in the spectrum of the 
Hamiltonian $G_{J_{\bold{k},\bold{q}}}$ above its ground-state eigenvalue is bounded away from zero, \textit{uniformly} in the 
size of the rectangle $J_{\bold{k},\bold{q}}$, when a suitable upper bound on the operators norms of the interaction potentials
is imposed.  The control of this gap (see Section \ref{gap-section}) relies on the fact that all the effective potentials 
appearing in  $G_{J_{\bold{k},\bold{q}}}$ have been block-diagonalized already in the previous steps.\\
These properties of the operator $G_{J_{\bold{k},\bold{q}}}$, combined with bounds on the norms of the effective potentials obtained at the previous conjugation step, enable us to construct the anti-symmetric matrix $S_{J_{\bold{k},\bold{q}}}$ used at the next conjugation step and control the norms of the effective potentials obtained after conjugation with
$\text{exp}[S_{J_{\bold{k},\bold{q}}}]$. This is described in more detail in Section \ref{algo}.

\subsection{The algorithm and the different regimes in the growth processes of rectangles}\label{algo}

Our strategy to control the norms of the effective potentials $V^{(\bold{k},\bold{q})}_{J_{\bold{r},\bold{i}}}$ is based on the following key ideas, which will give rise to a concrete algorithm.
\begin{itemize}
\item[I)] The number of shapes of connected sets of lattice sites arising in our construction is limited by making use of 
``minimal rectangles'' in such a way that, instead of two connected sets, only the minimal rectangle containing them will 
be recorded; (i.e., the rectangle with the property that any rectangle of smaller size cannot contain the union of those sets). 
Only keeping track of minimal rectangles reduces the combinatorial divergence, because the number of rectangles 
with a given circumference $k(:=\sum_{i=1}^d k_i)$ containing a specified site of the lattice grows polynomially in $k$, 
namely like $\mathcal{O}(k^{d-1})$. We then lump together all effective potential terms whose support is contained in
a given rectangle in such a way that no rectangle of smaller size can contain it. The sum of the norms of these terms 
is expected to be bounded above by $\mathcal{O}(t^{c\cdot k})$, where $c$ is a universal constant.
\item[II)] We will exploit some subtle mechanisms to identify and control the growth processes allowed by the algorithm introduced below. Depending on the relation between the size, $k$,  of $J_{\bold{k},\bold{q}}$  and the size, $r$, of $J_{\bold{r},\bold{i}}$, we will distinguish three different regimes for the growth processes that may give rise to 
the term $V^{(\bold{k},\bold{q})}_{J_{\bold{r},\bold{i} }}$ in (\ref{construction-conn}) below.
\end{itemize}

As implicitly indicated in the expression (\ref{K-1})-(\ref{K-2}) for the effective Hamiltonian $K_{\Lambda_N^{d}}^{(\bold{k},\bold{q})_{-1}}$, the potentials must be re-combined properly after each conjugation step $(\mathbf{k}, \mathbf{q})$ so as to determine a well defined flow of operators, $V^{(\bold{k},\bold{q})}_{J_{\bold{r},\bold{i}}}$, for every fixed support $J_{\bold{r},\bold{i}}$. This flow is obtained with the help of a specific algorithm described in Definition \ref{def-interactions-multi}, below. In Theorem \ref{th-potentials-multi}, we check that our algorithm is consistent with the conjugation in (\ref{def-K-0}). This amounts to showing that the r-h-s in (\ref{def-K-0}) has the form given in (\ref{K-1})-(\ref{K-2}), with $(\bold{k},\bold{q})_{-1}$ replaced by $(\bold{k},\bold{q})$ and effective potentials $V^{(\bold{k},\bold{q})}_{J_{\bold{l},\bold{i}}}$ as defined in 
Definition \ref{def-interactions-multi} formulated next.

\noindent
The algorithm is supposed to enable us to iteratively determine effective potentials $V^{(\bold{k},\bold{q})}_{J_{\bold{r},\bold{i}}}$
in terms of the potentials obtained at the previous step $(\bold{k},\bold{q})_{-1}$, starting from 
\begin{equation}
V_{J_{\bold{0},\bold{i}}}^{(\bold{0},\bold{N})}:= H_{\bold{i}}\,,\qquad
V_{J_{\bold{1}_j,\bold{q}}}^{(\bold{0},\bold{N})}:= V_{J_{\bold{1}_j,\bold{q}}}\,,\quad \text{and }\quad
V_{J_{\bold{k},\bold{i}}}^{(\bold{0},\bold{N})} =0\,, \,\,\, \text{for}\,\, |\bold{k}|\geq 2\,.
\end{equation}
\begin{defn}\label{def-interactions-multi}
Assuming that, at fixed $(\bold{k},\bold{q})_{-1}$ with $(\bold{k},\bold{q})_{-1} \succ (\bold{0},\bold{N})$, for any $\bold{r},\bold{i}$ the operators $V^{(\bold{k},\bold{q})_{-1}}_{J_{\bold{r},\bold{i}}}$ and $S_{J_{\bold{k},\bold{q}}}$ (defined as in (\ref{formula-S}), (\ref{formula-Sj}))  are well defined, or assuming $(\bold{k},\bold{q})=(\bold{1}_1,\bold{1})$ (where $\bold{1}_1=(1,0,\dots,0)$ and $\bold{1}=(1,\dots,1)$, respectively) and $S_{J_{\bold{1}_1,\bold{1}}}$ well defined, then we define:
\begin{itemize}
\item[a)]
if  $J_{\bold{k},\bold{q}} \nsubset J_{\bold{r},\bold{i}}$, 
\begin{equation}\label{a}
V^{(\bold{k},\bold{q})}_{J_{\bold{r},\bold{i}}}:=V^{(\bold{k},\bold{q})_{-1}}_{J_{\bold{r},\bold{i}}}\,;
\end{equation}
\item[b)]
if $J_{\bold{r},\bold{i}}=J_{\bold{k},\bold{q}}$,
\begin{equation}\label{L-S-series}
V^{(\bold{k},\bold{q})}_{J_{\bold{r},\bold{i}}}:= \sum_{j=1}^{\infty}t^{j-1}(V^{(\bold{k},\bold{q})_{-1}}_{J_{\bold{r},\bold{i}}})^{diag}_j \,
\end{equation}
where $(V^{(\bold{k},\bold{q})_{-1}}_{J_{\bold{r},\bold{i}}})^{diag}_j$ is defined like in (\ref{formula-v_j}), and $diag$ means diagonal part w.r.t. to the projections $P^{(-)}_{J_{\bold{r},\bold{i}}}$ and $P^{(+)}_{J_{\bold{r},\bold{i}}}$;
\item[c)]
if $J_{\bold{k},\bold{q}}\subset J_{\bold{r},\bold{i}}$,
\begin{eqnarray}\label{construction-conn}
V^{(\bold{k},\bold{q})}_{J_{\bold{r},\bold{i}}} &:= & e^{S_{J_{\bold{k},\bold{q}}}}\,V^{(\bold{k},\bold{q})_{-1}}_{J_{\bold{r},\bold{i}}}\,e^{-S_{J_{\bold{k},\bold{q}}}}\,+\sum_{J_{\bold{k}',\bold{q}'}\in \mathcal{G}^{(\bold{k},\bold{q})}_{J_{\bold{r},\bold{i}}}}\,\sum_{n=1}^{\infty}\frac{1}{n!}\,ad^{n}S_{J_{\bold{k},\bold{q}}}(V^{(\bold{k},\bold{q})_{-1}}_{J_{\bold{k}',\bold{q}'}})\,,\, \label{main-def-V-multi} 
\end{eqnarray}
where $ad$ is defined in (\ref{def-ad-1})-(\ref{def-ad-2}).
We observe that the set $\mathcal{G}^{(\bold{k},\bold{q})}_{J_{\bold{r},\bold{i}}}$ (see (\ref{def-Gcall})) is not empty only if the rectangle $J_{\bold{k},\bold{q}}$ has a nonempty intersection with the boundary of the rectangle $J_{\bold{r},\bold{i}}$.
\end{itemize}
\end{defn}

The rationale motivating the recombination of terms described in Definition \ref{def-interactions-multi} is explained in Section \ref{gap-section}.  Here a remark on item c) of Definition \ref{def-interactions-multi} may be helpful in order to understand the key ideas used to control the operator norms of the effective potentials.
\begin{rem}

\noindent
The sum on the r-h-s in (\ref{construction-conn}) accounts for all contributions to the term 
$V^{(\bold{k},\bold{q})}_{J_{\bold{r},\bold{i}}}$ with support $J_{\bold{r},\bold{i}}$ that correspond to ``growth processes'' of 
rectangles, i.e., to processes where the union of a rectangle $J_{\bold{k}',\bold{q}'}\neq J_{\bold{r},\bold{i}}$ and of the fixed 
rectangle $J_{\bold{k},\bold{q}}$ labelling the conjugation step in the block-diagonalization is a set with the property that 
$J_{\bold{r},\bold{i}}$ is the minimal rectangle associated to it, i.e., such that 
$[J_{\bold{k}',\bold{q}'}\cup J_{\bold{k},\bold{q}}]\equiv J_{\bold{r},\bold{i}}$. 
\end{rem}
To control the operator norms of the effective potentials, we begin by observing that, by construction, the potential $V^{(\bold{k},\bold{q})}_{J_{\bold{r},\bold{i}}}$ does not change anymore whenever $(\bold{k},\bold{q})\succ (\bold{r},\bold{i})$.  
Using this observation, we will prove by induction that, for every pair $(\mathbf{r},\mathbf{i})$, an upper bound 
of the following form
\begin{equation}
 \|V^{(\bold{k},\bold{q})}_{J_{\bold{r},\bold{i}}}\|\leq C_{j} \frac{t^{\frac{r-1}{3}}}{r^{\,\rho_{j}}}\quad,\quad j=1,2,3\,, \label{strat}
\end{equation} 
holds true, at all steps $(\bold{k},\bold{q})$ up to step $(\bold{r},\bold{i})$ (included), where $C_j$ and the exponent $\rho_{j}\equiv \rho_j(d)>0$ ($d$ being the space dimension) depend on the regime $\mathfrak{R}j$ introduced below, 
and the different regimes, $\mathfrak{R}1, \mathfrak{R}2$, and $\mathfrak{R}3$,  depend on 
the relative magnitude of the circumferences $k=|\bold{k}|$ and $r=|\bold{r}|$. 

We recall that, for quantum chains, control of the norms relies on a feature of formula (\ref{construction-conn}) that holds only 
in dimension $d=1$: An interval can only grow at the two end-points, hence at a number of vertices independent of the size of the interval. But in higher dimensions, $d>1$, the number of terms in the sum in  formula (\ref{main-def-V-multi}) labelled by rectangles,  $J_{\bold{k}',\bold{q}'}$, that intersect the rectangle $J_{\bold{k},\bold{q}}$ only at the boundary  grows like a positive power of $r$, (depending on the dimension $d$). This motivates the introduction of three different regimes, 
$\mathfrak{R}1,\mathfrak{R}2$, and $\mathfrak{R}3$, enabling us to exploit a different mechanism to estimate the number of terms in each of the regimes, as outlined below; see also Figure 1.

\begin{itemize}
\item[$\mathfrak{R}1$)] The first regime deals with rectangles labelled by $(\bold{k},\bold{q})$ that are ``small'' as compared
 to the rectangle labelled by $(\bold{r},\bold{i})$, namely with pairs $(\bold{k}, \bold{q})$ such that $k \leq \lfloor r^{\frac{1}{4}} \rfloor $. In order to establish the desired estimate (\ref{strat}), we iterate the re-expansion of the potential 
 $V^{(\bold{k},\bold{q})}_{J_{\bold{r},\bold{i}}}$ 
by applying formulae (\ref{main-def-V-multi}) and (\ref{a}).  As a consequence, each potential term resulting from the re-expansion can then be associated with a connected sequence of rectangles $J_{\bold{k}'',\bold{q}''}$ labelling the operators $S_{\bold{k}'',\bold{q}''}$, plus one labelling one of the potentials appearing in the Hamiltonian of definition (\ref{Hamiltonian}) or a potential of the type $V^{(\bold{k}',\bold{q}')}_{J_{\bold{k}',\bold{q}'}}$ (where $k' \leq \lfloor r^{\frac{1}{4}} \rfloor $), with the property that $J_{\bold{r},\bold{i}}$ is the minimal rectangle associated to this sequence. Roughly speaking, the result then holds for the following reasons: 

\noindent
1) At least $\mathcal{O}(r/\lfloor r^{\frac{1}{4}} \rfloor)$ rectangles $J_{\bold{k}'',\bold{q}''}$  are present in each connected set, and all the corresponding operators $S_{\bold{k}'',\bold{q}''}$ have norms of order $\vert t \vert \cdot \|V_{\bold{k}'',\bold{q}''}^{(\bold{k}'',\bold{q}'')_{-1}}\|$; apart from the resulting product of norms $\|V_{\bold{k}'',\bold{q}''}^{(\bold{k}'',\bold{q}'')_{-1}}\|$ which is also crucial in the argument, it is important that a total factor $|t|^{\mathcal{O}(r/\lfloor r^{\frac{1}{4}} \rfloor)}$  or smaller is gained from the re-expansion (due to the constraint $k\leq \lfloor r^{\frac{1}{4}} \rfloor $ that holds in this regime).

\noindent
2) 
 Notice that the rectangles contained in the considered connected set 
are ordered according to $\succ$, and, consequently,  only one growth process can yield each such a set. Due to this observation, the number of connected sets of rectangles resulting from the re-expansion, when each connected set is properly weighted in accordance with the inductive hypothesis on the norms of the potentials  $V^{(\bold{k}'',\bold{q}'')_{-1}}_{J_{\bold{k}'',\bold{q}''}}$,  provides an upper bound to $\| V^{(\bold{k},\bold{q})}_{J_{\bold{r},\bold{i}}}\|$. In fact, for $|t|$ small enough but independent of $N$,  this weighted number yields the sought bound (\ref{strat})  for $\|V^{(\bold{k},\bold{q})}_{J_{\bold{r},\bold{i}}}\|$.  \\
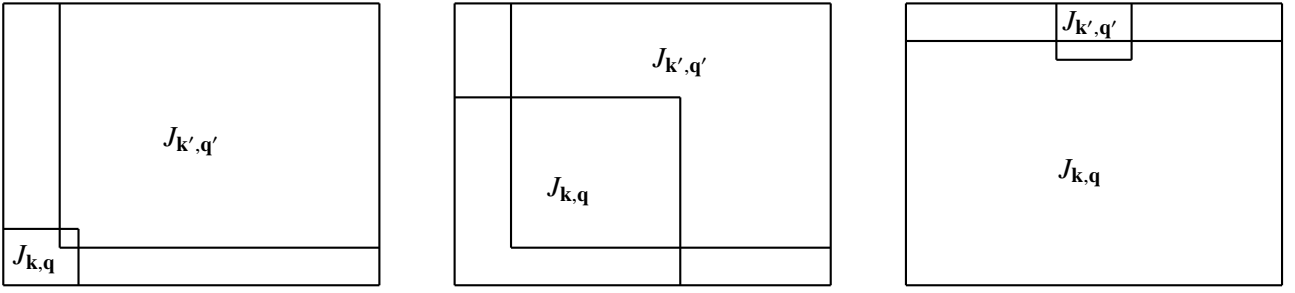
\begin{figure}
\begin{tikzpicture}[scale=0.5]

     \draw (5,4.5) node[anchor=north]  {\small{$J_{\bold{k}',\bold{q}'}$}};
     \draw (1.7,0.7) node[anchor=east]  {\small{$J_{\bold{k},\bold{q}}$}};

    \draw[thick] (0,0) -- +(0,7.5);
    \draw[thick] (0,0) -- +(10,0);
    \draw[thick] (10,0) -- +(0,7.5);
    \draw[thick] (0,7.5) -- +(10,0);
    \draw[thick] (2,0) -- +(0,1.5);
    \draw[thick] (0,1.5) -- +(2,0);
    \draw[thick] (1.5,1) -- +(0,6.5);
    \draw[thick] (1.5,1) -- +(8.5,0);

  \draw (18,6.6) node[anchor=north]  {\small{$J_{\bold{k}',\bold{q}'}$}};
     \draw (15.9, 2.5) node[anchor=east]  {\small{$J_{\bold{k},\bold{q}}$}};

  \draw[thick] (12,0) -- +(0,7.5);
    \draw[thick] (12,0) -- +(10,0);
    \draw[thick] (22,0) -- +(0,7.5);
    \draw[thick] (12,7.5) -- +(10,0);
    \draw[thick] (18,0) -- +(0,5);
    \draw[thick] (12,5) -- +(6,0);
    \draw[thick] (13.5,1) -- +(0,6.5);
    \draw[thick] (13.5,1) -- +(8.5,0);
    
    \draw (28.9,7.6) node[anchor=north]  {\small{$J_{\bold{k}',\bold{q}'}$}};
     \draw (29.5,3) node[anchor=east]  {\small{$J_{\bold{k},\bold{q}}$}};

  \draw[thick] (24,0) -- +(0,7.5);
    \draw[thick] (24,0) -- +(10,0);
    \draw[thick] (34,0) -- +(0,7.5);
    \draw[thick] (24,7.5) -- +(10,0);
    \draw[thick] (24,6.5) -- +(10,0);
    \draw[thick] (28,6) -- +(0,1.5);
    \draw[thick] (28,6) -- +(2,0);
    \draw[thick] (30,6) -- +(0,1.5);
       
\end{tikzpicture}
\caption{\footnotesize{Examples of configurations of $\mathfrak{R}1, \mathfrak{R}2, \mathfrak{R}3$, respectively.}}
\end{figure}

\item[$\mathfrak{R}2$)] The second regime is associated with pairs $(\bold{k},\bold{q})$ with the property that
$\lfloor  r^{\frac{1}{4}} \rfloor \leq k\leq r-\lfloor r^{\frac{1}{4}}\rfloor$. In this regime, thanks to the upper bound on $k$, the size of the rectangles $J_{\bold{k}',\bold{q}'}$ in  formula (\ref{main-def-V-multi}) is so large that it is enough to carry out only one re-expansion step and to then use the inductive hypotheses, similarly to the treatment of chains in \cite{FP}. In this regime we use a basic mechanism involving the use of the denominator 
$r^{\,\rho_{2}}$ in the inductive estimate (see (\ref{strat})) of the potential. If $k^{\,\rho_{2}}$ 
and $(r-k)^{\,\rho_{2}}$ are both large as it happens  in this regime, we can still control the polynomially growing number of terms in the sum of formula (\ref{main-def-V-multi}).
\item[$\mathfrak{R}3$)] The third regime is associated with ``large'' rectangles $(\bold{k},\bold{q})$, since 
$ r-\lfloor r^{\frac{1}{4}}\rfloor  \leq k \leq r$. In this regime, we  exploit a mechanism based on \emph{large denominators}. 
This means that we shall collect the contributions in (\ref{main-def-V-multi}) corresponding to potentials 
$V^{(\bold{k},\bold{q})_{-1}}_{J_{\bold{k}',\bold{q}'}}$  that are already block-diagonal and then estimate them in terms of a sum of projections $P^{(+)}_{J_{\bold{k}',\bold{q}'}}$ controlled,   through an induction, by 
the denominator appearing  in the expression of $(S_{J_{\bold{r}, \bold{i}}})_{1}$ (see formula (\ref{formula-Sj})); in the proof by induction  for this regime, we make use of the auxiliary quantities displayed in (\ref{R3-1}).\\
\end{itemize}


\section{Tree structure and paths of rectangles} \label{tree}

\noindent
In order to study regime $\mathfrak{R}1$ we shall re-expand the potentials $V^{(\bold{k},\bold{q})}_{J_{\bold{r},\bold{i}}}$, 
using the recursive Definition  \ref{def-interactions-multi} repeatedly. The method we develop to single out the terms in 
the re-expansion contributing to a certain effective potential, and to then count and \emph{weight} them, is of some
independent interest, irrespective of the crucial role it will play in our analysis of 
regime $\mathfrak{R}1$.  We therefore describe it carefully in this section.

For the purpose of re-expanding  $V^{(\bold{k},\bold{q})}_{J_{\bold{r},\bold{i}}}$, using Definition \ref{def-interactions-multi}, 
we observe that, for $r\gg 1$, case b) of Definition  \ref{def-interactions-multi} can occur only after many 
steps of the re-expansion, because $k \leq \lfloor r^{\frac{1}{4}} \rfloor $ in regime $\mathfrak{R}1$. In order to 
streamline our formulae, we introduce the notation
\begin{equation}\label{Acal}
\sum_{n=1}^{\infty}\frac{1}{n!}\,ad^{n}S_{J_{\bold{k},\bold{q}}}(\dots)=:\mathcal{A}_{J_{\bold{k},\bold{q}}}(\dots)\,.
\end{equation}
Depending on the relative position between $J_{\bold{k},\bold{q}}$ and $J_{\bold{r},\bold{i}}$,  we 
are instructed to use either formula
\begin{eqnarray}\label{exp-formula-bis} 
V^{(\bold{k},\bold{q})}_{J_{\bold{r},\bold{i}}}
& =&V^{(\bold{k},\bold{q})_{-1}}_{J_{\bold{r},\bold{i}}} \\
& &
+\mathcal{A}_{J_{\bold{k},\bold{q}}}(V^{(\bold{k},\bold{q})_{-1}}_{J_{\bold{r},\bold{i}}}) \label{exp-formula-bis-1}\\
& &+\sum_{J_{\bold{k}',\bold{q}'}\in \mathcal{G}^{(\bold{k},\bold{q})}_{J_{\bold{r},\bold{i}}}}\, \mathcal{A}_{J_{\bold{k},\bold{q}}}(V^{(\bold{k},\bold{q})_{-1}}_{J_{\bold{k}',\bold{q}'}})\,\label{exp-formula-bis-2}
\end{eqnarray}
or
\begin{equation}\label{cons}
V^{(\bold{k},\bold{q})}_{J_{\bold{r},\bold{i}}}=V^{(\bold{k},\bold{q})_{-1}}_{J_{\bold{r},\bold{i}}}\,,
\end{equation} 
corresponding to cases c) and a) in Definition  \ref{def-interactions-multi}, respectively.  
We will use formulae a) and c) of Definition  \ref{def-interactions-multi} iteratively for the potentials on the r-h-s  of 
(\ref{exp-formula-bis})-(\ref{exp-formula-bis-2}) and (\ref{cons}) when they apply, if it is the case all the way down to step  $(\bold{0},\bold{N})$, but do not re-expand potentials of the type $V^{(\bold{k}^{\prime\prime},\bold{q}^{\prime\prime})}_{J_{\bold{k}^{\prime\prime},\bold{q}^{\prime\prime}}}$ when they appear (i.e., we stop the re-expansion), which corresponds to case b) of Definition  \ref{def-interactions-multi}.
\\

The strategy can be summarized as consisting of the following steps.
\begin{itemize}
\item Introducing tree diagrams, we show that every contribution, $\mathfrak{b}$, to an effective potential -- where $\mathfrak{b}$ stands for ``branch-operator'', a notion that is motived by the tree structure described below -- of the re-expansion 
resulting from (\ref{exp-formula-bis})-(\ref{exp-formula-bis-2}) and (\ref{cons}) 
is determined by a set, $\mathcal{R}_\mathfrak{b}$, of rectangles that are ordered and whose union is connected.
\item We show that there is an injective map from $\{\mathcal{R}_\mathfrak{b}\}$ to a set, $\{\Gamma_\mathfrak{b}\}$, 
of paths of rectangles with certain properties.
\item By assigning suitable weights to the paths $\Gamma_\mathfrak{b}$ we will be able to derive upper bounds on the 
norms of the contributions $\mathfrak{b}$. This will allow us to estimate the norm 
$\|V^{(\bold{k},\bold{q})}_{J_{\bold{r},\bold{i}}}\|$ by counting (weighted) paths belonging to the set $\{\Gamma_\mathfrak{b}\}$.
\end{itemize}

\subsection{Tree expansion}

In order to find an efficient description (see Definition \ref{def-tree} below)  of the structure of contributions 
emerging from the re-expansion of $V^{(\bold{k},\bold{q})}_{J_{\bold{r},\bold{i}}}$,  
we study the type of terms we get after a few re-expansions steps. For example, if we assume that the relative positions 
of $J_{\bold{k},\bold{q}}$ and $J_{\bold{r},\bold{i}}$ are such that the first re-expansion step is of type c), followed by a re-expansion step of type a), then we get
\begin{eqnarray}
V^{(\bold{k},\bold{q})}_{J_{\bold{r},\bold{i}}}&=&
e^{S_{J_{\bold{k},\bold{q}}}}V^{(\bold{k},\bold{q})_{-2}}_{J_{\bold{r},\bold{i}}}e^{-S_{J_{\bold{k},\bold{q}}}}+\sum_{J_{\bold{k}',\bold{q}'}\in \mathcal{G}^{(\bold{k},\bold{q})}_{J_{\bold{r},\bold{i}}}}\,\mathcal{A}_{J_{\bold{k},\bold{q}}}(V^{(\bold{k},\bold{q})_{-2}}_{J_{\bold{k}',\bold{q}'}})\,,\quad \\
&=&V^{(\bold{k},\bold{q})_{-2}}_{J_{\bold{r},\bold{i}}}+\mathcal{A}_{J_{\bold{k},\bold{q}}}(V^{(\bold{k},\bold{q})_{-2}}_{J_{\bold{r},\bold{i}}})+\sum_{J_{\bold{k}',\bold{q}'}\in \mathcal{G}^{(\bold{k},\bold{q})}_{J_{\bold{r},\bold{i}}}}\,\mathcal{A}_{J_{\bold{k},\bold{q}}}(V^{(\bold{k},\bold{q})_{-2}}_{J_{\bold{k}',\bold{q}'}})\,.\label{example}
\end{eqnarray}
Notice that in (\ref{Acal}), and consequently in (\ref{example}), we interpret the sum over $n$ as a single contribution.
The re-expansion of every potential term alluded to above, iterated down either to the first level where case b) of Definition \ref{def-interactions-multi} applies, or, if this does not happen, to level $(\bold{0},\bold{N})$, can be described using an upside-down tree structure (see the first three levels in Figure 2), following the list of prescriptions described in the next definition.
\begin{figure}
\begin{center}
\begin{tikzpicture}[scale=0.5]

     \draw (6,0.5) node[anchor=north]  {\footnotesize{$(\bold{k},\bold{q})$}};
      \draw (6.3,-3.5) node[anchor=north]  {\footnotesize{$(\bold{k},\bold{q})_{-1}$}};
       \draw (6.3,-7.1) node[anchor=north]  {\footnotesize{$(\bold{k},\bold{q})_{-2}$}};
\draw[thin] node at (0,0)  {\textbullet};
\draw[thin] node at (-3,-4)  {\textbullet};
\draw[thick] node at (1,-4)  {\textbullet};
\draw[thick] node at (2,-4)  {\textbullet};
\draw[thick] node at (3,-4)  {\textbullet};
\draw[thick] node at (-4,-8)  {\textbullet};
\draw[thick] node at (-2,-8)  {\textbullet};
\draw[thick] node at (-2.35,-8)  {\textbullet};
\draw[thick] node at (-2.7,-8)  {\textbullet};
\draw[thick] node at (0.3,-8)  {\textbullet};
\draw[thick] node at (2.8,-8)  {\textbullet};
\draw[thick] node at (3.8,-8)  {\textbullet};
\draw[thick] node at (1.9,-8)  {\textbullet};
\draw[thick] node at (1.55,-8)  {\textbullet};
\draw[thick] node at (1.2,-8)  {\textbullet};

    \draw[thin] (0,0) -- +(-3,-4);
    \draw[thin] (0,0) -- +(3,-4);
    \draw[thin] (0,0) -- +(2,-4);
    \draw[thin] (0,0) -- +(1,-4);
    \draw[thin] (-3,-4) -- +(-1,-4);
    \draw[thin] (-3.0,-4) -- +(1,-4);
     \draw[thin] (-3,-4) -- +(0.65,-4);
      \draw[thin] (-3,-4) -- +(0.3,-4);
     \draw[thin] (1,-4) -- +(0.9,-4);
     \draw[thin] (1,-4) -- +(0.55,-4);
      \draw[thin] (1,-4) -- +(0.2,-4);
        \draw[thin] (1,-4) -- (0.3,-8);
      
      \draw[thin] (2,-4) -- (2.8,-8);
    \draw[thin] (3,-4) -- (3.8,-8);
 
    \end{tikzpicture}
    \caption{\footnotesize{Example of a tree associated with the first two steps of the re-expansion of $V^{(\bold{k},\bold{q})}_{J_{\bold{r},\bold{i}}}$}.}
    \end{center}
\end{figure}
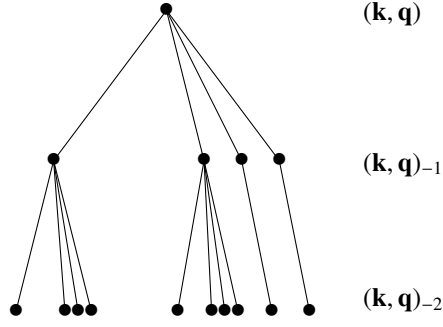

\begin{defn}\label{def-tree}

\noindent
\begin{enumerate}
\item The levels of a tree used to identify the contributions to the re-expansion of a potential $V^{(\bold{k},\bold{q})}_{J_{\bold{r},\bold{i}}}$ are labeled by  $(\bold{k^{\prime},\bold{q}^{\prime}})$, with  $(\bold{k^{\prime},\bold{q}^{\prime}})$ such that $ (\bold{k,\bold{q}}) \succeq (\bold{k^{\prime},\bold{q}^{\prime}})\succeq (\bold{0},\bold{N})$. We say that such a tree is \textit{rooted} at level $(\mathbf{k}, \mathbf{q})$.
\item There is a single vertex at the top of a tree rooted at level $(\bold{k},\bold{q})$; it is labeled by the symbol 
$V^{(\bold{k},\bold{q})}_{J_{\bold{r},\bold{i}}}$ of the potential.
\item The vertices at level $(\bold{k}^{\prime},\bold{q^{\prime}})_{-1}$ of a tree rooted at level $(\mathbf{k}, \mathbf{q})$ are determined by the vertices of the tree at level $(\bold{k}^{\prime},\bold{q^{\prime}})$ in the following way:
Each vertex $\mathfrak{v}\equiv \mathfrak{v}_{V_{J_{\bold{s},{u}}}^{(\bold{k}^{\prime},\bold{q}^{\prime})}}$  at level 
$(\bold{k^{\prime},\bold{q}^{\prime}})$, labeled by $V_{J_{\bold{s},{u}}}^{(\bold{k}^{\prime},\bold{q}^{\prime})}$, is linked 
to two sets of descendants (vertices) at level $(\bold{k^{\prime},\bold{q}^{\prime}})_{-1}$ with the following properties: The two sets of vertices 
are \textit{empty} if $(\bold{s},\bold{u})=(\bold{k}^{\prime},\bold{q}^{\prime})$; otherwise
\begin{itemize}
\item the leftmost set of vertices actually consists of a single vertex, which is labeled by the potential 
$V^{(\bold{k}^{\prime},\bold{q}^{\prime})_{-1}}_{J_{\bold{s},\bold{u}}}$;
\item the rightmost set of vertices is empty if $J_{\bold{k}^{\prime},\bold{q}^{\prime}}\nsubset J_{\bold{s},\bold{u}}$; otherwise it contains a vertex for each element $J_{\mathbf{s}', \mathbf{u}'}$ belonging to
$\mathcal{G}^{(\bold{k}^{\prime},\bold{q}^{\prime})}_{J_{\bold{s},\bold{u}}}\cup \{J_{\bold{s},\bold{u}}\}$, and this vertex is labeled by $V^{(\bold{k}^{\prime},\bold{q}^{\prime})_{-1}}_{J_{\bold{s}^{\prime},\bold{u}^{\prime}}}$.
\end{itemize}

\item Each vertex $\mathfrak{v}$ at level $(\bold{k}^{\prime},\bold{q}^{\prime})$ is connected by an edge to its descendants at level $(\bold{k}^{\prime},\bold{q}^{\prime})_{-1}$. Edges are labelled by rectangles, or carry no label, in the following way:
\begin{itemize}
\item [e-i)] the edge connecting a vertex $\mathfrak{v}$ at level $(\bold{k}^{\prime},\bold{q}^{\prime})$ to its leftmost descendant
 at level $(\bold{k}^{\prime},\bold{q}^{\prime})_{-1}$ has no label.
It stands for the map
$$V^{(\bold{k}^{\prime},\bold{q}^{\prime})}_{J_{\bold{s},\bold{u}}} \rightarrow V^{(\bold{k}^{\prime},\bold{q}^{\prime})_{-1}}_{J_{\bold{s},\bold{u}}}\,, $$ 
where $V^{(\bold{k}^{\prime},\bold{q}^{\prime})}_{J_{\bold{s},\bold{u}}}$ is the potential labelling $\mathfrak{v}$ and $V^{(\bold{k}^{\prime},\bold{q}^{\prime})_{-1}}_{J_{\bold{s},\bold{u}}} $ labels its leftmost descendant at level $(\bold{k}^{\prime},\bold{q}^{\prime})_{-1}$;
\item [e-ii)] each edge $\mathfrak{e}$ connecting the vertex $\mathfrak{v}$ at level $(\bold{k}^{\prime},\bold{q}^{\prime})$ to other descendants at level $(\bold{k}^{\prime},\bold{q}^{\prime})_{-1}$ is labeled by  a rectangle $J_{\bold{k}^{\prime},\bold{q}^{\prime}}$. It stands for the map
$$V^{(\bold{k}^{\prime},\bold{q}^{\prime})}_{J_{\bold{s},\bold{u}}} \rightarrow \mathcal{A}_{J_{\bold{k}^{\prime},\bold{q}^{\prime}}}(V^{(\bold{k}^{\prime},\bold{q}^{\prime})_{-1}}_{J_{\bold{s}^{\prime},\bold{u}^{\prime}}})\,,$$
where $V^{(\bold{k}^{\prime},\bold{q}^{\prime})}_{J_{\bold{s},\bold{u}}}$ labels the vertex $\mathfrak{v}$ and $V^{(\bold{k}^{\prime},\bold{q}^{\prime})_{-1}}_{J_{\bold{s}^{\prime},\bold{u}^{\prime}}}$ is the potential labelling the vertex connected to $\mathfrak{v}$ by the edge $\mathfrak{e}$.
\end{itemize}

\item A leaf of the tree is a vertex at some level $(\bold{k}^{\prime},\bold{q}^{\prime})$ that has no descendants, i.e., that is not connected to any vertex at level $(\bold{k}^{\prime},\bold{q}^{\prime})_{-1}$ by any edge. Note that a leaf of the tree 
is labeled by a potential of the type $V^{(\bold{k}^{\prime\prime},\bold{q}^{\prime\prime})}_{J_{\bold{k}^{\prime\prime},\bold{q}^{\prime\prime}}}$ for some $(\bold{k}^{\prime\prime},\bold{q}^{\prime\prime})\succeq (\bold{0},\bold{N})$.
%
\item A branch of a tree rooted at $(\mathbf{k}, \mathbf{q})$ is an ordered connected set of edges with the following properties:
\begin{itemize}
\item the first edge of a branch has the vertex at level $(\bold{k},\bold{q})$ as an endpoint;
\item the last edge of a branch has a leaf at some level $(\bold{k}^{\prime\prime},\bold{q}^{\prime\prime})$ as an endpoint (referred to as the leaf of the branch);
\item there is a single edge connecting vertices at levels $(\bold{k}^{\prime},\bold{q}^{\prime})$ and $(\bold{k}^{\prime},\bold{q}^{\prime})_{-1}$ for every $(\bold{k}^{\prime},\bold{q}^{\prime})$ with $(\bold{k},\bold{q})\succeq (\bold{k}^{\prime},\bold{q}^{\prime})\succ (\bold{k}^{\prime\prime},\bold{q}^{\prime\prime})$.
\end{itemize}




\item With each branch  $\mathfrak{b}$ of a tree 
we associate a set,   $\mathcal{R}_\mathfrak{b}$, of rectangles
consisting of i) those rectangles labelling the edges of $\mathfrak{b}$, and ii) the rectangle $J_{\bold{k}^{\prime\prime},\bold{q}^{\prime\prime}}$ indicating the support of the potential labelling the leaf of $\mathfrak{b}$. 

The set $\mathcal{R}_\mathfrak{b}$ inherits the ordering relation (\ref{ordering}), hence its elements can be enumerated  by a map 
$$i\in \big\{1,\cdots ,\vert \mathcal{R}_{\mathfrak{b}}\vert \big\}\rightarrow J_{\bold{k}^{(i)},\bold{q}^{(i)}}\in \mathcal{R}_\mathfrak{b}$$ 
with $(\bold{k}^{(i)},\bold{q}^{(i)})\succ (\bold{k}^{(i+1)},\bold{q}^{(i+1)})$ and where  $\vert \mathcal{R}_\mathfrak{b}\vert$ is the cardinality of the set $\mathcal{R}_\mathfrak{b}$. Note that $J_{\bold{k}^{(| \mathcal{R}_{\mathfrak{b}}|)},\bold{q}^{(|\mathcal{R}_{\mathfrak{b}}|)}}$ is the rectangle associated with the potential labelling the leaf of $\mathfrak{b}$. 

\item To every branch $\mathfrak{b}$ we can associate the \textit{``branch operator''}, also denoted by $\mathfrak{b}$,
\begin{equation}\label{opbranch-0}
\mathfrak{b}:=\mathcal{A}_{J_{\bold{k}^{(1)},\bold{q}^{(1)}}}(\,\mathcal{A}_{J_{\bold{k}^{(2)},\bold{q}^{(2)}}} (\cdots  \mathcal{A}_{J_{\bold{k}^{(|\mathcal{R}_b | -1)},\bold{q}^{(| \mathcal{R}_b| -1)}}}(V_{\mathcal{L}_\mathfrak{b}})\cdots )\, )\,,
\end{equation}
where $V_{\mathcal{L}_\mathfrak{b}}:= V^{(\bold{k}^{(| \mathcal{R}_{\mathfrak{b}}|)},\bold{q}^{(|\mathcal{R}_{\mathfrak{b}}|)})}_{J_{\bold{k}^{(| \mathcal{R}_{\mathfrak{b}}|)},\bold{q}^{(|\mathcal{R}_{\mathfrak{b}}|)}}}$ is the potential labelling the leaf of $\mathfrak{b}$.

The set of branches whose corresponding branch operators are non-zero is denoted by 
$\mathcal{B}_{V^{(\bold{k},\bold{q})}_{J_{\bold{r},\bold{i}}}}$.
\end{enumerate}

\end{defn}

\subsubsection{Properties of the branches $\mathfrak{b}\in \mathcal{B}_{V^{(\bold{k},\bold{q})}_{J_{\bold{r},\bold{i}}}}$}\label{properties}
Definition \ref{def-tree} implies the following properties of the elements of the set $\mathcal{B}_{V^{(\bold{k},\bold{q})}_{J_{\bold{r},\bold{i}}}}$ defined above:
\begin{itemize}
\item[P-i)]
For $\mathfrak{b}\in \mathcal{B}_{V^{(\bold{k},\bold{q})}_{J_{\bold{r},\bold{i}}}}$, the set 
$$\bigcup_{i\in \big\{1,\cdots,\vert \mathcal{R}_{\mathfrak{b}}\vert \big\}}J_{\bold{k}^{(i)},\bold{q}^{(i)}}$$ 
is connected, due to (\ref{opbranch-0}),  though $J_{\bold{k}^{(i)},\bold{q}^{(i)}}\cap J_{\bold{k}^{(i+1)},\bold{q}^{(i+1)}}$ might be empty for some $i$. Likewise, for any fixed $n\in \big\{1\cdots \vert \mathcal{R}_\mathfrak{b}\vert\big\}$, the set 
$\bigcup_{n\leq i\leq \vert\mathcal{R}_{\mathfrak{b}}\vert}J_{\bold{k}^{(i)},\bold{q}^{(i)}}$  is connected. Indeed, for any operator $\mathcal{O}$ and for any $m$,  $\mathcal{A}_{J_{\bold{k}^{(m)},\bold{q}^{(m)}}}(\mathcal{O})=0$ whenever the supports of $\mathcal{O}$ and $S_{J_{\bold{k}^{(m)},\bold{q}^{(m)}}}$ have empty intersection; see formula (\ref{Acal}).

\item[P-ii)]
For $\mathfrak{b}\in \mathcal{B}_{V^{(\bold{k},\bold{q})}_{J_{\bold{r},\bold{i}}}}$, the cardinality, $\vert \mathcal{R}_\mathfrak{b}\vert $, of the set $\mathcal{R}_\mathfrak{b}$ of rectangles 
is such that $\vert \mathcal{R}_\mathfrak{b}\vert \geq \mathcal{O}(\frac{r}{k})\geq \mathcal{O}( r^{\frac{3}{4}})$. This lower bound on 
$\vert \mathcal{R}_\mathfrak{b}\vert$ is a consequence of the restriction imposed on $k=\vert \mathbf{k} \vert$ and required in regime $\mathfrak{R}1$, (and it will turn out to be crucial to derive our estimate (\ref{est-r1-in-bis})-(\ref{est-r1-fin}) in Theorem \ref{th-norms}).

\item[P-iii)] The set $J_{\bold{r},\bold{i}}$ is the minimal rectangle associated with 
$\bigcup_{i\in \big\{1,\cdots,\vert\mathcal{R}_\mathfrak{b}\vert \big\}}J_{\bold{k}^{(i)},\bold{q}^{(i)}}$, for any branch 
$\mathfrak{b}\in \mathcal{B}_{V^{(\bold{k},\bold{q})}_{J_{\bold{r},\bold{i}}}}$.  Furthermore, if we amputate a branch at some vertex by keeping only the descendants of that vertex (i.e., the lower part only) then the same property holds for the rectangle associated with the potential labelling the (new) root vertex of the amputated branch that has been created. 
\item[P-iv)]
Two different branches $\mathfrak{b}, \mathfrak{b}' \in  \mathcal{B}_{V^{(\bold{k},\bold{q})}_{J_{\bold{r},\bold{i}}}}$ are associated with two different (ordered) sets of rectangles $\mathcal{R}_\mathfrak{b}$ and $\mathcal{R}_{\mathfrak{b}'}$.\\
Sketch of proof:  

\noindent
1) The two branches must cross at some vertex.

\noindent
2) Consider the first vertex (starting at the bottom of the tree) where they cross and the two (possibly) amputated branches corresponding to the two original branches that have this vertex as their root vertex.  

\noindent
3) Now, notice that there are two alternatives: 3-i) either the rectangles associated with the two edges linked to the root vertex (the vertex where they cross)  are different, in the sense that one edge is associated to a rectangle and the other to none; 3-ii) or some of  the remaining rectangles in the amputated branches must differ, due to property  P-iii),  since the potentials labelling the vertices at the level just below the common root vertex are different.
\item[P-v)]
Each term in the re-expansion is associated with a branch $\mathfrak{b}$ of the tree, and this correspondence  is bijective 
by construction. Thus, by property P-iv), two distinct non-zero terms in the re-expansion, corresponding to two different 
branches $\mathfrak{b}_1,\mathfrak{b}_2\in \mathcal{B}_{V^{(\bold{k},\bold{q})}_{J_{\bold{r},\bold{i}}}}$, are labelled by
two different sets of rectangles, $\mathcal{R}_{\mathfrak{b}_1}$ and $\mathcal{R}_{\mathfrak{b}_2}$, respectively. 
\end{itemize}

\subsection{Summing over the norms of branch-operator: \textit{weights} and \textit{paths}, $\Gamma_\mathfrak{b}$}\label{sum-branches}
Our task is to estimate the norms of the potentials $V^{(\bold{k}, \bold{q})}_{J_{\bold{r}, \bold{i}}}$,
which can be accomplished by taking the re-expansion of the potentials into account
according to the prescriptions of Definition \ref{def-tree}. More precisely, each potential  $V^{(\bold{k}, \bold{q})}_{J_{\bold{r}, \bold{i}}}$
can be expressed as the sum $\sum_{\mathfrak{b}\in  \mathcal{B}_{V^{(\bold{k},\bold{q})}_{J_{\bold{r},\bold{i}}}}}\mathfrak{b}$, where $\mathfrak{b}$ are the branch operators defined in point 8. of Definition \ref{def-tree}. Therefore, we are led  to estimating  the sum over the norms of branch operators, to wit
$$\sum_{\mathfrak{b}\in  \mathcal{B}_{V^{(\bold{k},\bold{q})}_{J_{\bold{r},\bold{i}}}}}\|\mathfrak{b}\|\,.$$ 
This can be done by assigning a ``weight"  to every set $\mathcal{R}_\mathfrak{b}$ of rectangles,  the weight being proportional 
to the product of operator norms of the potentials associated (in step $(\bold{k},\bold{q})_{-1}$) with each rectangle 
$J_{\bold{k},\bold{q}}$  in the set $\mathcal{R}_\mathfrak{b}$, i.e., 

\begin{equation}\label{tot-w}
\sum_{\mathfrak{b}\in  \mathcal{B}_{V^{(\bold{k},\bold{q})}_{J_{\bold{r},\bold{i}}}}}
(c\cdot t)^{\vert \mathcal{R}_\mathfrak{b}\vert -1}\,\|V_{\mathcal{L}_\mathfrak{b}}\|
\prod_{i\in \big\{1,\cdots,\vert \mathcal{R}_\mathfrak{b}\vert -1 \big\}}\|V^{(\bold{k}^{(i)},\bold{q}^{(i)})_{-1}}_{J_{\bold{k}^{(i)},\bold{q}^{(i)}}} \| \,,
\end{equation}
where $V_{\mathcal{L}_\mathfrak{b}}$ is the potential labelling the leaf of $\mathfrak{b}$, since a factor $(c\cdot t)\cdot \|V^{(\bold{k}^{(i)},\bold{q}^{(i)})_{-1}}_{J_{\bold{k}^{(i)},\bold{q}^{(i)}}} \|$  is associated with the map $ \mathcal{A}_{J_{\bold{k}^{(i)},\bold{q}^{(i)}}}$; here $c>0$ is a universal constant.

\noindent
In order to count the sets $\mathcal{R}_\mathfrak{b}$, we shall assign a path, $\Gamma_\mathfrak{b}$, to each 
$\mathfrak{b}$, where $\Gamma_\mathfrak{b}$ has the property to visit all the rectangles in the set 
$\mathcal{R}_\mathfrak{b}$. Since we must estimate the "weighted" number of sets $\mathcal{R}_\mathfrak{b}$, 
the paths must be weighted accordingly. 

\subsubsection{Paths of connected rectangles}
 The following definitions clarify what we mean by a path visiting rectangles.\\

\begin{defn} \label{pathsdef}
\noindent
\begin{itemize}
\item[i)] A  path $\Gamma$ is a finite sequence of rectangles $\{J_{\bold{s}^{(i)},\bold{u}^{(i)}}\}_{i=1}^n$, for 
some $n\in\mathbb{N}$, with the property that $J_{\bold{s}^{(i)},\bold{u}^{(i)}}\neq  J_{\bold{s}^{(i+1)},\bold{u}^{(i+1)}}$ 
and  $J_{\bold{s}^{(i)},\bold{u}^{(i)}}\cap J_{\bold{s}^{(i+1)},\bold{u}^{(i+1)}}\neq\emptyset$, for every $i=1\cdots n-1$.\\
Warning: In contrast to item 7 in Definition \ref{def-tree}, no relation is assumed here between the ordering labeled by the index $i$ and the ordering $\prec$.
\item[ii)] The set of ordered pairs,
$$\mathcal{S}_\Gamma:=\Big\{\big(J_{\bold{s}^{(i)},\bold{u}^{(i)}},J_{\bold{s}^{(i+1)},\bold{u}^{(i+1)}}\big) \vert \,i=1,\cdots, n-1\Big\},$$
is called the set of steps of the path $\Gamma\equiv \big\{J_{\bold{s}^{(i)},\bold{u}^{(i)}}\big\}_{i=1}^n$.
\item[iii)] The length, $l_{\Gamma}$, of the path $\Gamma\equiv \{J_{\bold{s}^{(i)},\bold{u}^{(i)}}\}_{i=1}^n$ is defined to be $l_{\Gamma}:=n-1$.
\item[iv)] The support, supp$(\Gamma)$,  of a path $\Gamma\equiv \{J_{\bold{s}^{(i)},\bold{u}^{(i)}}\}_{i=1}^n$ is defined to be $$\text{supp}(\Gamma):=\Big\{ J_{\bold{s}^{(i)},\bold{u}^{(i)}},\, i\in\{1\cdots n\}\Big\}.$$
\item[v)] A path $\Gamma\equiv \{J_{\bold{s}^{(i)},\bold{u}^{(i)}}\}_{i=1}^n$ is closed if $J_{\bold{s}^{(1)},\bold{u}^{(1)}}=J_{\bold{s}^{(n)},\bold{u}^{(n)}}$.
\end{itemize}
\end{defn}

Each rectangle $J_{\bold{k}^{(i)},\bold{q}^{(i)}}$ of the set $\mathcal{R}_{\mathfrak{b}}$ 
 contributes to the weight (\ref{tot-w}) of $\mathcal{R}_{\mathfrak{b}}$ through 
$c\cdot t\cdot \|V^{(\bold{k}^{(i)},\bold{q}^{(i)})_{-1}}_{J_{\bold{k}^{(i)},\bold{q}^{(i)}}} \| $ (except for $J_{\bold{k}^{(| \mathcal{R}_{\mathfrak{b}}|)},\bold{q}^{(|\mathcal{R}_{\mathfrak{b}}|)}}$ that contributes through $c\cdot  \|V_{\mathcal{L}_\mathfrak{b}}\|$), which (as it will be shown) decreases with 
the size of the rectangle. Thus, we have to make sure that the path $\Gamma_{\mathfrak{b}}$ does not visit 
small rectangles of $\mathcal{R}_{\mathfrak{b}}$, which have a ``big'' weight, repeatedly.
This motivates the requirements imposed on the paths $\Gamma_{\mathfrak{b}}$ considered henceforth, 
in particular property C) stated in the next section.

\subsubsection{Connected components, $\mathcal{Z}^{(j)}_{\rho}$, of rectangles, and definition of $\Gamma_{\mathfrak{b}}$}  \label{features}
Since the weight of a rectangle is a function of its size, it is convenient to write  the connected set $\bigcup_{i\in\{1,\cdots,|\mathcal{R}_{\mathfrak{b}}|\}}J_{\bold{k}^{(i)},\bold{q}^{(i)}}$ as the union 
$$\bigcup_{\rho=k_0}^{k} \Big(\bigcup_{j=1}^{j_{\rho}}\mathcal{Z}^{(j)}_{\rho}\Big)\,,$$  
where $\{\mathcal{Z}^{(j)}_{\rho}, \quad j=1,\dots,j_{\rho}\}$  are distinct connected components 
of (unions of) rectangles of a given size $\rho$, $k_0\leq \rho \leq k$, starting from the lowest one $k_0\geq 1$, with the following properties:

\noindent
1) $j_{k_0}=1$ (i.e., there is only one component for $\rho=k_0$);

\noindent
2) rectangles of the same size but belonging to different components do \textit{not} overlap, i.e., for any $\rho $, $\mathcal{Z}^{(j)}_{\rho}\cap \mathcal{Z}^{(j')}_{\rho}=\emptyset$\,, for $j\neq j'$.

\noindent
We call $\text{supp}(\mathcal{Z}^{(j)}_{\rho})$, $\rho=k_0, \dots,k\,,\,j=1,\dots, j_{\rho}$,  the set of rectangles of $\mathcal{Z}^{(j)}_{\rho}$, i.e.,
$$\text{supp}(\mathcal{Z}^{(j)}_{\rho}):=\Big\{J_{\bold{k}^{(i)},\bold{q}^{(i)}}: J_{\bold{k}^{(i)},\bold{q}^{(i)}}\subset \mathcal{Z}^{(j)}_{\rho}, 
i\in \big\{1,\cdots,\vert \mathcal{R}_{\mathfrak{b}}\vert \big\} \Big\}.$$ 
\\

Starting from a branch $\mathfrak{b}\in  \mathcal{B}_{V^{(\bold{k},\bold{q})}_{J_{\bold{r},\bold{i}}}}$, we shall  inductively construct a path, $\Gamma_{\mathfrak{b}}$, of length $l_{\Gamma_\mathfrak{b}}$ bounded by
$$l_{\Gamma_\mathfrak{b}} \leq 2(n_{k_0}+\sum_{j=1}^{j_2}n_{k_0+1}^{(j)}+\dots +\sum_{j=1}^{j_k}n_k^{(j)})-2\,,$$ 
with the following properties:
\begin{enumerate}
\item[A)]   the support of $\Gamma_\mathfrak{b}$ is  $\mathcal{R}_\mathfrak{\mathfrak{b}}$;
\item[B)] for each component $\mathcal{Z}^{(j)}_{\rho}$ consisting of the union of $n_{\rho}^{(j)}$ rectangles, at most $2n_{\rho}^{(j)}-2$ steps are made (i.e., there are at most $2n_{\rho}^{(j)}-2$ steps $\sigma\in\mathcal{S}_{\Gamma_\mathfrak{b}}$ for which $\sigma\in \text{supp}(\mathcal{Z}^{(j)}_{\rho})\times \text{supp}(\mathcal{Z}^{(j)}_{\rho})$);
\item[C)]  there are at most two steps connecting rectangles in $\text{supp}(\mathcal{Z}_\rho^{(j)})$ with rectangles
of lower size: more precisely, for every connected component $\mathcal{Z}_\rho^{(j)}$ there is at most one $J_{\bold{s},\bold{u}}$ in $\text{supp}(\mathcal{Z}_\rho^{(j)})$  such that $(J_{\bold{s}',\bold{u}'},J_{\bold{s},\bold{u}})\in\mathcal{S}_{\Gamma_\mathfrak{b}}$ with $s'<s$, and one $J_{\bold{s},\bold{u}}$ such that $(J_{\bold{s},\bold{u}},J_{\bold{s}',\bold{u}'})\in\mathcal{S}_{\Gamma_\mathfrak{b}}$ with $s>s'$.
\end{enumerate}

\noindent
The precise construction is carried out by induction in $k$ in Lemma \ref{conn-rect-2}, combined with Lemma \ref{conn-rect}; i.e.,   we assume that we have constructed a path $\Gamma_{\mathfrak{b}}^{(k'-1)}$, with $k_0\leq k'\leq k$, fulfilling A), B), and C) for the set $\cup_{\rho=k_0}^{k'-1} \cup_{j=1}^{j_{\rho}}\mathcal{Z}^{(j)}_{\rho}$, which is connected by Property P-i). 
Starting from this path, we construct a new one, denoted by $\Gamma_{\mathfrak{b}}^{(k')}$, with the desired properties.

\subsubsection{Weighted sums of paths}\label{sum-paths}
The features specified by A), B), and C), above, are used to distribute the total weight available, as shown in (\ref{tot-w}),  amongst the steps of the path $\Gamma_{\mathfrak{b}}$, in a way that is optimal to derive suitable 
bounds. In fact, we will associate a weight with the steps of the paths $\Gamma_{\mathfrak{b}}$ described in Section  \ref{features}, so as to estimate (\ref{tot-w}) in terms of a weighted sum of paths. The mechanism, which we shall illustrate below, is essentially the one used in Theorem \ref{th-norms} to control regime $\mathfrak{R}1$,  with some modifications that we omit here in order not to obscure the key ideas, and which are related to the proof by induction of Theorem \ref{th-norms}.

We observe that there are  $n_{\rho}^{(j)}$ rectangles in the set $\text{supp}(\mathcal{Z}^{(j)}_{\rho})$, and that, for the paths 
$\Gamma_\mathfrak{b}$, there are at most $2n_{\rho}^{(j)}-2$ steps between these rectangles; see property B) above.    
In addition, there are at most $2$ steps, from rectangles of lower size and back,  to be taken into account; see 
property C) above. Consequently, to each  step 
$\sigma=(J_{\bold{s}^{(i)},\bold{u}^{(i)}},J_{\bold{s}^{(i+1)},\bold{u}^{(i+1)}})\in \mathcal{S}_{\Gamma_{\mathfrak{b}}}$ 
we can assign the weight  
$$\mathfrak{w}_\sigma\equiv \mathfrak{w}_{s^{(i)}\to s^{(i+1)}}=((c+1) t)^{\frac{1}{2}}\cdot \min\,\Big\{\|V^{(\bold{s}^{(i)},\bold{u}^{(i)})_{-1}}_{J_{\bold{s}^{(i)},\bold{u}^{(i)}}} \|^{\frac{1}{2}}\,,\,\|V^{(\bold{s}^{(i+1)},\bold{u}^{(i+1)})_{-1}}_{J_{\bold{s}^{(i+1)},\bold{u}^{(i+1)}}} \|^{\frac{1}{2}}\Big\}\,,$$
where $t$ is sufficiently small such that 
$(c+1) t\cdot \|V^{(\bold{k}^{(i)},\bold{q}^{(i)})_{-1}}_{J_{\bold{k}^{(i)},\bold{q}^{(i)}}} \|<1$\,
, and the following estimate holds
\begin{equation}\label{weight-path}
(c\cdot t)^{ |R_b|-1}\,\|V_{\mathcal{L}_\mathfrak{b}}\|  \prod_{i=1}^{|R_b|-1} \|V^{(\bold{k}^{(i)}, \bold{q}^{(i)})_{-1}}_{J_{\bold{k}^{(i)}, \bold{q}^{(i)}}}\|  \leq \frac{1}{t}\prod_{\sigma\in\mathcal{S}_{\Gamma_{\mathfrak{b}}}}\mathfrak{w}_\sigma .
\end{equation}
The previous inequality is true because, if we denote by $\mathcal{S}_{\mathcal{Z}_\rho^{(j)}}$ the set of  at most $2n_\rho^{(j)}-2$ steps between rectangles of  $\text{supp}\mathcal{Z}_\rho^{(j)}$ and the additional at most $2$ steps from rectangles of lower size and back, then we have
$$
(c\cdot t)^{ |\text{supp}(\mathcal{Z}_\rho^{(j)})|}\,  \prod_{J_{\bold{s}, \bold{u}}\in\, \text{supp}(\mathcal{Z}_\rho^{(j)})}\|V^{(\bold{s},\bold{u})_{-1}}_{J_{\bold{s},\bold{u}}} \|\leq \prod_{\sigma\in\mathcal{S}_{\mathcal{Z}_\rho^{(j)}}}\mathfrak{w}_\sigma .
$$

Finally we use the estimate
\begin{eqnarray}\label{est-r1-in-intro}
& &\sum_{\mathfrak{b}\in  \mathcal{B}_{V^{(\bold{k},\bold{q})}_{J_{\bold{r},\bold{i}}}}}
c^{\vert \mathcal{R}_\mathfrak{b}\vert -1 }t^{|R_b|}\, \|V_{\mathcal{L}_\mathfrak{b}}\|
\prod_{i\in \big\{1,\cdots,\vert \mathcal{R}_\mathfrak{b}\vert -1 \big\}}\|V^{(\bold{k}^{(i)},\bold{q}^{(i)})_{-1}}_{J_{\bold{k}^{(i)},\bold{q}^{(i)}}} \| \\
& \leq  &  \sum_{\Gamma_{\mathfrak{b}},\, \mathfrak{b}\in \mathcal{B}_{V^{(\bold{k},\bold{q})}_{J_{\bold{r},\bold{i}}}}} \prod_{\sigma\in\mathcal{S}_{\Gamma_{\mathfrak{b}}}}\mathfrak{w}_\sigma \leq  C_d\cdot r^{2d-1}\cdot \sum_{j=\lfloor c_d\cdot \frac{r}{k}\rfloor}^{\infty} \Big(\sum_{\rho, \rho'=1}^{k} \mathfrak{w}_{\rho\to \rho'}\,D_{\rho,\rho'}\Big)^j\,,
\end{eqnarray}
where $\lfloor c_d\cdot \frac{r}{k} \rfloor$ is a lower bound for $\vert\mathcal{R}_{\mathfrak{b}}\vert$, and $C_d\cdot r^{2d-1}$,   is an upper bound on the possible positions of the rectangle $J_{\bold{k}^{(\vert \mathcal{R}_{\mathfrak{b}}\vert)},\bold{q}^{(\vert \mathcal{R}_{\mathfrak{b}}\vert)}}$ of the path, where $c_d, C_d$  are $d$-dependent constants; finally  
\begin{equation}\label{directions}
D_{s,s'}:=\mathfrak{C}_d \cdot s^d \cdot  s^{\prime d-1}\,,
\end{equation}
where $\mathfrak{C}_d$ is a $d$-dependent constant, is an upper bound on the number of possible directions 
of a path $\Gamma=\{J_{\bold{s}^{(i)},\bold{u}^{(i)}}\}_{i=1}^n$, extended by one more step as specified here: given the 
path $\Gamma=\{J_{\bold{s}^{(i)},\bold{u}^{(i)}}\}_{i=1}^n$, the number of paths 
$\Gamma^+=\{J_{\bold{s}^{'(i)},\bold{u}^{'(i)}}\}_{i=1}^{n+1}$ of length $l_{\Gamma^+}=n$, whose first $n$ elements
agree with $\Gamma$ (i.e., 
$ \{J_{\bold{s}^{(i)},\bold{u}^{(i)}}\}_{i=1}^{n}=\{J_{\bold{s}^{\prime (i)},\bold{u}^{\prime(i)}}\}_{i=1}^{n}$) 
and for which $s^{\prime (n+1)}:=s'$ and $s^{(n)}:=s$, is bounded from above by $D_{s,s'}$. \\
A minor modification of  the inequality provided in (3.44) will enable us to prove the result  of Theorem \ref{th-norms} concerning regime $\mathfrak{R}1$.

\section{The unitary conjugation $e^{S_{J_{\bold{k},\bold{q}}}}$ and the spectral gap of $G_{J_{\bold{k},\bold{q}}}$}\label{gap-section}
The operator  $e^{S_{J_{\bold{k},\bold{q}}}}$ is constructed so as to block-diagonalize the Hamiltonian 
$G_{J_{\bold{k},\bold{q}}}+tV^{(\bold{k},\bold{q})_{-1}}_{J_{\bold{k},\bold{q}}}$ w.r.t. the decomposition of the identity 
\begin{equation}\label{decomp-id}
\charf = P^{(+)}_{J_{\bold{k},\bold{q}}}+P^{(-)}_{J_{\bold{k},\bold{q}}}\,.
\end{equation} 
The operator
\begin{equation}\label{def-G-bis}
G_{J_{\bold{k},\bold{q}}}:=\sum_{\bold{i}\subset J_{\bold{k},\bold{q}} }H_{\bold{i}}+t\sum_{J_{\bold{k}_{(1)}',\bold{q}'}\subset J_{\bold{k},\bold{q}}} V^{(\bold{k},\bold{q})_{-1}}_{J_{\bold{k}_{(1)}',\bold{q}'}}+\dots+t\sum_{J_{\bold{k}'_{(|\bold{k}|-1)},\bold{q}' }\subset J_{\bold{k},\bold{q}}}V^{(\bold{k},\bold{q})_{-1}}_{J_{\bold{k}'_{(|\bold{k}|-1)},\bold{q}'}}\,
\end{equation}
is already block-diagonal with respect to (\ref{decomp-id}). \\
For this construction we refer the reader to the notation and results in  Sections 2 and 3 of \cite{DFFR}. We add the definition of $E_{J_{\bold{k},\bold{q}}}$, which is in fact the ground-state energy of the operator $G_{J_{\bold{k},\bold{q}}}$,
\begin{equation} \label{def-E-bis}
E_{J_{\bold{k},\bold{q}}}:=\langle \bigotimes_{\bold{j}\in J_{\bold{k},\bold{q}}} \Omega_{\bold{j}}\,,\, G_{J_{\bold{k},\bold{q}}}  \bigotimes_{\bold{j}\in J_{\bold{k},\bold{q}}} \Omega_{\bold{j}} \rangle \,,
\end{equation}
i.e., 
$$G_{J_{\bold{k},\bold{q}}}P^{(-)}_{J_{\bold{k},\bold{q}}}=E_{J_{\bold{k},\bold{q}}}P^{(-)}_{J_{\bold{k},\bold{q}}}\,.$$
We recall that 
\begin{equation}\label{def-ad-1}
ad\, A\,(B):=[A\,,\,B]\,
\end{equation}
where $A$ and $B$ are bounded operators, and, for $n\geq 2$,
\begin{equation}\label{def-ad-2}
ad^n A\,(B):=[A\,,\,ad^{n-1} A\,(B)]\,.
\end{equation}
To carry out the block-diagonalization step $(\bold{k},\bold{q})$, the operator $S_{J_{\bold{k},\bold{q}}}$
is defined by the series
\begin{equation}\label{formula-S}
S_{J_{\bold{k},\bold{q}}}:=\sum_{j=1}^{\infty}t^j(S_{J_{\bold{k},\bold{q}}})_j\,
\end{equation}
where
\begin{itemize}
\item
\begin{equation}\label{formula-Sj}
(S_{J_{\bold{k},\bold{q}}})_j:=ad^{-1}\,G_{J_{\bold{k},\bold{q}}}\,((V^{(\bold{k},\bold{q})_{-1}}_{J_{\bold{k},\bold{q}}})^{od}_j):=\frac{1}{G_{J_{\bold{k},\bold{q}}}-E_{J_{\bold{k},\bold{q}}}}P^{(+)}_{J_{\bold{k},\bold{q}}}\,(V^{(\bold{k},\bold{q})_{-1}}_{J_{\bold{k},\bold{q}}})_j\,P^{(-)}_{J_{\bold{k},\bold{q}}}-h.c.\,,
\end{equation}
where \emph{``od''} means \textit{off-diagonal} w.r.t. the decomposition of the identity (\ref{decomp-id}){\color{magenta};}

\item
$(V^{(\bold{k},\bold{q})_{-1}}_{J_{\bold{k},\bold{q}}})_1:=V^{(\bold{k},\bold{q})_{-1}}_{J_{\bold{k},\bold{q}}}$\,, and, for $j\geq 2$,
\begin{align}\label{formula-v_j}
&(V^{(\bold{k},\bold{q})_{-1}}_{J_{\bold{k},\bold{q}}})_j 
:= \sum_{p\geq 2, r_1\geq 1 \dots, r_p\geq 1\, ; \, r_1+\dots+r_p=j}\frac{1}{p!}\text{ad}\,(S_{J_{\bold{k},\bold{q}}})_{r_1}\Big(\text{ad}\,(S_{J_{\bold{k},\bold{q}}})_{r_2}\dots (\text{ad}\,(S_{J_{\bold{k},\bold{q}}})_{r_p}(G_{J_{\bold{k},\bold{q}}}))\dots \Big)\nonumber \\
&\quad +\sum_{p\geq 1, r_1\geq 1 \dots, r_p\geq 1\, ; \, r_1+\dots+r_p=j-1}\frac{1}{p!}\text{ad}\,(S_{J_{\bold{k},\bold{q}}})_{r_1}\Big(\text{ad}\,(S_{J_{\bold{k},\bold{q}}})_{r_2}\dots (\text{ad}\,(S_{J_{\bold{k},\bold{q}}})_{r_p}(V^{(\bold{k},\bold{q})_{-1}}_{J_{\bold{k},\bold{q}}}))\dots \Big)\,.\quad\quad\quad\quad
\end{align}
\end{itemize}
We recall that
\begin{equation}\label{def-K}
K_{\Lambda_N^{d}}^{(\bold{k},\bold{q})}:=e^{S_{J_{\bold{k},\bold{q}}}}\,K_{\Lambda_N^{d}}^{(\bold{k},\bold{q})_{-1}}\,e^{-S_{J_{\bold{k},\bold{q}}}}\,.
\end{equation}
The algorithm described in Definition \ref{def-interactions-multi} can be motivated by inspecting the proof of the next theorem, which establishes the consistency property alluded to in Sect. \ref{algo} before introducing Definition \ref{def-interactions-multi}.
\begin{thm}\label{th-potentials-multi}
The Hamiltonian $K_{\Lambda_N^{d}}^{(\bold{k},\bold{q})}:=e^{S_{J_{\bold{k},\bold{q}}}}\,K_{\Lambda_N^{d}}^{(\bold{k},\bold{q})_{-1}}\,e^{-S_{J_{\bold{k},\bold{q}}}}$ can be written in the form given in (\ref{K-tranf-1}),  where the terms $\{V^{(\bold{k},\bold{q})}_{J_{\bold{l},\bold{i}}}\}$ are obtained from the terms $\{V^{(\bold{k},\bold{q})_{-1}}_{J_{\bold{l},\bold{i}}}\}$  according to the algorithm described in Definition \ref{def-interactions-multi}.
\end{thm}

\noindent
\emph{Proof.}

\noindent
In the expression
\begin{align}\label{K-tranf-1-bis}
e^{S_{J_{\bold{k},\bold{q}}}}\,K_{\Lambda_N^{d}}^{(\bold{k},\bold{q})_{-1}}\,e^{-S_{J_{\bold{k},\bold{q}}}} 
&=e^{S_{J_{\bold{k},\bold{q}}}}\,\Big\{\sum_{\bold{i}\in \Lambda^{(d)}_N}H_{\bold{i}}+t\sum_{\bold{k}_{(1)}'\,,\,\bold{q}'}V^{(\bold{k},\bold{q})}_{J_{\bold{k}_{(1)}',\bold{q}'}}+t\sum_{\bold{k}_{(2)}'\,,\,\bold{q}'}V^{(\bold{k},\bold{q})}_{J_{\bold{k}_{(2)}',\bold{q}'}}+\dots+\nonumber \\
+ &t\sum_{\bold{k}_{(|\bold{k}|)}'\,,\,\bold{q}'}V^{(\bold{k},\bold{q})}_{J_{\bold{k}'_{(|\bold{k}|)},\bold{q}'}} + t\sum_{\bold{k}_{(|\bold{k}|+1)}'\,,\,\bold{q}'}
V^{(\bold{k},\bold{q})}_{J_{\bold{k}'_{(|\bold{k}|+1)}},\bold{q}'}+
\dots+tV^{(\bold{k},\bold{q})}_{J_{\bold{N},\bold{1}}}\,\Big\}e^{-S_{J_{\bold{k},\bold{q}}}}\,,
\end{align}
we observe that:

\begin{itemize}
\item
For all rectangles $J_{\bold{l},\bold{i}}$ such that $J_{\bold{l},\bold{i}} \cap J_{\bold{k},\bold{q}}=\emptyset$ we have that
\begin{equation}
e^{S_{J_{\bold{k},\bold{q}}}}\,V^{(\bold{k},\bold{q})_{-1}}_{J_{\bold{l},\bold{i}}}\,e^{-S_{J_{\bold{k},\bold{q}}}}=V^{(\bold{k},\bold{q})_{-1}}_{J_{\bold{l},\bold{i}}}=V^{(\bold{k},\bold{q})}_{J_{\bold{l},\bold{i}}}\,,
\end{equation}
where the last identity is due to item a) in Definition \ref{def-interactions-multi}.
\item
Regarding  the terms constituting $G_{J_{\bold{k}, \bold{q}}}$ (see the definition in (\ref{def-G})) we note that if we add $V^{(\bold{k},\bold{q})_{-1}}_{J_{\bold{k},\bold{q}}}$ we get
\begin{align}
e^{S_{J_{\bold{k},\bold{q}}}}\, &(G_{J_{\bold{k}, \bold{q}}}+V^{(\bold{k},\bold{q})_{-1}}_{J_{\bold{k},\bold{q}}})\,
e^{-S_{J_{\bold{k},\bold{q}}}} =\, \nonumber
\\
= &\sum_{\bold{i}\subset J_{\bold{k},\bold{q}} }H_{\bold{i}}+t\sum_{J_{\bold{k}_{(1)}',\bold{q}'}\subset J_{\bold{k},\bold{q}}} V^{(\bold{k},\bold{q})_{-1}}_{J_{\bold{k}_{(1)}',\bold{q}'}}+\dots+t\sum_{J_{\bold{k}'_{(|\bold{k}|-1)}},\bold{q}' \subset J_{\bold{k},\bold{q}}}V^{(\bold{k},\bold{q})_{-1}}_{J_{\bold{k}'_{(|\bold{k}|-1)},\bold{q}'}} +\sum_{j=1}^{\infty}t^{j-1}(V^{(\bold{k},\bold{q})_{-1}}_{J_{\bold{k},\bold{q}}})^{diag}_j  \nonumber \\
= &\sum_{\bold{i}\subset J_{\bold{k},\bold{q}} }H_{\bold{i}}+t\sum_{J_{\bold{k}_{(1)}',\bold{q}'}\subset J_{\bold{k},\bold{q}}} V^{(\bold{k},\bold{q})_{-1}}_{J_{\bold{k}_{(1)}',\bold{q}'}}+\dots+t\sum_{J_{\bold{k}'_{(|\bold{k}|-1)}},\bold{q}' \subset J_{\bold{k},\bold{q}}}V^{(\bold{k},\bold{q})_{-1}}_{J_{\bold{k}'_{(|\bold{k}|-1)},\bold{q}'}}+tV^{(\bold{k},\bold{q})}_{J_{\bold{k},\bold{q}}}\,,
\end{align}
where the first equation results from the Lie-Schwinger procedure and the second one follows from Definition \ref{def-interactions-multi}, items a) and b). 
\item
For the terms $V^{(\bold{k},\bold{q})_{-1}}_{J_{\bold{l},\bold{i}}}$ with $J_{\bold{l},\bold{i}}\cap J_{\bold{k},\bold{q}}\neq\emptyset$, but   $J_{\bold{k},\bold{q}} \nsubset J_{\bold{l},\bold{i}}$ and   $J_{\bold{l},\bold{i}} \nsubset J_{\bold{k},\bold{q}}$, we write
\begin{equation}
e^{S_{J_{\bold{k},\bold{q}}}}\,V^{(\bold{k},\bold{q})_{-1}}_{J_{\bold{l},\bold{i}}}\,e^{-S_{J_{\bold{k},\bold{q}}}}=V^{(\bold{k},\bold{q})_{-1}}_{J_{\bold{l},\bold{i}}}+\sum_{n=1}^{\infty}\frac{1}{n!}\,ad^{n}S_{J_{\bold{k},\bold{q}}}(V^{(\bold{k},\bold{q})_{-1}}_{J_{\bold{l},\bold{i}}}) \,,\label{growth}
\end{equation}
where the first term on the r-h-s is  $V^{(\bold{k},\bold{q})}_{J_{\bold{l},\bold{i}}}$ by definition (see item a) in Definition \ref{def-interactions-multi}), and the second term contributes to the potential $V^{(\bold{k},\bold{q})}_{J_{\bold{r},\bold{j}}}$, where $J_{\bold{r}, \bold{j}} \equiv [J_{\bold{l},\bold{i}}\cup J_{\bold{k}, \bold{q}}]$, along with analogous terms contained 
in the second sum on the r-h-s of formula (\ref{construction-conn}) (where $\bold{i}$ is replaced by $\bold{j}$), and with
\begin{equation}\label{extra}
e^{S_{J_{\bold{k},\bold{q}}}}\,V^{(\bold{k},\bold{q})_{-1}}_{J_{\bold{r},\bold{j}}}\,e^{-S_{J_{\bold{k},\bold{q}}}}\,.
\end{equation}
 Notice that the term in (\ref{extra}) corresponds to the first term in (\ref{construction-conn}) (where $\bold{i}$ is replaced by $\bold{j}$). 
\end{itemize} 
\hspace{14cm} \qed

In the remainder of this section we reproduce a key result, established in \cite{FP}, which enables us to estimate the 
spectral gap above the ground-state energy of the Hamiltonian $G_{J_{\bold{k},\bold{q}}}$. The proof is included for 
the convenience of the reader; but the arguments are essentially identical to those used in \cite{FP}.
As for chains ($d=1$, see \cite{FP}), it is not difficult to prove that, under the assumption that
\begin{equation}\label{ass-2-multi}
\|V^{(\bold{k},\bold{q})_{-1}}_{J_{\bold{l},\bold{i}}}\| \leq t^{\frac{l-1}{4}}\,\quad,\quad l=|\bold{l}|:=l_1+l_2+\cdots+l_d\,,\quad \text{for  }\,\, 0\leq t {\color{brown}< }t_d\,,
\end{equation} 
the Hamiltonian $G_{J_{\bold{k}, \bold{q}}}$ has a gap $\Delta_{J_{\bold{k}, \bold{q}}} \geq \frac{1}{2}$, for all 
$ t \in [0, t_{d})$, where $t_d$ depends on the lattice dimension but is independent of $(\bold{k},\bold{q})$ and $N$. The main ingredients for the proof can be found in Lemma \ref{op-ineq-1} and Corollary \ref{op-ineq-2}; namely
\begin{equation}\label{gen-lemmaA-1}
P^{(+)}_{J_{{\bold{l}, \bold{i}}}}\leq \sum_{\bold{j}\in J_{\bold{l},\bold{i}} } P^{\perp}_{\Omega_{\bold{j}}} \,,
\end{equation}
and
\begin{eqnarray}\label{gen-lemmaA-2}
\sum_{\bold{i}\,:\,J_{\bold{l},\bold{i}}\subset  J_{\bold{k},\bold{q}}}P^{(+)}_{J_{\bold{l},\bold{i}}}&\leq &\Big\{\prod_{j=1}^{d}(l_j+1)\Big\} \sum_{\bold{i}\in  J_{\bold{k},\bold{q}}} P^{\perp}_{\Omega_{\bold{i}}}\,\\
&\leq& (l+1)^d\sum_{\bold{i}\in  J_{\bold{k},\bold{q}}} P^{\perp}_{\Omega_{\bold{i}}}\,.
\end{eqnarray}
\begin{rem}\label{shapes}
Observe that the number of shapes\footnote{By shape we mean an equivalence class of rectangles that can be obtained from one another by translation on the lattice.} of rectangles $J_{\bold{l},\bold{i}}$ at fixed $|\bold{l}|=l$ is bounded above by  $(l+1)^{d-1}= O(l^{d-1})$. As a consequence:
\begin{itemize}
\item[a)] the number of rectangles
$J_{\bold{k}, \bold{q}}\subset J_{\bold{r}, \bold{i}}$ with fixed circumference $k$ is bounded by\\
$(r+1)^d (k+1)^{d-1}= O(r^d k^{d-1})$;
\item[b)]  the number of rectangles
$J_{\bold{k}', \bold{q}'}\subset J_{\bold{r}, \bold{i}}$ is then  bounded by $(r+1)^d\sum_{k=1}^r (k+1)^{d-1}= O (r^{2d})$;
 \item[c)] the number of rectangles in $\mathcal{G}^{(\bold{k},\bold{q})}_{J_{\bold{r},\bold{i}}}$ is bounded by
$2d(r+1)^{d-1}\sum_{k=1}^r (k+1)^{d-1}= O(r^{2d-1})$.\footnote{The factor $2d(r+1)^{d-1}$ is an upper bound to the number of sites that sit on  one of the faces of the rectangle $J_{\bold{r}, \bold{i}}$. By definition, the rectangles in
$\mathcal{G}^{(\bold{k},\bold{q})}_{J_{\bold{r},\bold{i}}}$ have non-empty intersection with at least one of the faces of
$J_{\bold{r}, \bold{i}}$. }
\end{itemize}
\end{rem}
\begin{rem}\label{remark-decomp}
Our block-diagonalization procedure relies on the following crucial property: If $V^{(\bold{k},\bold{q})}_{J_{\bold{l},\bold{i}}}$ is block-diagonal w.r.t. the decomposition of the identity into 
$$\charf = P^{(+)}_{J_{\bold{l},\bold{i}}}+P^{(-)}_{J_{\bold{l},\bold{i}}}\,,$$
i.e., if
$$V^{(\bold{k},\bold{q})}_{J_{\bold{l},\bold{i}}}=P^{(+)}_{J_{\bold{l},\bold{i}}}V^{(\bold{k},\bold{q})}_{J_{\bold{l},\bold{i}}}P^{(+)}_{J_{\bold{l},\bold{i}}}+P^{(-)}_{J_{\bold{l},\bold{i}}}V^{(\bold{k},\bold{q})}_{J_{\bold{l},\bold{i}}}P^{(-)}_{J_{\bold{l},\bold{i}}}\,\,,$$  then we have that 
$$P^{(+)}_{J_{\bold{l}',\bold{i}'}}\Big[P^{(+)}_{J_{\bold{l},\bold{i}}}V^{(\bold{k},\bold{q})}_{J_{\bold{l},\bold{i}}}P^{(+)}_{J_{\bold{l},\bold{i}}}+P^{(-)}_{J_{\bold{l},\bold{i}}}V^{(\bold{k},\bold{q})}_{J_{\bold{l},\bold{i}}}P^{(-)}_{J_{\bold{l},\bold{i}}}\Big]P^{(-)}_{J_{\bold{l}',\bold{i}'}}=0\,,$$
for  $J_{\bold{l}',\bold{i}'}$ with $J_{\bold{l},\bold{i}}\subset J_{\bold{l}',\bold{i}'}$. This is seen 
by using
\begin{equation}
P^{(+)}_{J_{\bold{l},\bold{i}}}\,P^{(-)}_{J_{\bold{l}',\bold{i}'}}=0\,
\end{equation}
in the first term, and 
\begin{equation}
P^{(-)}_{J_{\bold{l},\bold{i}}}V^{(\bold{k},\bold{q})}_{J_{\bold{l},\bold{i}}}P^{(-)}_{J_{\bold{l},\bold{i}}}\,P^{(-)}_{J_{{\bold{l}',\bold{i}'}}}=P^{(-)}_{J_{\bold{l}',\bold{i}'}}P^{(-)}_{J_{\bold{l},\bold{i}}}V^{(\bold{k},\bold{q})}_{J_{\bold{l},\bold{i}}}P^{(-)}_{J_{\bold{l},\bold{i}}}P^{(-)}_{J_{\bold{l}',\bold{i}'}}\,,
\end{equation}
combined with
\begin{equation}
P^{(+)}_{J_{\bold{l}',\bold{i}'}}P^{(-)}_{J_{\bold{l}',\bold{i}'}}=0\,,
\end{equation}
in the second term.
\end{rem}

\begin{lem}\label{bound-lemma-gap}
Assuming (\ref{ass-2-multi}), the following bound on the operator $G_{J_{\bold{k},\bold{q}}}$ holds{\color{magenta}:}
\begin{align}\label{specG}
P^{(+)}_{J_{\bold{k},\bold{q}}} G_{J_{\bold{k},\bold{q}}} P^{(+)}_{J_{\bold{k},\bold{q}}}
&\geq  \Big(1-2\cdot C_d\cdot t \sum_{l=1}^{\infty}(l+1)^{2d}\cdot t^{\frac{l-1}{4}} \Big)\,P^{(+)}_{J_{\bold{k},\bold{q}}} + \nonumber \\
&+P^{(+)}_{J_{\bold{k},\bold{q}}}\,\Big[t\sum_{J_{\bold{k}_{(1)}',\bold{q}'}\subset J_{\bold{k},\bold{q}}} \langle V^{(\bold{k},\bold{q})_{-1}}_{J_{\bold{k}_{(1)}',\bold{q}'}}\rangle +\dots+t\sum_{J_{\bold{k}'_{(|\bold{k}|-1)},\bold{q}'} \subset J_{\bold{k},\bold{q}}}\langle V^{(\bold{k},\bold{q})_{-1}}_{J_{\bold{k}'_{(|\bold{k}|-1)}},\bold{q}'} \rangle \Big] P^{(+)}_{J_{\bold{k},\bold{q}}}
\end{align}
for $t\in [0, t_d)$, with $t_d$ independent of $(\bold{k},\bold{q})$ and $N$, 
where $C_d$ is the $d$-dependent constant implicit in the estimate of the number of shapes in Remark \ref{shapes}.
\end{lem}

\noindent
\emph{Proof} 
\noindent
We observe that{\color{magenta},}  due to Remark \ref{remark-decomp}, for $1\leq j \leq |\bold{k}|-1$ we can write
\begin{eqnarray}
& &P^{(+)}_{J_{\bold{k},\bold{q}}}\sum_{J_{\bold{k}_{(j)}',\bold{q}'}\subset J_{\bold{k},\bold{q}}}  V^{(\bold{k},\bold{q})_{-1}}_{J_{\bold{k}_{(j)}',\bold{q}'}} P^{(+)}_{J_{\bold{k},\bold{q}}} \label{gap-1}\\
&=&P^{(+)}_{J_{\bold{k},\bold{q}}}\sum_{J_{\bold{k}_{(j)}',\bold{q}'}\subset J_{\bold{k},\bold{q}}} P^{(+)}_{J_{\bold{k}_{(j)}',\bold{q}'}} V^{(\bold{k},\bold{q})_{-1}}_{J_{\bold{k}_{(j)}',\bold{q}'}}P^{(+)}_{J_{\bold{k}_{(j)}',\bold{q}'}}\, P^{(+)}_{J_{\bold{k},\bold{q}}}+P^{(+)}_{J_{\bold{k},\bold{q}}}\sum_{J_{\bold{k}_{(j)}',\bold{q}'}\subset J_{\bold{k},\bold{q}}} \langle  V^{(\bold{k},\bold{q})_{-1}}_{J_{\bold{k}_{(j)}',\bold{q}'}}\rangle P^{(-)}_{J_{\bold{k}_{(j)}',\bold{q}'}}\, P^{(+)}_{J_{\bold{k},\bold{q}}}\nonumber
\end{eqnarray}
where $$\langle V^{(\bold{k},\bold{q})_{-1}}_{J_{\bold{l},\bold{i}}}\rangle:=\langle  \Big(\otimes \prod_{\bold{j}\in J_{\bold{l},\bold{i}}} P_{\Omega_{\bold{j}}}\Big)\,,\,  V^{(\bold{k},\bold{q})_{-1}}_{J_{\bold{l}, \bold{i}}} \Big(\otimes \prod_{\bold{j}\in J_{\bold{l},\bold{i}}}P_{\Omega_{\bold{j}}}\Big)\rangle\,.$$
Furthermore,  we can estimate
\begin{eqnarray}
& &
\pm P^{(+)}_{J_{\bold{k},\bold{q}}}\sum_{J_{\bold{k}_{(j)}',\bold{q}'}\subset J_{\bold{k},\bold{q}}} P^{(+)}_{J_{\bold{k}_{(j)}',\bold{q}'}} V^{(\bold{k},\bold{q})_{-1}}_{J_{\bold{k}_{(j)}',\bold{q}'}}P^{(+)}_{J_{\bold{k}_{(j)}',\bold{q}'}}\, P^{(+)}_{J_{\bold{k},\bold{q}}}\\
&\leq &P^{(+)}_{J_{\bold{k},\bold{q}}}\sum_{J_{\bold{k}_{(j)}',\bold{q}'}\subset J_{\bold{k},\bold{q}}} \|V^{(\bold{k},\bold{q})_{-1}}_{J_{\bold{k}_{(j)}',\bold{q}'}}\|P^{(+)}_{J_{\bold{k}_{(j)}',\bold{q}'}}\,P^{(+)}_{J_{\bold{k},\bold{q}}}\label{gap-2}\\
&\leq &P^{(+)}_{J_{\bold{k},\bold{q}}} \, t^{\frac{j-1}{4}}\,\sum_{J_{\bold{k}_{(j)}',\bold{q}'}\subset J_{\bold{k},\bold{q}}} P^{(+)}_{J_{\bold{k}_{(j)}',\bold{q}'}}\, P^{(+)}_{J_{\bold{k},\bold{q}}}\label{gap-3}\\
&\leq &P^{(+)}_{J_{\bold{k},\bold{q}}}C_d\cdot (j+1)^{2d}\cdot t^{\frac{j-1}{4}} \sum_{\bold{i}\in  J_{\bold{k}, \bold{q}}} P^{\perp}_{\Omega_{\bold{i}}} \, P^{(+)}_{J_{\bold{k},\bold{q}}}\label{gap-4}
\end{eqnarray}
where we have used
\begin{itemize}
\item[1)] the bound in (\ref{ass-2-multi}) for the step from  (\ref{gap-2}) to (\ref{gap-3});
\item[2)] the property in (\ref{gen-lemmaA-2})  combined with Remark \ref{shapes} for the step from  (\ref{gap-3}) to (\ref{gap-4}).
\end{itemize}
Hence, we can combine the inequality  (due to (\ref{gaps}))
\begin{equation}\label{ineq-free}
\sum_{\bold{i}\subset J_{\bold{k},\bold{q}} }H_{\bold{i}} \geq  \sum_{\bold{i}\in J_{\bold{k},\bold{q}} }  P^{\perp}_{\Omega_{\bold{i}}} \,
\end{equation}
with (\ref{gap-1})-(\ref{gap-4}), and we get
\begin{eqnarray}
& &P^{(+)}_{J_{\bold{k},\bold{q}}}G_{J_{\bold{k},\bold{q}}}P^{(+)}_{J_{\bold{k},\bold{q}}}\\
&\geq &P^{(+)}_{J_{\bold{k},\bold{q}}}\,\Big[ (1- C_d\cdot t \sum_{l=1}^{\infty}(l+1)^{2d}\cdot t^{\frac{l-1}{4}} \Big)\,  \sum_{\bold{i}\in J_{\bold{k},\bold{q}} } P^{\perp}_{\Omega_{\bold{i}}} \Big]P^{(+)}_{J_{\bold{k},\bold{q}}}\quad\quad\quad\quad \label{gap-eq-1} \\
& &+P^{(+)}_{J_{\bold{k},\bold{q}}}\,\Big[t\sum_{J_{\bold{k}_{(1)}',\bold{q}'}\subset J_{\bold{k},\bold{q}}} \langle V^{(\bold{k},\bold{q})_{-1}}_{J_{\bold{k}_{(1)}',\bold{q}'}}\rangle P^{(-)}_{J_{\bold{k}_{(1)}',\bold{q}'}}  +\dots+t\sum_{J_{\bold{k}'_{(|\bold{k}|-1)},\bold{q}'} \subset J_{\bold{k},\bold{q}}}\langle V^{(\bold{k},\bold{q})_{-1}}_{J_{\bold{k}'_{(|\bold{k}|-1)}},\bold{q}'}}\rangle P^{(-)}_{J_{\bold{k}_{(|\bold{k}|-1)}',\bold{q}'}} \Big]P^{(+)}_{J_{\bold{k},\bold{q}}\,.\quad \quad\quad 
\end{eqnarray}
Next, we use the identity$$ \charf=P^{(-)}_{J_{\bold{k}_{(j)}',\bold{q}'}} +P^{(+)}_{J_{\bold{k}_{(j)}',\bold{q}'}} $$
in the r-h-s of (\ref{gap-eq-1}), and we get
\begin{eqnarray}
& &P^{(+)}_{J_{\bold{k},\bold{q}}}G_{J_{\bold{k},\bold{q}}}P^{(+)}_{J_{\bold{k},\bold{q}}}\\
&\geq &P^{(+)}_{J_{\bold{k},\bold{q}}}\,\Big[ (1- C_d\cdot t \sum_{l=1}^{\infty}(l+1)^{2d}\cdot t^{\frac{l-1}{4}} \Big)\,  \sum_{\bold{i}\in J_{\bold{k},\bold{q}} }  P^{\perp}_{\Omega_{\bold{i}}} \Big]P^{(+)}_{J_{\bold{k},\bold{q}}}\quad\quad\quad\quad \\
& &-P^{(+)}_{J_{\bold{k},\bold{q}}}\,\Big[t\sum_{J_{\bold{k}_{(1)}',\bold{q}'}\subset J_{\bold{k},\bold{q}}} \langle V^{(\bold{k},\bold{q})_{-1}}_{J_{\bold{k}_{(1)}',\bold{q}'}}\rangle P^{(+)}_{J_{\bold{k}_{(1)}',\bold{q}'}}  +\dots+t\sum_{J_{\bold{k}'_{(|\bold{k}|-1)},\bold{q}'} \subset J_{\bold{k},\bold{q}}}\langle V^{(\bold{k},\bold{q})_{-1}}_{J_{\bold{k}'_{(|\bold{k}|-1)}},\bold{q}'}}\rangle P^{(+)}_{J_{\bold{k}_{(|\bold{k}|-1)}',\bold{q}'}} \Big]P^{(+)}_{J_{\bold{k},\bold{q}}\quad\quad\quad \\
& &+P^{(+)}_{J_{\bold{k},\bold{q}}}\,\Big[t\sum_{J_{\bold{k}_{(1)}',\bold{q}'}\subset J_{\bold{k},\bold{q}}} \langle V^{(\bold{k},\bold{q})_{-1}}_{J_{\bold{k}_{(1)}',\bold{q}'}}\rangle  +\dots+t\sum_{J_{\bold{k}'_{(|\bold{k}|-1)},\bold{q}'} \subset J_{\bold{k},\bold{q}}}\langle V^{(\bold{k},\bold{q})_{-1}}_{J_{\bold{k}'_{(|\bold{k}|-1)}},\bold{q}'}}\rangle  \Big]P^{(+)}_{J_{\bold{k},\bold{q}}\,.
\end{eqnarray}
By invoking the obvious bound $$|\langle V^{(\bold{k},\bold{q})_{-1}}_{J_{\bold{k}'_{(j)},\bold{q}'}}\rangle|\leq \|\langle V^{(\bold{k},\bold{q})_{-1}}_{J_{\bold{k}'_{(j)}},\bold{q}'}\rangle\|$$ we finally get
\begin{eqnarray}
& &P^{(+)}_{J_{\bold{k},\bold{q}}}G_{J_{\bold{k},\bold{q}}}P^{(+)}_{J_{\bold{k},\bold{q}}}\label{gap-in}\\
&\geq &P^{(+)}_{J_{\bold{k},\bold{q}}}\,\Big[ (1-2\cdot C_d\cdot t \sum_{l=1}^{\infty}(l+1)^{2d}\cdot t^{\frac{l-1}{4}} \Big)\,  \sum_{\bold{i}\in J_{\bold{k},\bold{q}} }  P^{\perp}_{\Omega_{\bold{i}}} \Big]P^{(+)}_{J_{\bold{k},\bold{q}}}\quad\quad\quad\quad \\
& &+P^{(+)}_{J_{\bold{k},\bold{q}}}\,\Big[t\sum_{J_{\bold{k}_{(1)}',\bold{q}'}\subset J_{\bold{k},\bold{q}}} \langle V^{(\bold{k},\bold{q})_{-1}}_{J_{\bold{k}_{(1)}',\bold{q}'}}\rangle  +\dots+t\sum_{J_{\bold{k}'_{(|\bold{k}|-1)},\bold{q}'} \subset J_{\bold{k},\bold{q}}}\langle V^{(\bold{k},\bold{q})_{-1}}_{J_{\bold{k}'_{(|\bold{k}|-1)}},\bold{q}'}}\rangle  \Big]P^{(+)}_{J_{\bold{k},\bold{q}}\\
&\geq & 
\Big(1-2\cdot C_d\cdot t \sum_{l=1}^{\infty}(l+1)^{2d}\cdot t^{\frac{l-1}{4}} \Big)\,P^{(+)}_{J_{\bold{k},\bold{q}}} \\
& &+P^{(+)}_{J_{\bold{k},\bold{q}}}\,\Big[t\sum_{J_{\bold{k}_{(1)}',\bold{q}'}\subset J_{\bold{k},\bold{q}}} \langle V^{(\bold{k},\bold{q})_{-1}}_{J_{\bold{k}_{(1)}',\bold{q}'}}\rangle  +\dots+t\sum_{J_{\bold{k}'_{(|\bold{k}|-1)},\bold{q}'} \subset J_{\bold{k},\bold{q}}}\langle V^{(\bold{k},\bold{q})_{-1}}_{J_{\bold{k}'_{(|\bold{k}|-1)}},\bold{q}'}}\rangle  \Big]P^{(+)}_{J_{\bold{k},\bold{q}}\label{gap-fin}
\end{eqnarray}
where Lemma \ref{op-ineq-1} is used for the last inequality, and $t(>0)$ is assumed small enough such that
\begin{equation}
1-2\cdot C_d\cdot t \sum_{l=1}^{\infty}(l+1)^{2d}\cdot t^{\frac{l-1}{4}} >0\,.
\end{equation}
\qed


Lemma \ref{bound-lemma-gap} implies that under the assumption in (\ref{ass-2-multi}) the Hamiltonian $G_{J_{\bold{k},\bold{q}}}$ has a gap that can be estimated from below by $\frac{1}{2}$ for $t>0$ sufficiently small but independent of $N$ and $(\bold{k},\bold{q})$, as  stated in the Corollary below.
\begin{cor}\label{cor-gap}
Assuming Lemma \ref{bound-lemma-gap},  for $t>0$ sufficiently small, dependent on $d$ but independent of $N$ and $(\bold{k},\bold{q})$, the Hamiltonian $G_{J_{\bold{k},\bold{q}}}$ has a gap $\Delta_{J_{\bold{k},\bold{q}}}\geq \frac{1}{2}$ above the ground state energy $$E_{J_{\bold{k},\bold{q}}}=t\sum_{J_{\bold{k}_{(1)}',\bold{q}'}\subset J_{\bold{k},\bold{q}}} \langle V^{(\bold{k},\bold{q})_{-1}}_{J_{\bold{k}_{(1)}',\bold{q}'}}\rangle +\dots+t\sum_{J_{\bold{k}'_{(j_{\bold{k}-1})},\bold{q}'} \subset J_{\bold{k},\bold{q}}} \langle V^{(\bold{k},\bold{q})_{-1}}_{J_{\bold{k}'_{(j_{\bold{k}-1})},\bold{q}'}} \rangle 
$$ corresponding to the  ground state vector $\bigotimes_{\bold{i}\in J_{\bold{k},\bold{q}}}\Omega_{\bold{i}}\,$, due to the identity
\begin{eqnarray}
& &P^{(-)}_{J_{\bold{k},\bold{q}}}G_{J_{\bold{k},\bold{q}}}P^{(-)}_{J_{\bold{k},\bold{q}}}\\
&= &P^{(-)}_{J_{\bold{k},\bold{q}}}\,\Big[t\sum_{J_{\bold{k}_{(1)}',\bold{q}'}\subset J_{\bold{k},\bold{q}}} \langle V^{(\bold{k},\bold{q})_{-1}}_{J_{\bold{k}_{(1)}',\bold{q}'}}\rangle P^{(-)}_{J_{\bold{k}_{(1)}',\bold{q}'}}  +\dots+t\sum_{J_{\bold{k}'_{(|\bold{k}|-1)},\bold{q}'} \subset J_{\bold{k},\bold{q}}}\langle V^{(\bold{k},\bold{q})_{-1}}_{J_{\bold{k}'_{(|\bold{k}|-1)}},\bold{q}'}}\rangle P^{(-)}_{J_{\bold{k}_{(|\bold{k}|-1)}',\bold{q}'}} \Big] P^{(-)}_{J_{\bold{k},\bold{q}}\nonumber \\
&= &P^{(-)}_{J_{\bold{k},\bold{q}}}\,\Big[t\sum_{J_{\bold{k}_{(1)}',\bold{q}'}\subset J_{\bold{k},\bold{q}}} \langle V^{(\bold{k},\bold{q})_{-1}}_{J_{\bold{k}_{(1)}',\bold{q}'}}\rangle  +\dots+t\sum_{J_{\bold{k}'_{(|\bold{k}|-1)},\bold{q}'} \subset J_{\bold{k},\bold{q}}}\langle V^{(\bold{k},\bold{q})_{-1}}_{J_{\bold{k}'_{(|\bold{k}|-1)}},\bold{q}'}}\rangle  \Big] P^{(-)}_{J_{\bold{k},\bold{q}}\,.\nonumber 
\end{eqnarray}
\end{cor}

\section{Control of $\|V^{(\bold{k},\bold{q})}_{J_{\bold{r},\bold{i}}}\|$} \label{proofs}

The next theorem is the key result of the paper and is based on a lengthy analysis  of the different regimes (outlined  in Sect. \ref{algo})  to control the potentials yielded, step by step,  by the algorithm in Definition \ref{def-interactions-multi}.

\begin{thm}\label{th-norms}
There exists $t_d>0$ such that for $0\leq t < t_{d}$ the Hamiltonians $G_{J_{\bold{k},\bold{q}}}$ and $K_{N^d}^{(\bold{k},\bold{q})}$ are well defined, and for any rectangle $J_{\bold{r},\bold{i}}$, with $r=|\bold{r}|\geq 1$, and for $x_d:=20d$, we have: 

\noindent
S1) 

\noindent
Let $(\bold{k},\bold{q})_* := (\bold{k}_*,\bold{q}_*)$ be defined for some $(\bold{k}_*,\bold{q}_*)$ such that $ |\bold{k}_*|=\lfloor r^{\frac{1}{4}} \rfloor$, where $\lfloor \cdot \rfloor$ is the integer part. If  $(\bold{k},\bold{q})\prec (\bold{k},\bold{q})_*$, then
 \begin{equation}\label{inductive-reg-1}
 \|V^{(\bold{k},\bold{q})}_{J_{\bold{r},\bold{i}}}\|\leq \frac{t^{\frac{r-1}{3}}}{r^{x_d+2d}}\,;
 \end{equation}

\noindent
 Let $ (\bold{k},\bold{q})_{**}:= (\bold{k}_{**},\bold{q}_{**})$ be defined for some $(\bold{k}_{**},\bold{q}_{**})$  such that $|\bold{k}_{**}| = r-\lfloor r^{\frac{1}{4}} \rfloor $. If $ (\bold{k},\bold{q})_{**}\succ (\bold{k},\bold{q})\succeq (\bold{k},\bold{q})_{*} $, then 
\begin{equation}
\|V^{(\bold{k},\bold{q})}_{J_{\bold{r},\bold{i}}}\|\leq 2 \cdot \frac{t^{\frac{r-1}{3}}}{r^{x_d+2d}}\,;\label{inductive-reg-2}
\end{equation}
 
\noindent
  If $ (\bold{r},\bold{i})\succ (\bold{k},\bold{q}) \succeq  (\bold{k},\bold{q})_{**}$, then
\begin{equation}
\| \frac{1}{\sum_{\bold{j}\in J_{\bold{r},\bold{i}}} P^{\perp}_{\Omega_{\bold{j}}}+1}\,P^{(\#)}_{J_{\bold{r},\bold{i}}}V^{(\bold{k},\bold{q})}_{J_{\bold{r},\bold{i}}}P^{(\hat{\#})}_{J_{\bold{r},\bold{i}}}\,\frac{1}{\sum_{\bold{j}\in J_{\bold{r},\bold{i}}} P^{\perp}_{\Omega_{\bold{j}}}+1}\|\leq  3\cdot\frac{t^{\frac{r-1}{3}}}{r^{x_d+2d}}\,, \quad \#, \hat{\#}=\pm \,,   \label{R3-1}
\end{equation}
and
\begin{equation}
\| V^{(\bold{k},\bold{q})}_{J_{\bold{r},\bold{i}}}\|\leq  48  \cdot \frac{t^{\frac{r-1}{3}}}{r^{x_d}}\,; \label{R3-2}
\end{equation}

\noindent
 If $(\bold{k},\bold{q})\succeq  (\bold{r},\bold{i})$, then
\begin{equation}
\| V^{(\bold{k},\bold{q})}_{J_{\bold{r},\bold{i}}}\|\leq  96  \cdot \frac{t^{\frac{r-1}{3}}}{r^{x_d}}\,.  \label{R3-3}
\end{equation}

\noindent
S2)  

\noindent
$G_{J_{(\bold{k},\bold{q})_{+1}}}$ has spectral gap  $\Delta_{J_{(\bold{k},\bold{q})_{+1}}}\geq \frac{1}{2}$ above its ground state energy,  where $G_{J_{\bold{k},\bold{q}}}$ is defined in (\ref{def-G})  for $|\bold{k}|\geq 2$,  and $$G_{J_{(\bold{1}_j,\bold{q})_{+1}}}:=H^{(0)}_{J_{(\bold{1}_j,\bold{q})_{+1}}}:=\sum_{\bold{i}\in J_{(\bold{1}_j,\bold{q})_{+1}}}H_{\bold{i}}$$  provided $(\bold{1}_j,\bold{q})_{+1}$ is of the form $(\bold{1}_{j'},\bold{q}')$ for some $j'$ and $\bold{q}'$; $ (\bold{1}_j,\bold{q})$ is defined in (\ref{def-1_j}).
\end{thm}

\noindent
\emph{Proof.}

The proof is by induction  in the diagonalization step $(\bold{k},\bold{q})$. Hence for each $(\bold{r},\bold{i})$ we shall prove S1) and S2) from $(\bold{k},\bold{q})=(\bold{0},\bold{N})$ up to $(\bold{k},\bold{q})=(\bold{N-1},\bold{1})$; (notice that in step $(\bold{k}, \bold{q})$  S2) concerns the Hamiltonian $G_{J_{(\bold{k},\bold{q})_{+1}}}$, and it  is not defined for $(\bold{k},\bold{q})=(\bold{N-1},\bold{1})$). That is we assume that S1) holds for all $V^{(\bold{k}',\bold{q}')}_{J_{\bold{r},\bold{i}}}$ with $(\bold{k}',\bold{q}') \prec (\bold{k},\bold{q})$ and  S2) for all $(\bold{k}',\bold{q}') \prec (\bold{k},\bold{q})$. Then we show that they hold for all $V^{(\bold{k},\bold{q})}_{J_{\bold{r},\bold{i}}}$ and for $G_{J_{(\bold{k},\bold{q})_{+1}}}$.   By Lemma \ref{control-LS}, this implies that $S_{J_{\bold{k},\bold{q}}}$, and, consequently, that $K_{N^d}^{(\bold{k},\bold{q})}$ are well defined operators (see (\ref{def-K})).

\noindent
For $(\bold{k},\bold{q})= (\bold{0},\bold{N})$,  S1) can be verified by direct computation,  because 
$$\|V_{J_{\bold{1}_j,\bold{q}}}^{(\bold{0},\bold{N})}\|=\| V_{J_{\bold{1}_j,\bold{q}}}\|\leq1\,,$$
and
$V_{J_{\bold{r},\bold{i}}}^{(\bold{0},\bold{N})}=0$ otherwise; S2) holds trivially since, by definition, $(\bold{0},\bold{N})_{+1}=(\bold{1}_1, \bold{1})$ and $G_{J_{\bold{1}_1,\bold{1}}}=H^{(0)}_{J_{\bold{1}_1,\bold{1}}}$ (recall that $\bold{1}_j$ is defined in (\ref{def-1_j})).

At each stage of our proof we choose $t(\geq 0)$ in an interval such that the previous stages  and Lemma \ref{control-LS} are verified. Hence by this procedure we may progressively restrict such an interval until we determine a $t_d>0$ for which all the stages hold true for $0\leq t <  t_d$. 

\emph{Warning}: Throughout the proof several positive constants are introduced. We shall use the symbols $c,C$ for those that are universal and the symbols $c_d, C_d$ for those that depend on the dimension $d$, and their value may change from line to line.
\\

\noindent
\emph{Induction step in the proof of S1)}

\noindent
Starting from Definition  \ref{def-interactions-multi} we consider the following cases:
\\

\noindent
\emph{Case $r=1$.}

\noindent
Let $k>1(=r)$ or $k=1(=r)$ but $J_{\bold{r},\bold{i}}$ such that $\bold{i} \neq \bold{q}$.  Then the possible cases are described in a), see Definition  \ref{def-interactions-multi}, and we have that
\begin{equation}
\|V^{(\bold{k},\bold{q})}_{J_{\bold{r},\bold{i}}}\|=\|V^{(\bold{k},\bold{q})_{-1}}_{J_{\bold{r},\bold{i}}}\|\,.
\end{equation} 

\noindent
Let $k=1$ and  assume that $J_{\bold{r},\bold{i}}$ is equal to $J_{\bold{k},\bold{q}}$. Then  we refer to case b) and find that
\begin{equation}
\|V^{(\bold{k},\bold{q})}_{J_{\bold{k},\bold{q}}}\|\leq 2\|V^{(\bold{k},\bold{q})_{-1}}_{J_{\bold{k},\bold{q}}}\|\leq 2\,,
\end{equation}
where:
\begin{itemize}
\item[i)] the inequality  $\|V^{(\bold{k},\bold{q})}_{J_{\bold{k},\bold{q}}}\|\leq 2\|V^{(\bold{k},\bold{q})_{-1}}_{J_{\bold{k},\bold{q}}}\|$ holds for $t(\geq 0)$ sufficiently small uniformly in $\bold{q}$ and $N$,  thanks to  Lemma \ref{control-LS} which can be applied since we assume S1) and S2) in step $(\bold{k},\bold{q})_{-1}$;
\item[ii)]  we use $ \|V^{(\bold{k},\bold{q})_{-1}}_{J_{\bold{k}, \bold{q}}}\|=\|V^{(\bold{k},\bold{q})_{-2}}_{J_{\bold{k}, \bold{q}}}\|=\dots =\|V^{(\bold{0},\bold{N})}_{J_{\bold{k}, \bold{q}}}\|\leq 1$. 
\end{itemize}
Inequality (\ref{R3-1}) follows trivially by using $\| \frac{1}{\sum_{\bold{j}\in J_{\bold{r},\bold{i}}} P^{\perp}_{\Omega_{\bold{j}}}+1}\|\leq 1$ and  $\|P^{(\#)}_{J_{\bold{r},\bold{i}}}V^{(\bold{k},\bold{q})}_{J_{\bold{r},\bold{i}}}P^{(\hat{\#})}_{J_{\bold{r},\bold{i}}}\|\leq \|V^{(\bold{k},\bold{q})}_{J_{\bold{r},\bold{i}}}\|\,.$\\

\noindent
\emph{Case $r=2$.}\\
This case is not much different from the one corresponding to $r=1$ with the exception that  also formula 
\begin{equation}\label{normest-c}
\|V^{(\bold{k}',\bold{q}')}_{J_{\bold{r},\bold{i}}} \| \leq  \|V^{(\bold{k}',\bold{q}')_{-1}}_{J_{\bold{r},\bold{i}}}\|\,+\|\sum_{J_{\bold{k}'',\bold{q}''}\in \mathcal{G}^{(\bold{k}',\bold{q}')}_{J_{\bold{r},\bold{i}}}}\,\sum_{n=1}^{\infty}\frac{1}{n!}\,ad^{n}S_{J_{\bold{k}',\bold{q}'}}(V^{(\bold{k}',\bold{q}')_{-1}}_{J_{\bold{k}'',\bold{q}''}})\|
\end{equation}
must be used  in the re-expansion, for some $(\bold{k}',\bold{q}')$ with $k'=1$, and then iterated for the first term of the r-h-s of (\ref{normest-c}) if the conditions of c) in Definition \ref{def-interactions-multi} are fulfilled. The second term in (\ref{normest-c}) is a remainder that, however, is produced  along the re-expansion only for a finite number of steps, and this number is bounded by a constant independent of $(\bold{k},\bold{q})$, $\bold{i}$, and $N$. Note also that, for $t>0$ sufficiently small,  the norm of the last term in (\ref{normest-c}) can be bounded by a constant multiplied by a factor $t$, using Lemma \ref{control-LS} and the inductive hypotheses S1), S2) for $r=1$. For $t(\geq 0)$ sufficiently small, these observations suffice to state S1) for rectangles with $r=2$, provided S1) and S2) hold for $r=1$.
\\

\noindent
\emph{Case $r> 2$.}

\noindent
As explained in Sect. \ref{algo}, in order to control the norm $\|V^{(\bold{k},\bold{q})}_{J_{\bold{r},\bold{i}}} \|$   we distinguish three regimes  depending on the relative magnitude between  $k=|\bold{k}|$ and $r=|\bold{r}|$. They are  associated with (\ref{inductive-reg-1}), (\ref{inductive-reg-2}), and (\ref{R3-1})-(\ref{R3-2})-(\ref{R3-3}), respectively. 
For the convenience of the reader, we recall how the inductive hypotheses are used in the following analysis of the three regimes. By assuming that (\ref{inductive-reg-1}), (\ref{inductive-reg-2}), (\ref{R3-1}), (\ref{R3-2}), and (\ref{R3-3}) are true for the potentials associated with any rectangle $J_{\bold{l}',\bold{i}'}$, with $(\bold{l}', \bold{i}')\prec (\bold{r}, \bold{i})$,  in steps $(\bold{k}', \bold{q}')\prec (\bold{k}, \bold{q})$, we prove  that, depending on the considered regime, (\ref{inductive-reg-1}), (\ref{inductive-reg-2}), and  (\ref{R3-1}) hold, respectively,   in step $(\bold{k}, \bold{q})$ for the potential associated with $J_{\bold{r},\bold{i}}$; but if (\ref{R3-1}) is verified then, consequently,   also (\ref{R3-2}) and (\ref{R3-3}) hold true (in step $(\bold{k}, \bold{q})$)\,.
\\

\noindent
\underline{\emph{Regime $\mathfrak{R}1)$} }
\\

Here we apply the argument explained in Section \ref{sum-branches} in order to show that S1) holds for $V^{(\bold{k},\bold{q})}_{J_{\bold{r},\bold{i}}}$ with $(\bold{k}, \bold{q})$  belonging to the first regime, provided S1)  and S2) hold for all potentials in step $(\bold{k}',\bold{q}')\prec (\bold{k},\bold{q})$. Given the assumption,  we can exploit (\ref{bound-S}) in Lemma \ref{control-LS} so as to conclude that  for any (bounded) operator $V$ 
\begin{equation}\label{norm-comm}
\|\mathcal{A}_{J_{\bold{k}'',\bold{q}''}}(\,V \, )\|\leq c\cdot t \cdot  \|V^{(\bold{k}'',\bold{q}'')_{-1}}_{J_{\bold{k}'',\bold{q}''}} \|\cdot \|V\|
\end{equation}
if $(\bold{k}'',\bold{q}'')\preceq (\bold{k},\bold{q})$, where $c$ is a universal constant.
We recall that, as explained in Section 3, the strategy is to re-expand the potential
$V^{(\bold{k}, \bold{q})}_{J_{\bold{r}, \bold{i}}}$
according to the prescriptions of Definition \ref{def-tree}. Consequently, the potential 
can be expressed as the sum $\sum_{\mathfrak{b}\in  \mathcal{B}_{V^{(\bold{k},\bold{q})}_{J_{\bold{r},\bold{i}}}}}\mathfrak{b}$, where $\mathfrak{b}$ are the branch operators defined in point 8. of Definition \ref{def-tree}.
Due to property P-v) in Section \ref{properties}, the number of summands coincides with the number of sets $\mathcal{R}_{\mathfrak{b}}$ that are associated with  $V^{(\bold{k},\bold{q})}_{J_{\bold{r},\bold{i}}}$. Furthermore, in order to estimate the norm of the sum of the operators resulting from the re-expansion,  it is enough to use (\ref{norm-comm}) repeatedly, i.e.,
\begin{equation}
\|\mathcal{A}_{J_{\bold{k}^{(1)},\bold{q}^{(1)}}}(\,\mathcal{A}_{J_{\bold{k}^{(2)},\bold{q}^{(2)}}} (\cdots  \mathcal{A}_{J_{\bold{k}^{(\vert \mathcal{R}_{\mathfrak{b}}\vert-1)},\bold{q}^{(\vert \mathcal{R}_b\vert -1)}}}(V_{\mathcal{L}_\mathfrak{b}})\cdots )\, )\|\leq (c\cdot t)^{\vert \mathcal{R}_{\mathfrak{b}}\vert  -1}  \|V_{\mathcal{L}_\mathfrak{b}} \|  \prod_{i=1}^{\vert \mathcal{R}_{\mathfrak{b}}\vert -1}\|V^{(\bold{k}^{(i)},\bold{q}^{(i)})_{-1}}_{J_{\bold{k}^{(i)},\bold{q}^{(i)}}} \|\,\label{r1-estimate--1}
\end{equation} 
where $V_{\mathcal{L}_\mathfrak{b}}$ is the potential labelling the leaf of $\mathfrak{b}$, and  compute the ``weighted" number of sets $\{\,J_{\bold{k}^{(i)},\bold{q}^{(i)}},\, i\in\{1,\cdots,\vert \mathcal{R}_{\mathfrak{b}}\vert \} \}$, weighted in the sense that each rectangle $J_{\bold{k}^{(i)},\bold{q}^{(i)}} $ is given the weight $c \cdot t \cdot \| V^{(\bold{k}^{(i)},\bold{q}^{(i)})_{-1}}_{J_{\bold{k}^{(i)},\bold{q}^{(i)}} }\| $ except for the one associated with the leaf of the branch, that is given the weight $\|V_{\mathcal{L}_\mathfrak{b}}\|$. 

\noindent Following the scheme described in Sect. \ref{sum-branches}, we estimate the weighted sum of sets $\mathcal{R}_{\mathfrak{b}}$ in terms of a weighted sum of paths $\Gamma_{\mathfrak{b}}$. Differently from Sect. \ref{sum-paths}, here we assign the weight to each step after extracting from (\ref{r1-estimate--1}) what is needed to provide the bound in (\ref{inductive-reg-1}). The overall control will be ensured by the pre-factor $(c\cdot t)^{\vert \mathcal{R}_{\mathfrak{b}}\vert -1}$ that is small enough due to  the upper bound on $k$, $k\leq \lfloor  r^{\frac{1}{4}} \rfloor$,  in regime $\frak{R}1$. Indeed the latter implies the lower bound $\vert \mathcal{R}_{\mathfrak{b}}\vert\geq \lfloor c_d \cdot r/ k \rfloor$.

\noindent
In detail, concerning the powers of $t$, notice that from the product 
\begin{equation}
(c\cdot t)^{\vert \mathcal{R}_{\mathfrak{b}}\vert -1} \| V_{\mathcal{L}_\mathfrak{b}}\|\prod_{i=1}^{\vert \mathcal{R}_{\mathfrak{b}}\vert-1}\|V^{(\bold{k}^{(i)},\bold{q}^{(i)})_{-1}}_{J_{\bold{k}^{(i)},\bold{q}^{(i)}}} \|\label{product}
\end{equation}
we get at least $t^{\frac{r-1}{3}}$ due to: 1) the requirement that $J_{\bold{r},\bold{i}}$ is the minimal rectangle associated with $\cup_{i\in  \{1,\cdots,\vert \mathcal{R}_{\mathfrak{b}}\vert\}}J_{\bold{k}^{(i)},\bold{q}^{(i)}}$; 2) borrowing a power $t^{\frac{2}{3}}$ from each factor $t$ in $(c\cdot t)^{\vert \mathcal{R}_{\mathfrak{b}}\vert -1}$. Hence, in the product (\ref{product})  we can factor out $t^{\frac{r-1}{3}}$ and keep a power $t^{\frac{1}{3}}$ for each rectangle of $\mathcal{R}_{\mathfrak{b}}$ except the one associated with the leaf of the branch. This also means that we can assign at least a factor
\begin{equation}\label{weight-rect}
(c+1)\frac{t^{1/6}}{\rho^{x_d}}
\end{equation}
to each rectangle of size $\rho$ in $\mathcal{R}_{\mathfrak{b}}$.

Consider the rectangles of the set $\text{supp}(\mathcal{Z}^{(j)}_{\rho})$  (see Sect. \ref{features}): there are $n_{\rho}^{(j)}$ such rectangles, and, for the constructed paths $\Gamma_{\mathfrak{b}}$, there are at most $2n_{\rho}^{(j)}-2$ steps between them. In addition there are at most $2$ steps, from rectangles of lower size and back,  to be taken into account. To each  step $\mathcal{S}_{\Gamma_{\mathfrak{b}}}\ni\sigma=(J_{\bold{s}^{(i)},\bold{u}^{(i)}},J_{\bold{s}^{(i+1)},\bold{u}^{(i+1)}})$ we  assign the weight  
$$w_\sigma:=\Big((c+1)\frac{t^{1/6}}{s_\sigma^{x_d}}\Big)^{1/2}$$
where $s_\sigma:=\max\{s^{(i)},s^{(i+1)}\}$, with $w_\sigma<1$ for $t>0$ sufficiently small. 

\noindent
From the considerations regarding (\ref{product}) and (\ref{weight-rect}), we get the first inequality in the next formula (\ref{weight-path})
\begin{equation}\label{weight-path}
(c\cdot t)^{\vert \mathcal{R}_{\mathfrak{b}}\vert -1} \|V_{\mathcal{L}_\mathfrak{b}}\|\prod_{i=1}^{\vert \mathcal{R}_{\mathfrak{b}}\vert-1}\|V^{(\bold{k}^{(i)},\bold{q}^{(i)})_{-1}}_{J_{\bold{k}^{(i)},\bold{q}^{(i)}}} \|
\leq  t^{\frac{r-1}{3}} \prod_{\rho=1\,; 
\\ j_\rho\neq 0}^k\,\prod_{j=1}^{j_{\rho}}\Big((c+1) \frac{t^{1/6}}{\rho^{x_d}}\Big)^{n_\rho^{(j)}} \leq t^{\frac{r-1}{3}}\cdot \prod_{\sigma\in\mathcal{S}_{\Gamma_{\mathfrak{b}}}}w_\sigma ,
\end{equation}
whereas for the second inequality we use the following observation: if we denote by $\mathcal{S}_{\mathcal{Z}_\rho^{(j)}}$ the set consisting  of  at most $2n_\rho^{(j)}-2$ steps between rectangles of  $\text{supp}\mathcal{Z}_\rho^{(j)}$ and the additional at most $2$ steps from rectangles of lower size and back, then we have
$$ \Big((c+1)\frac{t^{\frac{1}{6}}}{\rho^{x_d}}\Big)^{n_\rho^{(j)}}\leq \prod_{\sigma\in\mathcal{S}_{\mathcal{Z}_\rho^{(j)}} } w_\sigma\,,$$
since $w_\sigma$, $\sigma\in\mathcal{S}_{\mathcal{Z}_\rho^{(j)}} $, coincides with $\Big((c+1)\frac{t^{\frac{1}{6}}}{\rho^{x_d}}\Big)^{\frac{1}{2}}<1$ and $|\mathcal{S}_{\mathcal{Z}_\rho^{(j)}}|\leq 2n_\rho^{(j)} $, by construction.

%
\noindent
Hence the total weighted number of rectangles 
\begin{equation}
\sum_{\mathfrak{b}\in \mathcal{B}_{V^{(\bold{k},\bold{q})}_{J_{\bold{r},\bold{i}}}}}
(c\cdot t)^{\vert \mathcal{R}_{\mathfrak{b}}\vert -1} \|V_{\mathcal{L}_\mathfrak{b}}\|\prod_{i=1}^{\vert \mathcal{R}_{\mathfrak{b}}\vert-1}\|V^{(\bold{k}^{(i)},\bold{q}^{(i)})_{-1}}_{J_{\bold{k}^{(i)},\bold{q}^{(i)}}} \|
 \leq \sum_{\Gamma_{\mathfrak{b},\, \mathfrak{b}\in \mathcal{B}_{V^{(\bold{k},\bold{q})}_{J_{\bold{r},\bold{i}}}}}}t^{\frac{r-1}{3}}\cdot \prod_{\sigma\in\mathcal{S}_{\Gamma_{\mathfrak{b}}}}w_\sigma
\end{equation}
can be bounded from above by estimating the number of weighted paths $\Gamma_{\mathfrak{b}}$ as follows
\begin{eqnarray}
& & \sum_{\Gamma_{\mathfrak{b}},\, \mathfrak{b}\in \mathcal{B}_{V^{(\bold{k},\bold{q})}_{J_{\bold{r},\bold{i}}}}}t^{\frac{r-1}{3}}\cdot \prod_{\sigma\in\mathcal{S}_{\Gamma_{\mathfrak{b}}}}w_\sigma\\
&\leq&C_d\cdot r^{2d-1}\cdot t^{\frac{r-1}{3}}\cdot \sum_{j= \lfloor c_d \cdot r/k \rfloor}^{\infty} \Big(\sum_{\rho, \rho'=1}^{k} \Big((c+1)\frac{t^{1/6}}{(\max\{\rho,\rho^\prime\})^{x_d}}\Big)^{1/2} D_{\rho,\rho'}\Big)^j\,\,\label{est-r1-in}
\end{eqnarray}

where:
\begin{itemize}
\item
$$\sum_{\rho'=1}^{k} \Big((c+1)\frac{t^{1/6}}{(\max\{\rho,\rho^\prime\})^{x_d}} \Big)^{1/2} D_{\rho,\rho'} $$
accounts for all the weighted directions for a step from a rectangle of size $\rho$, where $D_{\rho,\rho'}$ has been defined in (\ref{directions}); notice that the weight for the number of directions is due to the restriction of the class of paths used in  the argument that culminates in (\ref{weight-path});
\item
the term $C_d\cdot r^{2d-1}$ is a bound\footnote{ It is enough to consider the volume of the rectangle $J_{\bold{r},\bold{i}}$ and Remark \ref{shapes}.} on the number of possible initial rectangles of a fixed path $\Gamma_{\mathfrak{b}}$;
\item
the sum over $j$ is the sum over the number of steps of  $\Gamma_{\mathfrak{b}}$ which by construction is bounded from below by $\lfloor c_d \cdot r/k \rfloor$.
 
\end{itemize}

\noindent
Next, we bound
\begin{eqnarray}
(\ref{est-r1-in}) \label{est-r1-in-bis}&\leq &C_d\cdot r^{2d-1}\cdot t^{\frac{r-1}{3}}\cdot \sum_{j= \lfloor c_d \cdot r/k \rfloor}^{\infty}\Big((c+1)^{1/2} \cdot t^{\frac{1}{12}}\cdot   2 \sum_{\rho=1}^{k} 
 \frac{\rho \cdot D_{\rho,\rho}}{\rho^{x_d/ 2} }  \Big)^j\,\\\
&\leq &C_d\cdot r^{2d-1}\cdot t^{\frac{r-1}{3}}\cdot t^{\frac{1}{24}\cdot \lfloor c_d \cdot r/k \rfloor}\cdot \sum_{j= \lfloor c_d \cdot r/k \rfloor}^{\infty}\Big((c+1) \cdot t^{\frac{1}{24}}\cdot 2 \sum_{\rho=1}^{k}  \frac{\rho \cdot D_{\rho,\rho}}{\rho^{x_d/2}}\Big)^j\,\quad\nonumber \\
&\leq &\frac{t^{\frac{r-1}{3}}}{r^{x_d+2d}}\,, \label{est-r1-fin}
\end{eqnarray}
where $t\geq 0$ has been chosen small enough such that (recall $k\leq \lfloor r^{\frac{1}{4}} \rfloor$)
\begin{equation}
C_d\cdot r^{4d-1+x_d} \cdot  t^{\frac{1}{24}\cdot \lfloor c_d \cdot r/k \rfloor}\cdot \sum_{j=  \lfloor c_d \cdot r/k \rfloor}^{\infty}\,\Big((c+1)^{1/2} \cdot t^{\frac{1}{24}}\cdot 2 \sum_{\rho=1}^{k}  \frac{\rho \cdot D_{\rho,\rho}}{\rho^{x_d/2}}\Big)^j<1.\,
\end{equation}

%

\medskip

\noindent
\underline{\emph{Regime $\mathfrak{R}2)$} }
\\

For $(\bold{k},\bold{q})$ in this regime, starting from the inequality
\begin{eqnarray}
\|V^{(\bold{k},\bold{q})}_{J_{\bold{r},\bold{i}}}\|
&\leq &\|V^{(\bold{k},\bold{q})_{-1}}_{J_{\bold{r},\bold{i}}}\|+\|\sum_{J_{\bold{k}',\bold{q}'}\in \mathcal{G}^{(\bold{k},\bold{q})}_{J_{\bold{r},\bold{i}}}}\,\sum_{n=1}^{\infty}\frac{1}{n!}\,ad^{n}S_{J_{\bold{k},\bold{q}}}(V^{(\bold{k},\bold{q})_{-1}}_{J_{\bold{k}',\bold{q}'}})\|\,,
\end{eqnarray}
we only keep expanding the first potential on the r-h-s. Then, using the inductive hypotheses (\ref{inductive-reg-1}), (\ref{inductive-reg-2}),  (\ref{R3-2}),  and (\ref{R3-3}), for $t\geq 0$ sufficiently small, we can estimate
\begin{eqnarray}
\|V^{(\bold{k},\bold{q})}_{J_{\bold{r},\bold{i}}}\|
&\leq&\|V^{(\bold{k}_*,\bold{q}_*)}_{J_{\bold{r},\bold{i}}}\|\\
& &+\sum_{s=\lfloor r^\frac{1}{4} \rfloor}^{ r-\lfloor r^\frac{1}{4} \rfloor }\sum_{s_{1}=0}^{s}\sum_{s_{2}=0}^{s-s_{1}}\dots \sum_{s_{d}=0}^{s-s_{1}-\dots -s_{d-1}}\delta_{s_1+s_2+\dots +s_d-s}\, \cdot c_d \cdot r^{2d} \cdot t\cdot \frac{t^{\frac{s-1}{3}} }{s^{x_d}}\cdot \frac{t^{\frac{r-s-1}{3}} }{(r-s)^{x_d}}\quad\quad\quad  \label{crudesum}
\end{eqnarray}
where: 
\begin{itemize}
\item $(\bold{k}_*,\bold{q}_*)$ is the greatest rectangle  of regime $\frak{R}1$ with respect to the ordering $\succ$, and by construction $k_*=\lfloor r^\frac{1}{4}\rfloor$;
\item the factor $$c_d\cdot r^{2d-1}\cdot t \cdot \frac{t^{\frac{s-1}{3}}}{s^{x_d}}\cdot \frac{ t^{\frac{r-s-1}{3}} }{(r-s)^{x_d}}\,$$
is an upper bound to the sum of the products of the type $\|\mathcal{A}_{J_{(\bold{k},\bold{q}})_{-j}}(\,V^{(\bold{k},\bold{q})_{-j-1}}_{J_{\bold{k}',\bold{q}'}} \, )\|$ for some $j$ and  where the size of the rectangle associated with $(\bold{k},\bold{q})_{-j}$ is equal to $s$;
\item the multiplicative factor $\mathcal{O}(r^{2d-1}) $ is an upper bound estimate (see Remark \ref{shapes})
to the number of rectangles $J_{\bold{k}', \bold{q}'}\subset J_{\bold{r}, \bold{i}}$ such that  $[ J_{\bold{k}',\bold{q}'} \cup J_{\bold{k},\bold{q}} ]=J_{\bold{r},\bold{i}}$.
\end{itemize}
Now for any $s$ with $\lfloor r^\frac{1}{4} \rfloor\leq s\leq r-\lfloor r^\frac{1}{4} \rfloor$ we have
\begin{eqnarray}
&&\sum_{s_{1}=0}^{s}\sum_{s_{2}=0}^{s-s_{1}}\dots \sum_{s_{d}=0}^{s-s_{1}-\dots -s_{d-1}}\delta_{s_1+s_2+\dots +s_d-s}\cdot  c_d \cdot r^{2d} \cdot t\cdot \frac{t^{\frac{s-1}{3}}\cdot }{s^{x_d}}\frac{t^{\frac{r-s-1}{3}} }{(r-s)^{x_d}}\\
&\leq & s^d \cdot c_d \cdot r^{2d-1}\cdot  t^\frac{2}{3}\cdot \frac{t^{\frac{r-1}{3}}}{s^{x_d}\cdot (r-s)^{x_d}}\\
&\leq & r^d \cdot c_d \cdot r^{2d-1}\cdot t^\frac{2}{3}\cdot \frac{t^{\frac{r-1}{3}}}{s^{x_d} \cdot (r-s)^{x_d}}\\
&\leq & 2^{x_d} \cdot c_d \cdot r^{2d-1}\cdot  t^\frac{2}{3}\cdot \frac{t^{\frac{r-1}{3}}}{r^{x_d} \cdot r^{x_d/4}}
\end{eqnarray}
as 
$$\max_{\lfloor r^\frac{1}{4} \rfloor\leq s\leq r-\lfloor r^\frac{1}{4} \rfloor}\, \frac{1}{s^{x_d}\cdot (r-s)^{x_d}} \leq \frac{1}{r^{x_d/4}\cdot (r-r^\frac{1}{4})^{x_d}}\leq \frac{2^{x_d}}{r^{x_d}\cdot r^{x_d/4}}$$
since $r-\lfloor r^\frac{1}{4} \rfloor\geq\frac{r}{2}$. But then, using the inductive hypothesis for $\|V^{(\bold{k},\bold{q})_*}_{J_{\bold{r},\bold{i}}}\|$,
\begin{eqnarray}
\|V^{(\bold{k},\bold{q})}_{J_{\bold{r},\bold{i}}}\|&\leq &\|V^{(\bold{k},\bold{q})_*}_{J_{\bold{r},\bold{i}}}\|+\sum_{s=\lfloor r^\frac{1}{4} \rfloor}^{r-\lfloor r^\frac{1}{4} \rfloor} r^d \cdot 2^{x_d} \cdot c_d \cdot r^{2d-1}\cdot  t^\frac{2}{3}\cdot \frac{t^{\frac{r-1}{3}}}{r^{x_d} \cdot r^{x_d/4}}\\
&\leq &\frac{t^\frac{r-1}{3}}{r^{x_d+2d}} +2^{x_d} \cdot c_d \cdot t^\frac{2}{3} \cdot \frac{t^\frac{r-1}{3}}{r^{\frac{5x_d}{4}-3d}}\\
&\leq & 2\cdot \frac{t^\frac{r-1}{3}}{r^{x_d+2d}}
\end{eqnarray}
since  $x_d= 20d$ and $t\geq 0$ is small enough.

\medskip
\noindent
\underline{\emph{Regime $\mathfrak{R}3)$}}
\\


\noindent
\underline{\emph{Proof of  (\ref{R3-1})}}
\\

\noindent
For $(\bold{k},\bold{q})_{**}\prec (\bold{k},\bold{q})\prec (\bold{r},\bold{i})$ we first consider
\begin{eqnarray}
& &\frac{1}{\sum_{\bold{j}\in J_{\bold{r},\bold{i}}} P^{\perp}_{\Omega_{\bold{j}}}+1}\, P^{(+)}_{J_{\bold{r},\bold{i}}}\,V^{(\bold{k},\bold{q})}_{J_{\bold{r},\bold{i}}}\,P^{(-)}_{J_{\bold{r},\bold{i}}}\,\frac{1}{\sum_{\bold{j}\in J_{\bold{r},\bold{i}}} P^{\perp}_{\Omega_{\bold{j}}}+1}\\
&=&\frac{1}{\sum_{\bold{j}\in J_{\bold{r},\bold{i}}} P^{\perp}_{\Omega_{\bold{j}}}+1}\, P^{(+)}_{J_{\bold{r},\bold{i}}}\,V^{(\bold{k},\bold{q})}_{J_{\bold{r},\bold{i}}}\,P^{(-)}_{J_{\bold{r},\bold{i}}}\,.
\end{eqnarray}

\noindent
We recall that for $(\bold{k},\bold{q})\prec (\bold{r},\bold{i})$ the two types of re-expansion that have to be considered correspond to a) and c) in Definition \ref{def-interactions-multi}. Notice that the re-expansion of type a) is trivial since it does not change the potential. Using the re-expansion of type c), that is associated with formula  (\ref{exp-formula-bis})-(\ref{exp-formula-bis-2}), we get
\begin{eqnarray}
& &\frac{1}{\sum_{\bold{j}\in J_{\bold{r},\bold{i}}} P^{\perp}_{\Omega_{\bold{j}}}+1}\,P^{(+)}_{J_{\bold{r},\bold{i}}}\,V^{(\bold{k},\bold{q})}_{J_{\bold{r},\bold{i}}}\,P^{(-)}_{J_{\bold{r},\bold{i}}}\\
&=&\frac{1}{\sum_{\bold{j}\in J_{\bold{r},\bold{i}}} P^{\perp}_{\Omega_{\bold{j}}}+1}\,P^{(+)}_{J_{\bold{r},\bold{i}}}\,V^{(\bold{k},\bold{q})_{-1}}_{J_{\bold{r},\bold{i}}}\,P^{(-)}_{J_{\bold{r},\bold{i}}}\label{moving}\\
& &+\frac{1}{\sum_{\bold{j}\in J_{\bold{r},\bold{i}}} P^{\perp}_{\Omega_{\bold{j}}}+1}\,P^{(+)}_{J_{\bold{r},\bold{i}}}\,\Big\{\,\sum_{n=1}^{\infty}\frac{1}{n!}\,ad^{n}S_{J_{\bold{k},\bold{q}}}(V^{(\bold{k},\bold{q})_{-1}}_{J_{\bold{r},\bold{i}}})\,\Big\}\,P^{(-)}_{J_{\bold{r},\bold{i}}}\label{rest-1}\\
& &+\frac{1}{\sum_{\bold{j}\in J_{\bold{r},\bold{i}}} P^{\perp}_{\Omega_{\bold{j}}}+1}\,P^{(+)}_{J_{\bold{r},\bold{i}}}\,\Big\{\sum_{J_{\bold{k}',\bold{q}'}\in \mathcal{G}^{(\bold{k},\bold{q})}_{J_{\bold{r},\bold{i}}}}\,\sum_{n=1}^{\infty}\frac{1}{n!}\,ad^{n}S_{J_{\bold{k},\bold{q}}}(V^{(\bold{k},\bold{q})_{-1}}_{J_{\bold{k}',\bold{q}'}})\Big\}\,P^{(-)}_{J_{\bold{r},\bold{i}}}\,. \label{rest-2}
\end{eqnarray}
We shall keep re-expanding the terms analogous to $V^{(\bold{k},\bold{q})_{-1}}_{J_{\bold{r},\bold{i}}}$ in (\ref{moving}),  from $(\bold{k},\bold{q})_{-1}$ down to $(\bold{k}_{**},\bold{q}_{**})$. The pair $(\bold{k}_{**},\bold{q}_{**})$ represents the greatest rectangle with respect to the ordering $\succ$ in regime $\frak{R}2$,  and by construction has $k_{**}=r-\lfloor r^\frac{1}{4}\rfloor$. 

\noindent
On the contrary, at each step we estimate the terms of the type (\ref{rest-1}) and (\ref{rest-2}) that are produced by the iteration, without further expanding the potentials analogous to $V^{(\bold{k},\bold{q})_{-1}}_{J_{\bold{r},\bold{i}}}$ and  $V^{(\bold{k},\bold{q})_{-1}}_{J_{\bold{k}',\bold{q}'}}$ that are contained in them.
\\

\noindent
\emph{Estimate of  (\ref{rest-1})}
\\

\noindent
Concerning  (\ref{rest-1}), we observe that using the inductive hypotheses (\ref{inductive-reg-2})-(\ref{R3-3})
along with Lemma (\ref{control-LS}) we can bound
\begin{equation}\label{first-nonleading}
\|(\ref{rest-1})\|\leq  c\cdot t\cdot \frac{ t^{\frac{k-1}{3}}}{k^{x_d}}\cdot \frac{ t^{\frac{r-1}{3}}}{r^{x_d}}\,.
\end{equation}
At fixed $k$ the number of contributions of type  (\ref{rest-1}) can be estimated from above by $\mathcal{O}(r^d\cdot k^{d-1})$; see Remark \ref{shapes}.
Being $k\geq r-\lfloor r^{\frac{1}{4}}\rfloor $ in regime $\mathfrak{R}3$, the power $ t^{\frac{k-1}{3}}$  will be used to control the number of this type of contributions produced along the way down to $(\bold{k}_{**},\bold{q}_{**})$. 
\\

\noindent
\emph{Estimate of (\ref{rest-2})}
\\

\noindent
It is convenient to split the corresponding term, $(\ref{rest-2})$, into
\begin{eqnarray}
& &(\ref{rest-2})\\
&= &\frac{1}{\sum_{\bold{j}\in J_{\bold{r},\bold{i}}} P^{\perp}_{\Omega_{\bold{j}}}+1}\,\Big\{\sum_{J_{\bold{k}',\bold{q}'}\in \mathcal{G}^{(\bold{k},\bold{q})}_{J_{\bold{r},\bold{i}}}}\,\,ad\,S_{J_{\bold{k},\bold{q}}}(V^{(\bold{k},\bold{q})_{-1}}_{J_{\bold{k}',\bold{q}'}})\Big\}\,P^{(-)}_{J_{\bold{r},\bold{i}}}\label{rest-2-first}\\
& &+\frac{1}{\sum_{\bold{j}\in J_{\bold{r},\bold{i}}} P^{\perp}_{\Omega_{\bold{j}}}+1}\,\Big\{\sum_{J_{\bold{k}',\bold{q}'}\in \mathcal{G}^{(\bold{k},\bold{q})}_{J_{\bold{r},\bold{i}}}}\,\sum_{n=2}^{\infty}\frac{1}{n!}\,ad^{n}S_{J_{\bold{k},\bold{q}}}(V^{(\bold{k},\bold{q})_{-1}}_{J_{\bold{k}',\bold{q}'}})\Big\}\,P^{(-)}_{J_{\bold{r},\bold{i}}}\,.\label{rest-2-second}
\end{eqnarray}

\noindent
In (\ref{rest-2-first}) we distinguish $J_{\bold{k}',\bold{q}'}$ \emph{small} and \emph{large} depending on whether $(\bold{k}',\bold{q}')\prec (\bold{k},\bold{q})$ or  $(\bold{k},\bold{q})\preceq (\bold{k}',\bold{q}')$, respectively, and denote by $(\mathcal{G}^{(\bold{k},\bold{q})}_{J_{\bold{r},\bold{i}}})_{small}$ the subset formed by the \emph{small} $J_{\bold{k}',\bold{q}'}$ belonging to the set  $\mathcal{G}^{(\bold{k},\bold{q})}_{J_{\bold{r},\bold{i}}}$. We call 
$$(\ref{rest-2-first})_{small}\quad \text{and}\quad (\ref{rest-2-first})_{large}\,,$$
respectively, the corresponding contributions to (\ref{rest-2-first}).
Next, we study some commutators that enter the expression  $(\ref{rest-2-first})_{small}$ estimated below. We observe that
\begin{eqnarray}
& &[S_{J_{\bold{k},\bold{q}}}\,,\,V^{(\bold{k},\bold{q})_{-1}}_{J_{\bold{k}',\bold{q}'}}]\\
& =&[S_{J_{\bold{k},\bold{q}}}\,,\,P^{(+)}_{J_{\bold{k}',\bold{q}'}}\,V^{(\bold{k},\bold{q})_{-1}}_{J_{\bold{k}',\bold{q}'}}\,P^{(+)}_{J_{\bold{k}',\bold{q}'}}+P^{(-)}_{J_{\bold{k}',\bold{q}'}}\,V^{(\bold{k},\bold{q})_{-1}}_{J_{\bold{k}',\bold{q}'}}\,P^{(-)}_{J_{\bold{k}',\bold{q}'}}]\,\\
& =&[S_{J_{\bold{k},\bold{q}}}\,,\,P^{(+)}_{J_{\bold{k}',\bold{q}'}}\,V^{(\bold{k},\bold{q})_{-1}}_{J_{\bold{k}',\bold{q}'}}\,P^{(+)}_{J_{\bold{k}',\bold{q}'}}+\langle V^{(\bold{k},\bold{q})_{-1}}_{J_{\bold{k}',\bold{q}'}}\rangle\,P^{(-)}_{J_{\bold{k}',\bold{q}'}}]\,\\
& =&[S_{J_{\bold{k},\bold{q}}}\,,\,P^{(+)}_{J_{\bold{k}',\bold{q}'}}\,V^{(\bold{k},\bold{q})_{-1}}_{J_{\bold{k}',\bold{q}'}}\,P^{(+)}_{J_{\bold{k}',\bold{q}'}}]-[S_{J_{\bold{k},\bold{q}}}\,,\,<V^{(\bold{k},\bold{q})_{-1}}_{J_{\bold{k}',\bold{q}'}}>P^{(+)}_{J_{\bold{k}',\bold{q}'}}]\,,
\end{eqnarray}
where we have exploited that $V^{(\bold{k},\bold{q})_{-1}}_{J_{\bold{k}',\bold{q}'}}$ is block-diagonalized since, by definition, \emph{small} means    $(\bold{k}',\bold{q}')\prec (\bold{k},\bold{q})$. We also observe that $P^{(+)}_{J_{\bold{k}',\bold{q}'}}P^{(-)}_{J_{\bold{r},\bold{i}}}=0$  since  $J_{\bold{k}',\bold{q}'} \subset J_{\bold{r},\bold{i}}$ by construction, hence
 \begin{eqnarray}
& &P^{(+)}_{J_{\bold{r},\bold{i}}}\,[S_{J_{\bold{k},\bold{q}}}\,,\,P^{(+)}_{J_{\bold{k}',\bold{q}'}}\,V^{(\bold{k},\bold{q})_{-1}}_{J_{\bold{k}',\bold{q}'}}\,P^{(+)}_{J_{\bold{k}',\bold{q}'}}] \,P^{(-)}_{J_{\bold{r},\bold{i}}}\\
& &-P^{(+)}_{J_{\bold{r},\bold{i}}}\, [S_{J_{\bold{k},\bold{q}}}\,,\,<V^{(\bold{k},\bold{q})_{-1}}_{J_{\bold{k}',\bold{q}'}}>P^{(+)}_{J_{\bold{k}',\bold{q}'}}]\,P^{(-)}_{J_{\bold{r},\bold{i}}}\\
&=&-P^{(+)}_{J_{\bold{r},\bold{i}}}\,P^{(+)}_{J_{\bold{k}',\bold{q}'}}\,V^{(\bold{k},\bold{q})_{-1}}_{J_{\bold{k}',\bold{q}'}}\,P^{(+)}_{J_{\bold{k}',\bold{q}'}}S_{J_{\bold{k},\bold{q}}}P^{(-)}_{J_{\bold{r},\bold{i}}}\\
& &+P^{(+)}_{J_{\bold{r},\bold{i}}}\,<V^{(\bold{k},\bold{q})_{-1}}_{J_{\bold{k}',\bold{q}'}}>P^{(+)}_{J_{\bold{k}',\bold{q}'}}S_{J_{\bold{k},\bold{q}}}\,P^{(-)}_{J_{\bold{r},\bold{i}}}\,.
\end{eqnarray}

\noindent
We recall that for $j\geq 1$
\begin{equation}\label{def-Sj}
(S_{J_{\bold{k},\bold{q}}})_j:=\frac{1}{G_{J_{\bold{k},\bold{q}}}-E_{J_{\bold{k},\bold{q}}}}P^{(+)}_{J_{\bold{k},\bold{q}}}\,(V^{(\bold{k},\bold{q})_{-1}}_{J_{\bold{k},\bold{q}}})_j\,P^{(-)}_{J_{\bold{k},\bold{q}}}-h.c.\,{\color{magenta},}
\end{equation}
and  from Lemma  \ref{control-LS} we get
\begin{equation}\label{two-factors}
\|\sum_{j=2}^{\infty}t^{j}(S_{J_{\bold{k},\bold{q}}})_j\|\leq C\cdot t^2\cdot \|(V^{(k,q-1)}_{J_{\bold{k},\bold{q}}})_1\|^2
\end{equation}
for $j\geq 2$ and $t\geq 0$ sufficiently small.
Hence we split  $(\ref{rest-2-first})_{small}$ into two contributions: 

\noindent
1) the leading order term
\begin{eqnarray}
& &-\frac{1}{\sum_{\bold{j}\in J_{\bold{r},\bold{i}}} P^{\perp}_{\Omega_{\bold{j}}}+1}\,P^{(+)}_{J_{\bold{r},\bold{i}}}\times \label{lead-2.115}\\
& &\quad \times\Big\{\sum_{J_{\bold{k}',\bold{q}'}\in (\mathcal{G}^{(\bold{k},\bold{q})}_{J_{\bold{r};\bold{i}}})_{small}   }\,\,\,P^{(+)}_{J_{\bold{k}',\bold{q}'}}\,\Big(V^{(\bold{k},\bold{q})_{-1}}_{J_{\bold{k}',\bold{q}'}}-\langle V^{(\bold{k},\bold{q})_{-1}}_{J_{\bold{k}',\bold{q}'}}\rangle\Big)\,P^{(+)}_{J_{\bold{k}',\bold{q}'}}\,\Big(\frac{t}{G_{J_{\bold{k},\bold{q}}}-E_{J_{\bold{k},\bold{q}}}}P^{(+)}_{J_{\bold{k},\bold{q}}}\,V^{(\bold{k},\bold{q})_{-1}}_{J_{\bold{k},\bold{q}}}\,P^{(-)}_{J_{\bold{k},\bold{q}}}-h.c.\Big)\Big\}\,P^{(-)}_{J_{\bold{r},\bold{i}}}\nonumber \\
&= &-\frac{1}{\sum_{\bold{j}\in J_{\bold{r},\bold{i}}} P^{\perp}_{\Omega_{\bold{j}}}+1}\,P^{(+)}_{J_{\bold{r},\bold{i}}}\times\\
& &\quad\times \Big\{\sum_{J_{\bold{k}',\bold{q}'}\in (\mathcal{G}^{(\bold{k},\bold{q})}_{J_{\bold{r};\bold{i}}})_{small}}\,\,\,P^{(+)}_{J_{\bold{k}',\bold{q}'}}\,\Big(V^{(\bold{k},\bold{q})_{-1}}_{J_{\bold{k}',\bold{q}'}}-\langle V^{(\bold{k},\bold{q})_{-1}}_{J_{\bold{k}',\bold{q}'}}\rangle\Big)\,P^{(+)}_{J_{\bold{k}',\bold{q}'}}\,\frac{t}{G_{J_{\bold{k},\bold{q}}}-E_{J_{\bold{k},\bold{q}}}}P^{(+)}_{J_{\bold{k},\bold{q}}}\,V^{(\bold{k},\bold{q})_{-1}}_{J_{\bold{k},\bold{q}}}\,P^{(-)}_{J_{\bold{k},\bold{q}}}\Big\}\,P^{(-)}_{J_{\bold{r},\bold{i}}}\nonumber
\end{eqnarray}
where  we have used $P^{(+)}_{J_{\bold{k},\bold{q}}}\,P^{(-)}_{J_{\bold{r},\bold{i}}}=0$; 

\noindent
2) the remainder term
\begin{equation}
-\frac{1}{\sum_{\bold{j}\in J_{\bold{r},\bold{i}}} P^{\perp}_{\Omega_{\bold{j}}}+1}\,P^{(+)}_{J_{\bold{r},\bold{i}}}\,\Big\{\sum_{J_{\bold{k}',\bold{q}'}\in (\mathcal{G}^{(\bold{k},\bold{q})}_{J_{\bold{r};\bold{i}}})_{small}}\,\,\,P^{(+)}_{J_{\bold{k}',\bold{q}'}}\,\Big(V^{(\bold{k},\bold{q})_{-1}}_{J_{\bold{k}',\bold{q}'}}-\langle V^{(\bold{k},\bold{q})_{-1}}_{J_{\bold{k}',\bold{q}'}}\rangle\Big)\,P^{(+)}_{J_{\bold{k}',\bold{q}'}}\,\sum_{j=2}^{\infty}t^{j}(S_{J_{\bold{k},\bold{q}}})_j\,\Big\}\,P^{(-)}_{J_{\bold{r},\bold{i}}}\,.\label{rem-2.115}
\end{equation}
In order to estimate the leading order term  (\ref{lead-2.115}) we make use of the inequality
\begin{eqnarray}
& &\|\frac{1}{\sum_{\bold{j}\in J_{\bold{r},\bold{i}}} P^{\perp}_{\Omega_{\bold{j}}}+1}\,\Big\{\sum_{J_{\bold{k}',\bold{q}'}\in (\mathcal{G}^{(\bold{k},\bold{q})}_{J_{\bold{r};\bold{i}}})_{small}}\,\,\,P^{(+)}_{J_{\bold{k}',\bold{q}'}}\,\Big(V^{(\bold{k},\bold{q})_{-1}}_{J_{\bold{k}',\bold{q}'}}-\langle V^{(\bold{k},\bold{q})_{-1}}_{J_{\bold{k}',\bold{q}'}}\rangle\Big)\,P^{(+)}_{J_{\bold{k}',\bold{q}'}}\,\frac{t}{G_{J_{\bold{k},\bold{q}}}-E_{J_{\bold{k},\bold{q}}}}P^{(+)}_{J_{\bold{k},\bold{q}}}\,V^{(\bold{k},\bold{q})_{-1}}_{J_{\bold{k},\bold{q}}}\,P^{(-)}_{J_{\bold{k},\bold{q}}}\Big\}\,P^{(-)}_{J_{\bold{r},\bold{i}}}\|\nonumber\\
&\leq&\|\sum_{J_{\bold{k}',\bold{q}'}\in (\mathcal{G}^{(\bold{k},\bold{q})}_{J_{\bold{r};\bold{i}}})_{small}}\,\frac{1}{\sum_{\bold{j}\in J_{\bold{r},\bold{i}}} P^{\perp}_{\Omega_{\bold{j}}}+1}\,P^{(+)}_{J_{\bold{r},\bold{i}}}\,\,P^{(+)}_{J_{\bold{k}',\bold{q}'}}\Big(V^{(\bold{k},\bold{q})_{-1}}_{J_{\bold{k}',\bold{q}'}}-\langle V^{(\bold{k},\bold{q})_{-1}}_{J_{\bold{k}',\bold{q}'}}\rangle\Big)\,P^{(+)}_{J_{\bold{k}',\bold{q}'}}\|
\,\\
& &\quad \times \, t\cdot \|\,\frac{1}{G_{J_{\bold{k},\bold{q}}}-E_{J_{\bold{k},\bold{q}}}}P^{(+)}_{J_{\bold{k},\bold{q}}}\,(\sum_{\bold{j}\in J_{\bold{k},\bold{q}}} P^{\perp}_{\Omega_{\bold{j}}}+1)\|\cdot \|\frac{1}{\sum_{\bold{j}\in J_{\bold{k},\bold{q}}} P^{\perp}_{\Omega_{\bold{j}}}+1}P^{(+)}_{J_{\bold{k},\bold{q}}}\,V^{(\bold{k},\bold{q})_{-1}}_{J_{\bold{k},\bold{q}}}\,P^{(-)}_{J_{\bold{k},\bold{q}}}\|\,. \label{2.158}
\end{eqnarray}
Now  we introduce the notation
  \begin{equation}
\overline{\sum}_{J_{\bold{k}',\bold{q}' }\,,\,J_{\bold{k}'',\bold{q}''}} :=\sum_{J_{\bold{k}',\bold{q}'}\,,\,J_{\bold{k}'',\bold{q}''}\,\in (\mathcal{G}^{(\bold{k},\bold{q})}_{J_{\bold{r};\bold{i}}})_{small}\,;\, J_{\bold{k}',\bold{q}'}\cap J_{\bold{k}'',\bold{q}''}=\emptyset}
  \end{equation}
  and
   \begin{equation}\label{sum-prime}
  \sum'_{J_{\bold{k}',\bold{q}' }\,,\,J_{\bold{k}'',\bold{q}''}} :=\sum_{J_{\bold{k}',\bold{q}'}\,,\,J_{\bold{k}'',\bold{q}''}\,\in (\mathcal{G}^{(\bold{k},\bold{q})}_{J_{\bold{r};\bold{i}}})_{small}\,;\, J_{\bold{k}',\bold{q}'}\cap J_{\bold{k}'',\bold{q}''}\neq\emptyset}\,.
  \end{equation}
We can write
  \begin{eqnarray}
 & &\|\sum_{J_{\bold{k}',\bold{q}'}\in (\mathcal{G}^{(\bold{k},\bold{q})}_{J_{\bold{r};\bold{i}}})_{small}}\,\frac{1}{\sum_{\bold{j}\in J_{\bold{r},\bold{i}}} P^{\perp}_{\Omega_{\bold{j}}}+1}\,P^{(+)}_{J_{\bold{r},\bold{i}}}\,\,P^{(+)}_{J_{\bold{k}',\bold{q}'}}\Big(V^{(\bold{k},\bold{q})_{-1}}_{J_{\bold{k}',\bold{q}'}}-\langle V^{(\bold{k},\bold{q})_{-1}}_{J_{\bold{k}',\bold{q}'}}\rangle\Big)\,P^{(+)}_{J_{\bold{k}',\bold{q}'}}\|^2\\ 
  &\leq &\sup_{\| \psi\|=1}\,\overline{\sum}_{J_{\bold{k}',\bold{q}'} \,,\,J_{\bold{k}'',\bold{q}''}}  \,\langle \frac{1}{\sum_{\bold{j}\in J_{\bold{r},\bold{i}}} P^{\perp}_{\Omega_{\bold{j}}}+1}\psi, \label{nonitersect}\\
 & &\quad\quad\quad P^{(+)}_{J_{\bold{k}',\bold{q}'}}\Big(V^{(\bold{k},\bold{q})_{-1}}_{J_{\bold{k}',\bold{q}'}}-\langle V^{(\bold{k},\bold{q})_{-1}}_{J_{\bold{k}',\bold{q}'}}\rangle\Big)\,P^{(+)}_{J_{\bold{k}',\bold{q}'}}\,P^{(+)}_{J_{\bold{k}'',\bold{q}''}}\Big(V^{(\bold{k},\bold{q})_{-1}}_{J_{\bold{k}'',\bold{q}''}}-\langle V^{(\bold{k},\bold{q})_{-1}}_{J_{\bold{k}'',\bold{q}''}}\rangle\Big)\,P^{(+)}_{J_{\bold{k}'',\bold{q}''}}\frac{1}{\sum_{\bold{j}\in J_{\bold{r},\bold{i}}} P^{\perp}_{\Omega_{\bold{j}}}+1}\psi \rangle \nonumber \,\\
 & &+\sup_{\| \psi\|=1}\,  \sum'_{J_{\bold{k}',\bold{q}' }\,,\,J_{\bold{k}'',\bold{q}''}}   \,\langle \frac{1}{\sum_{\bold{j}\in J_{\bold{r},\bold{i}}} P^{\perp}_{\Omega_{\bold{j}}}+1}\psi, \label{intersect}\\
 & &\quad\quad\quad P^{(+)}_{J_{\bold{k}',\bold{q}'}}\Big(V^{(\bold{k},\bold{q})_{-1}}_{J_{\bold{k}',\bold{q}'}}-\langle V^{(\bold{k},\bold{q})_{-1}}_{J_{\bold{k}',\bold{q}'}}\rangle\Big)\,P^{(+)}_{J_{\bold{k}',\bold{q}'}}\,P^{(+)}_{J_{\bold{k}'',\bold{q}''}}\Big(V^{(\bold{k},\bold{q})_{-1}}_{J_{\bold{k}'',\bold{q}''}}-\langle V^{(\bold{k},\bold{q})_{-1}}_{J_{\bold{k}'',\bold{q}''}}\rangle\Big)\,P^{(+)}_{J_{\bold{k}'',\bold{q}''}}\frac{1}{\sum_{\bold{j}\in J_{\bold{r},\bold{i}}} P^{\perp}_{\Omega_{\bold{j}}}+1}\psi \rangle\,. \nonumber
\end{eqnarray}
    
\noindent
\emph{Leading terms in $(\ref{rest-2})$: Contribution proportional to (\ref{nonitersect})}

\noindent
 We observe that for $J_{\bold{k}',\bold{q}'}\cap J_{\bold{k}'',\bold{q}''}=\emptyset$ we have
\begin{eqnarray}
& &P^{(+)}_{J_{\bold{k}',\bold{q}'}}\Big(V^{(\bold{k},\bold{q})_{-1}}_{J_{\bold{k}',\bold{q}'}}-\langle V^{(\bold{k},\bold{q})_{-1}}_{J_{\bold{k}',\bold{q}'}}\rangle\Big)\,P^{(+)}_{J_{\bold{k}',\bold{q}'}}P^{(+)}_{J_{\bold{k}'',\bold{q}''}}\Big(V^{(\bold{k},\bold{q})_{-1}}_{J_{\bold{k}'',\bold{q}''}}-\langle V^{(\bold{k},\bold{q})_{-1}}_{J_{\bold{k}'',\bold{q}''}}\rangle\Big)\,P^{(+)}_{J_{\bold{k}'',\bold{q}''}}\\
&=&P^{(+)}_{J_{\bold{k}',\bold{q}'}}P^{(+)}_{J_{\bold{k}'',\bold{q}''}}\\
& & \times \Big(V^{(\bold{k},\bold{q})_{-1}}_{J_{\bold{k}',\bold{q}'}}-\langle V^{(\bold{k},\bold{q})_{-1}}_{J_{\bold{k}',\bold{q}'}}\rangle\Big)\,P^{(+)}_{J_{\bold{k}',\bold{q}'}}P^{(+)}_{J_{\bold{k}'',\bold{q}''}}\Big(V^{(\bold{k},\bold{q})_{-1}}_{J_{\bold{k}'',\bold{q}''}}-\langle V^{(\bold{k},\bold{q})_{-1}}_{J_{\bold{k}'',\bold{q}''}}\rangle\Big)\,\\
& & \times  \,P^{(+)}_{J_{\bold{k}',\bold{q}'}}P^{(+)}_{J_{\bold{k}'',\bold{q}''}}\, 
 \end{eqnarray}
since 
 $$[P^{(+)}_{J_{\bold{k}',\bold{q}'}}\,,\,V^{(\bold{k},\bold{q})_{-1}}_{J_{\bold{k}'',\bold{q}''}}-\langle V^{(\bold{k},\bold{q})_{-1}}_{J_{\bold{k}'',\bold{q}''}}\rangle]=[V^{(\bold{k},\bold{q})_{-1}}_{J_{\bold{k}',\bold{q}'}}-\langle V^{(\bold{k},\bold{q})_{-1}}_{J_{\bold{k}',\bold{q}'}}\rangle\,,\,P^{(+)}_{J_{\bold{k}'',\bold{q}''}}]=0\,.$$
On the contrary we notice that
\begin{equation}
[P^{(+)}_{J_{\bold{k}',\bold{q}'}}\,,\,P^{(+)}_{J_{\bold{k}'',\bold{q}''}}]=[P^{(-)}_{J_{\bold{k}',\bold{q}'}}\,,\,P^{(-)}_{J_{\bold{k}'',\bold{q}''}}]=0
\end{equation}
even if $J_{\bold{k}',\bold{q}'}\cap J_{\bold{k}'',\bold{q}''}\neq\emptyset$. Indeed,
 $$P^{(-)}_{J_{\bold{k}',\bold{q}'}}=P^{(-)}_{J_{\bold{k}',\bold{q}'}\setminus J_{\bold{k}'',\bold{q}''}}\otimes P^{(-)}_{J_{\bold{k}',\bold{q}'}\cap J_{\bold{k}'',\bold{q}''}}\quad,\quad P^{(-)}_{J_{\bold{k}'',\bold{q}''}}=P^{(-)}_{J_{\bold{k}'',\bold{q}''}\setminus J_{\bold{k}',\bold{q}'}}\otimes P^{(-)}_{J_{\bold{k}',\bold{q}'}\cap J_{\bold{k}'',\bold{q}''}}$$
 hence
\begin{eqnarray}
P^{(-)}_{J_{\bold{k}',\bold{q}'}}P^{(-)}_{J_{\bold{k}'',\bold{q}''}}
&=&(P^{(-)}_{J_{\bold{k}',\bold{q}'}\setminus J_{\bold{k}'',\bold{q}''}}\otimes P^{(-)}_{J_{\bold{k}',\bold{q}'}\cap J_{\bold{k}'',\bold{q}''}})\,(P^{(-)}_{J_{\bold{k}'',\bold{q}''}\setminus J_{\bold{k}',\bold{q}'}}\otimes P^{(-)}_{J_{\bold{k}',\bold{q}'}\cap J_{\bold{k}'',\bold{q}''}})\\
& =&P^{(-)}_{J_{\bold{k}',\bold{q}'}\setminus J_{\bold{k}'',\bold{q}''}}\otimes P^{(-)}_{J_{\bold{k}',\bold{q}'}\cap J_{\bold{k}'',\bold{q}''}}\otimes P^{(-)}_{J_{\bold{k}'',\bold{q}''}\setminus J_{\bold{k}',\bold{q}'}}\\
&=&(P^{(-)}_{J_{\bold{k}'',\bold{q}''}\setminus J_{\bold{k}',\bold{q}'}}\otimes P^{(-)}_{J_{\bold{k}',\bold{q}'}\cap J_{\bold{k}'',\bold{q}''}})\,(P^{(-)}_{J_{\bold{k}',\bold{q}'}\setminus J_{\bold{k}'',\bold{q}''}}\otimes P^{(-)}_{J_{\bold{k}',\bold{q}'}\cap J_{\bold{k}'',\bold{q}''}})\\
&=&P^{(-)}_{J_{\bold{k}'',\bold{q}''}}P^{(-)}_{J_{\bold{k}',\bold{q}'}}\,.
\end{eqnarray}
Hence we can estimate
\begin{eqnarray}
& &\sup_{\| \psi\|=1}\,\overline{\sum}_{J_{\bold{k}',\bold{q}' \,,\,J_{\bold{k}'',\bold{q}''}}}  \,\Big|\langle \frac{1}{\sum_{\bold{j}\in J_{\bold{r},\bold{i}}} P^{\perp}_{\Omega_{\bold{j}}}+1}\psi\,,\,\\
& &\quad\quad P^{(+)}_{J_{\bold{k}',\bold{q}'}}\Big(V^{(\bold{k},\bold{q})_{-1}}_{J_{\bold{k}',\bold{q}'}}-\langle V^{(\bold{k},\bold{q})_{-1}}_{J_{\bold{k}',\bold{q}'}}\rangle\Big)\,P^{(+)}_{J_{\bold{k}',\bold{q}'}}P^{(+)}_{J_{\bold{k}'',\bold{q}''}}\Big(V^{(\bold{k},\bold{q})_{-1}}_{J_{\bold{k}'',\bold{q}''}}-\langle V^{(\bold{k},\bold{q})_{-1}}_{J_{\bold{k}'',\bold{q}''}}\rangle\Big)\,P^{(+)}_{J_{\bold{k}'',\bold{q}''}}\frac{1}{\sum_{\bold{j}\in J_{\bold{r},\bold{i}}} P^{\perp}_{\Omega_{\bold{j}}}+1}\psi \rangle\Big| \nonumber \\
&\leq  & \sup_{\| \psi\|=1}\,\overline{\sum}_{J_{\bold{k}',\bold{q}' }\,,\,J_{\bold{k}'',\bold{q}''}}  \,\Big\|\frac{P^{(+)}_{J_{\bold{k}',\bold{q}'}}P^{(+)}_{J_{\bold{k}'',\bold{q}''}}}{\sum_{\bold{j}\in J_{\bold{r},\bold{i}}} P^{\perp}_{\Omega_{\bold{j}}}+1}\,\psi\Big\|^2 \cdot  \| V^{(\bold{k},\bold{q})_{-1}}_{J_{\bold{k}',\bold{q}'}}-\langle V^{(\bold{k},\bold{q})_{-1}}_{J_{\bold{k}',\bold{q}'}}\rangle\|\cdot  \|  V^{(\bold{k},\bold{q})_{-1}}_{J_{\bold{k}'',\bold{q}''}}-\langle V^{(\bold{k},\bold{q})_{-1}}_{J_{\bold{k}'',\bold{q}''}}\rangle\|\nonumber \\
&\leq &\sup_{\| \psi\|=1}\,\overline{\sum}_{J_{\bold{k}',\bold{q}'} \,,\,J_{\bold{k}'',\bold{q}''}}  \, 4 \|V^{(\bold{k},\bold{q})_{-1}}_{J_{\bold{k}',\bold{q}'}}\|\cdot \|V^{(\bold{k},\bold{q})_{-1}}_{J_{\bold{k}'',\bold{q}''}}\|
\cdot  \Big\|\frac{P^{(+)}_{J_{\bold{k}',\bold{q}'}}P^{(+)}_{J_{\bold{k}'',\bold{q}''}}}{\sum_{\bold{j}\in J_{\bold{r},\bold{i}}} P^{\perp}_{\Omega_{\bold{j}}}+1}\,\psi\Big\|^2 \,\\
&\leq &\sup_{\| \psi\|=1}\,\sum_{J_{\bold{k}',\bold{q}'} \,,\,J_{\bold{k}'',\bold{q}''}\,\in \,\mathcal{G}^{(\bold{k},\bold{q})}_{\bold{r},\bold{i}}}  \, 4 \|V^{(\bold{k},\bold{q})_{-1}}_{J_{\bold{k}',\bold{q}'}}\|\cdot \|V^{(\bold{k},\bold{q})_{-1}}_{J_{\bold{k}'',\bold{q}''}}\|
\cdot  \Big\|\frac{P^{(+)}_{J_{\bold{k}',\bold{q}'}}P^{(+)}_{J_{\bold{k}'',\bold{q}''}}}{\sum_{\bold{j}\in J_{\bold{r},\bold{i}}} P^{\perp}_{\Omega_{\bold{j}}}+1}\,\psi\Big\|^2 \,\label{nonint-1}\\
&=&\sup_{\| \psi\|=1}\,\langle \sum_{J_{\bold{k}',\bold{q}'}\,\in \,\mathcal{G}^{(\bold{k},\bold{q})}_{J_{\bold{r},\bold{i}}}}2\|V^{(\bold{k},\bold{q})_{-1}}_{J_{\bold{k}',\bold{q}'}}\|\frac{P^{(+)}_{J_{\bold{k}',\bold{q}'}}}{\sum_{\bold{j}\in J_{\bold{r},\bold{i}}} P^{\perp}_{\Omega_{\bold{j}}}+1}\psi\,,\,\sum_{J_{\bold{k}'',\bold{q}''}\,\in \,\mathcal{G}^{(\bold{k},\bold{q})}_{J_{\bold{r},\bold{i}}}}2\|V^{(\bold{k},\bold{q})_{-1}}_{J_{\bold{k}'',\bold{q}''}}\|\frac{P^{(+)}_{J_{\bold{k}'',\bold{q}''}}}{\sum_{\bold{j}\in J_{\bold{r},\bold{i}}} P^{\perp}_{\Omega_{\bold{j}}}+1}\psi\rangle \quad\quad\quad   \label{nonint-2}\\
&=&\Big\|\sum_{J_{\bold{k}',\bold{q}'}\,\in \,\mathcal{G}^{(\bold{k},\bold{q})}_{J_{\bold{r},\bold{i}}}}2\|V^{(\bold{k},\bold{q})_{-1}}_{J_{\bold{k}',\bold{q}'}}\|\frac{P^{(+)}_{J_{\bold{k}',\bold{q}'}}}{\sum_{\bold{j}\in J_{\bold{r},\bold{i}}} P^{\perp}_{\Omega_{\bold{j}}}+1}\Big\|^2\,,
\end{eqnarray}
where in the step from (\ref{nonint-1}) to  (\ref{nonint-2}) we have used $[P^{(+)}_{J_{\bold{k}',\bold{q}'}}\,,\,P^{(+)}_{J_{\bold{k}'',\bold{q}''}}]=0$.

Now suppose that there are $1\leq l\leq d$ components of $\bold{k}$ different from the corresponding ones in $\bold{r}$, without loss of generality we can assume that they are the first $l$ components; for $l\leq d-1$  we get
\begin{eqnarray}
& &\Big\|\sum_{J_{\bold{k}',\bold{q}'}\,\in \,\mathcal{G}^{(\bold{k},\bold{q})}_{J_{\bold{r},\bold{i}}}}2\|V^{(\bold{k},\bold{q})_{-1}}_{J_{\bold{k}',\bold{q}'}}\|\frac{P^{(+)}_{J_{\bold{k}',\bold{q}'}}}{\sum_{\bold{j}\in J_{\bold{r},\bold{i}}} P^{\perp}_{\Omega_{\bold{j}}}+1}\Big\|  \label{3-0}
\\ \label{3-1}
&= &\Big\|\sum_{\bold{s}:\exists \bold{u} \text{ with }J_{\bold{s},\bold{u}} \,\in \,\mathcal{G}^{(\bold{k},\bold{q})}_{J_{\bold{r},\bold{i}}}} \,\,\sum_{\bold{u}\,:\,\, J_{\bold{s},\bold{u}} \,\in \,\mathcal{G}^{(\bold{k},\bold{q})}_{J_{\bold{r},\bold{i}}}} 2\|V^{(\bold{k},\bold{q})_{-1}}_{J_{\bold{s},\bold{u}}}\|\frac{P^{(+)}_{J_{\bold{s},\bold{u}}}}{\sum_{\bold{j}\in J_{\bold{r},\bold{i}}} P^{\perp}_{\Omega_{\bold{j}}}+1}\Big\|\\\label{3-2}
&\leq &C \cdot \Big\{ \sum_{s_{1}=r_1-k_1}^{r} \dots \sum_{s_{l}=r_l - k_l}^{r} \sum_{s_{l+1}=0}^{r}\dots \sum_{s_{d}=0}^{r} \frac{t^{\left(\sum_{j=1}^{d}\frac{s_j}{3}\right)-\frac{1}{3}}}{(s_1+\dots +s_d)^{x_d}}\cdot \Big[  \prod_{j=l+1}^{d}(s_j+1)\Big]\Big\}, 
\end{eqnarray}
 (we call $s_1,\dots ,s_d$ the components of $\bold{k}'$) where in the step from (\ref{3-1}) to  (\ref{3-2}) we use:
\begin{itemize}
\item an upper bound for $\|V^{(\bold{k},\bold{q})_{-1}}_{J_{\bold{s},\bold{u}}}\|$ that is independent of $\bold{u}$ by means of the inductive hypothesis (\ref{R3-3}),
 \begin{equation}
 \|V^{(\bold{k},\bold{q})_{-1}}_{J_{\bold{s},\bold{u}}}\|\leq 96\cdot \frac{t^{\frac{s-1}{3}}}{s^{x_d}\,;} 
 \end{equation}
\item
the fact that, for fixed $\bold{k}^\prime$, if $k_j\neq r_j$ for $j=1,\cdots ,l$ then $q^\prime_1,\cdots, q^\prime_l$ are uniquely\footnote{Note that, for $1\leq j\leq l$, if $q_j\neq i_j$ and $q_j\neq i_j+k_j$ then $k^\prime_j$ must coincide with $r_j$,  thus $q^\prime_j$ is fixed.  Otherwise if $q_j= i_j$ or $q_j= i_j+k_j$, then, for fixed $k^{\prime}_j$, $q^{\prime}_j=i_j+r_j-k^{\prime}_j$ or $q^{\prime}_j=i_j$,  respectively.} determined by the condition $[ J_{\bold{k}',\bold{q}'} \cup J_{\bold{k},\bold{q}} ]=J_{\bold{r},\bold{i}}$;
\item
the estimate
 $$\sum_{\bold{u}\,:\,u_{1},\dots,u_l=\text{fixed}\,,\, J_{\bold{s},\bold{u}} \,\in \,\mathcal{G}^{(\bold{k},\bold{q})}_{J_{\bold{r},\bold{i}}}}\,\,\,P^{(+)}_{J_{\bold{s},\bold{u}}}\,\leq \Big\{\prod_{j=l+1}^{d}(s_j+1)\Big\}\,\sum_{\bold{j}\in J_{\bold{r},\bold{i}}}  P^{\perp}_{\Omega_{\bold{j}}}$$
that can be proved following the same reasoning of Corollary \ref{op-ineq-2}.
\end{itemize}
When $l=d$ the estimate of (\ref{3-0}) written above holds  with the product 
$\prod_{j=l+1}^d (s_j+1)$ replaced by $1$.

\noindent
Next, for $j=1,\dots,l$, we set 
\begin{equation}\label{def-rho} 
\rho_j:=s_j-(r_j-k_j)\quad \Rightarrow \quad s_j=\rho_j+(r_j-k_j)\,,
\end{equation}
and we observe that since $s_j\geq r_j-k_j$ for  $j=1,\dots,l$, and $s_j\geq 0$ for $j=l+1,\dots,d$, we have
\begin{equation}
(s_1+\dots +s_d)^{x_d}\geq (r_1-k_1+\dots +r_l-k_l)^{x_d}\,.
\end{equation}
Hence we can estimate
\begin{eqnarray}
&&(\ref{3-2})\nonumber \\
& \leq &C\cdot t^{-1/3}\cdot t^{\sum_{j=1}^{l}\frac{(r_j-k_j)}{3}} \cdot \Big\{ \sum_{s_{l+1}=0}^{r}\dots \sum_{s_{d}=0}^{r} \frac{1}{(r_1-k_1+\dots +r_l-k_l)^{x_d}}\cdot \Big[ t^{\sum_{j=l+1}^{d}\frac{s_{j}}{3}}\cdot \prod_{j=l+1}^{d}(s_j+1)\Big]  \sum_{\rho_{1}=0}^{\infty}\dots \sum_{\rho_{l}=0}^{\infty} t^{\sum_{j=1}^{l}\frac{\rho_{j}}{3}} \Big\}\nonumber \\
&= & C\cdot t^{-1/3}\cdot t^{\sum_{j=1}^{l}\frac{(r_j-k_j)}{3}}  \cdot  \frac{1}{(r_1-k_1+\dots +r_l-k_l)^{x_d}}\times \label{3-2,5}\\
& &\quad\quad\quad \times  \Big\{\sum_{s_{l+1}=0}^{r}\dots \sum_{s_{d}=0}^{r} \Big[ t^{\sum_{j=l+1}^{d}\frac{s_{j}}{3}}\cdot \prod_{j=l+1}^{d}(s_j+1)\Big] \sum_{\rho_{1}=0}^{\infty}\dots \sum_{\rho_{l}=0}^{\infty} t^{\sum_{j=1}^{l}\frac{\rho_{j}}{3}}  \Big\}\nonumber \\
&\leq &C_d\cdot t^{-1/3} \Big(\frac{t^{\frac{r-1}{3}}}{t^{\frac{k-1}{3}}}\Big) \cdot \frac{1}{(r_1-k_1+\dots +r_l-k_l)^{x_d}} \label{3-3}
\end{eqnarray}
where we have exploited:
\begin{itemize}
\item the quantity
\begin{equation}
\sum_{s_{l+1}=0}^{r}\dots \sum_{s_{d}=0}^{r} \Big[ t^{\sum_{j=l+1}^{d}\frac{s_{j}}{3}}\cdot \prod_{j=l+1}^{d}(s_j+1)\Big] \sum_{\rho_{1}=0}^{\infty}\dots \sum_{\rho_{l}=0}^{\infty} t^{\sum_{j=1}^{l}\frac{\rho_{j}}{3}}
\end{equation}
is bounded from above by a $d$-dependent constant;
\item
for the considered $\bold{k}$ 
 \begin{equation}
t^{\sum_{j=1}^{l}\frac{(r_j-k_j)}{3}}=t^{\frac{r-k}{3}}
 \end{equation}
since $k_{j}=r_j$ for $j=l+1,\dots,d$, by assumption.
\end{itemize}

\noindent
\emph{Leading terms in $(\ref{rest-2})$: Contribution proportional to (\ref{intersect})}
\\

\noindent
By the Schwarz inequality and the trivial bound $ab\leq \frac{a^2}{2}+\frac{b^2}{2}$, we estimate (recall the notation $\sum'$ in (\ref{sum-prime}))
\begin{eqnarray}
& &\sup_{\| \psi\|=1}\,  \sum'_{J_{\bold{k}',\bold{q}' }\,,\,J_{\bold{k}'',\bold{q}''}}  \,\Big|\langle \frac{1}{\sum_{\bold{j}\in \sigma_{\bold{r},\bold{i}}} P^{\perp}_{\Omega_{\bold{j}}}+1}\psi, \label{intersect-bis}\\
 & &\quad\quad\quad P^{(+)}_{J_{\bold{k}',\bold{q}'}}\Big(V^{(\bold{k},\bold{q})_{-1}}_{J_{\bold{k}',\bold{q}'}}-\langle V^{(\bold{k},\bold{q})_{-1}}_{J_{\bold{k}',\bold{q}'}}\rangle\Big)\,P^{(+)}_{J_{\bold{k}',\bold{q}'}}\,P^{(+)}_{J_{\bold{k}'',\bold{q}''}}\Big(V^{(\bold{k},\bold{q})_{-1}}_{J_{\bold{k}'',\bold{q}''}}-\langle V^{(\bold{k},\bold{q})_{-1}}_{J_{\bold{k}'',\bold{q}''}}\rangle\Big)\,P^{(+)}_{J_{\bold{k}'',\bold{q}''}}\frac{1}{\sum_{\bold{j}\in J_{\bold{r},\bold{i}}} P^{\perp}_{\Omega_{\bold{j}}}+1}\psi \rangle\Big| \nonumber\\
 &\leq &\sup_{\| \psi\|=1}\,  \sum'_{J_{\bold{k}',\bold{q}' }\,,\,J_{\bold{k}'',\bold{q}''}}  \,\, 4 \|V^{(\bold{k},\bold{q})_{-1}}_{J_{\bold{k}',\bold{q}'}}\|\cdot \|V^{(\bold{k},\bold{q})_{-1}}_{J_{\bold{k}'',\bold{q}''}}\|
\cdot \Big\{\frac{1}{2} \Big\|\frac{P^{(+)}_{J_{\bold{k}',\bold{q}'}}}{\sum_{\bold{j}\in J_{\bold{r},\bold{i}}} P^{\perp}_{\Omega_{\bold{j}}}+1}\,\psi\Big\|^2 +\frac{1}{2}\Big\|\frac{P^{(+)}_{J_{\bold{k}'',\bold{q}''}}}{\sum_{\bold{j}\in J_{\bold{r},\bold{i}}} P^{\perp}_{\Omega_{\bold{j}}}+1}\,\psi\Big\|^2 \Big\}\,.\quad\quad\quad\quad  \label{fin-intersect}
\end{eqnarray}
Since the expression in (\ref{fin-intersect}) is symmetric under the permutation of $J_{\bold{k}',\bold{q}'}$ with $J_{\bold{k}'',\bold{q}''}$, we can write
\begin{eqnarray}
&& (\ref{intersect-bis})\\
&\leq &\sup_{\| \psi\|=1}\,  \sum'_{J_{\bold{k}',\bold{q}' }\,,\,J_{\bold{k}'',\bold{q}''}}  \,\, 4 \|V^{(\bold{k},\bold{q})_{-1}}_{J_{\bold{k}',\bold{q}'}}\|\cdot \|V^{(\bold{k},\bold{q})_{-1}}_{J_{\bold{k}'',\bold{q}''}}\|
\cdot \Big\{ \Big\|\frac{P^{(+)}_{J_{\bold{k}',\bold{q}'}}}{\sum_{\bold{j}\in J_{\bold{r},\bold{i}}} P^{\perp}_{\Omega_{\bold{j}}}+1}\,\psi\Big\|^2 \Big\}\nonumber \\
&= &\sup_{\| \psi\|=1}\,  \sum'_{J_{\bold{k}',\bold{q}' }\,,\,J_{\bold{k}'',\bold{q}''}}  \,\, 4 \|V^{(\bold{k},\bold{q})_{-1}}_{J_{\bold{k}',\bold{q}'}}\|\cdot \|V^{(\bold{k},\bold{q})_{-1}}_{J_{\bold{k}'',\bold{q}''}}\|
\cdot \Big\{\langle\frac{1}{\sum_{\bold{j}\in J_{\bold{r},\bold{i}}} P^{\perp}_{\Omega_{\bold{j}}}+1}\,\psi, \frac{P^{(+)}_{J_{\bold{k}',\bold{q}'}}}{\sum_{\bold{j}\in J_{\bold{r},\bold{i}}} P^{\perp}_{\Omega_{\bold{j}}}+1}\,\psi\rangle \Big\}\nonumber \\
&= &\sup_{\| \psi\|=1}\,  \Big\{\langle\frac{1}{\sum_{\bold{j}\in J_{\bold{r},\bold{i}}} P^{\perp}_{\Omega_{\bold{j}}}+1}\,\psi, \sum'_{J_{\bold{k}',\bold{q}' }\,,\,J_{\bold{k}'',\bold{q}''}}  \,\, 4 \|V^{(\bold{k},\bold{q})_{-1}}_{J_{\bold{k}',\bold{q}'}}\|\cdot \|V^{(\bold{k},\bold{q})_{-1}}_{J_{\bold{k}'',\bold{q}''}}\|\frac{P^{(+)}_{J_{\bold{k}',\bold{q}'}}}{\sum_{\bold{j}\in J_{\bold{r},\bold{i}}} P^{\perp}_{\Omega_{\bold{j}}}+1}\,\psi\rangle \Big\}\nonumber\\
&\leq &\Big\|\sum'_{J_{\bold{k}',\bold{q}' }\,,\,J_{\bold{k}'',\bold{q}''}}  \,\, 4 \|V^{(\bold{k},\bold{q})_{-1}}_{J_{\bold{k}',\bold{q}'}}\|\cdot \|V^{(\bold{k},\bold{q})_{-1}}_{J_{\bold{k}'',\bold{q}''}}\|\frac{P^{(+)}_{J_{\bold{k}',\bold{q}'}}}{\sum_{\bold{j}\in J_{\bold{r},\bold{i}}} P^{\perp}_{\Omega_{\bold{j}}}+1}\,\Big\|\,.  \label{3-4}
\end{eqnarray}

With steps similar to (\ref{3-1})-(\ref{3-3}), assuming that there are $1\leq l\leq d$ components of $\bold{k}$ different from the corresponding ones in $\bold{r}$ (without loss of generality we identify them with the first $l$ components). Then we can bound (\ref{3-4}) as described below (warning: for $l= d$, $ \prod_{j=l+1}^{d}(s_j+1)\,,\, \prod_{j=l+1}^{d}s_j $ must be replaced by $1$ in (\ref{3-2-1}) and related fomulae)
 \begin{eqnarray}
(\ref{3-4})
&\leq&  \Big\|\sum_{J_{\bold{k}',\bold{q}'}\,\in \,\mathcal{G}^{(\bold{k},\bold{q})}_{J_{\bold{r},\bold{i}}}}  \,\, 4 \|V^{(\bold{k},\bold{q})_{-1}}_{J_{\bold{k}',\bold{q}'}}\|\frac{P^{(+)}_{J_{\bold{k}',\bold{q}'}}}{\sum_{\bold{j}\in J_{\bold{r},\bold{i}}} P^{\perp}_{\Omega_{\bold{j}}}+1} \sum_{J_{\bold{k}'',\bold{q}''}\,\in \,\mathcal{G}^{(\bold{k},\bold{q})}_{J_{\bold{r},\bold{i}}}\,\, :\,\,J_{\bold{k}',\bold{q}'}\cap J_{\bold{k}'',\bold{q}''}\neq \emptyset} \|V^{(\bold{k},\bold{q})_{-1}}_{J_{\bold{k}'',\bold{q}''}}\|\, \Big\|\nonumber\\
&\leq & C \cdot \Big\{ \sum_{s_{1}=r_1-k_1}^{r} \dots \sum_{s_{l}=r_l - k_l}^{r} \sum_{s_{l+1}=0}^{r}\dots \sum_{s_{d}=0}^{r} \frac{t^{(\sum_{j=1}^{d}\frac{s_j}{3})-\frac{1}{3}}}{(s_1+\dots +s_d)^{x_d}}\cdot \Big[  \prod_{j=l+1}^{d}(s_j+1)\Big]\Big\} \label{3-2-1} \\
& &\quad\times  \Big(\sum_{w=r-k}^r \frac{t^{\frac{w-1}{3}}}{w^{x_d}}\cdot \Big(\prod_{j=1}^{d}s_j\Big)\cdot w^{d-1}\Big)\nonumber
\end{eqnarray}
due to the estimate
$$\sum_{J_{\bold{k}'',\bold{q}''}\,\in \,\mathcal{G}^{(\bold{k},\bold{q})}_{J_{\bold{r},\bold{i}}}\, :\,J_{\bold{s},\bold{q}'}\cap J_{\bold{k}'',\bold{q}''}\neq \emptyset} \|V^{(\bold{k},\bold{q})_{-1}}_{J_{\bold{k}'',\bold{q}''}}\| \leq \mathcal{O}\Big( \sum_{w=r-k}^r \frac{t^{\frac{w-1}{3}}}{w^{x_d}}\cdot \Big(\prod_{j=1}^{d}s_j\Big)\cdot w^{d-1}\Big)$$
where:
\begin{itemize}
\item[i)]
$\mathcal{O}(\Big(\prod_{j=1}^{d}s_j\Big)\cdot w^{d-1})$ bounds from above the number of rectangles $J_{\bold{w}, \bold{q}''}$ overlapping with the rectangle $J_{\bold{s}, \bold{q}'}$;
\item[ii)]
$ \mathcal{O}( \frac{t^{\frac{w-1}{3}}}{w^{x_d}})$ is the bound to $ \|V^{(\bold{k},\bold{q})_{-1}}_{J_{\bold{w},\bold{q}''}}\| $ provided by the inductive hypotheses.
\end{itemize}
Next, using the definition in (\ref{def-rho})  and arguments as in (\ref{3-2,5})-(\ref{3-3}), we write
\begin{eqnarray}
& &(\ref{3-2-1})\nonumber \\
&\leq & C\cdot  (t^{-1/3} \frac{t^{\frac{r-1}{3}}}{t^{\frac{k-1}{3}}})^2 \cdot \Big\{ \frac{ 1}{(r_1-k_1+\dots +r_l-k_l)^{x_d}}\cdot \sum_{s_{l+1}=0}^{r}\dots \sum_{s_{d}=0}^{r}   \sum_{w=r-k}^{r}\big[w^{d-1}\frac{1}{w^{x_d}}\big]\label{inters-bis}\\
& &\quad \times  \Big[  \Big(\prod_{j=l+1}^{d}s_j\Big)\cdot t^{\sum_{j=l+1}^{d}s_{j}/3}\cdot \prod_{j=l+1}^{d}(s_j+1)\Big]\sum_{\rho_{1}=0}^{\infty}\dots \sum_{\rho_{l}=0}^{\infty} t^{\sum_{j=1}^{l}\frac{\rho_{j}}{3}} \prod_{j=1}^l\Big(\rho_j+r_j-k_j\Big)\Big\}\,. \nonumber 
\end{eqnarray}
Now we multiply the r-h-s of (\ref{inters-bis}) by
$$\frac{ (r_1-k_1+\dots +r_l-k_l)^d}{(r_1-k_1+\dots +r_l-k_l)^{l}}\geq 1\,$$
and we get
\begin{eqnarray}
& &(\ref{3-2-1})\\
&\leq &C\cdot  (t^{-1/3} \frac{t^{\frac{r-1}{3}}}{t^{\frac{k-1}{3}}})^2 \cdot \Big\{ \frac{ (r_1-k_1+\dots +r_l-k_l)^d}{(r_1-k_1+\dots +r_l-k_l)^{x_d}}\cdot \sum_{s_{l+1}=0}^{r}\dots \sum_{s_{d}=0}^{r}   \sum_{w=r-k}^{r}\big[w^{d-1}\frac{1}{w^{x_d}}\big] \label{inters-bisbis}\\
&&\quad \times  \Big[  \Big(\prod_{j=l+1}^{d}s_j\Big)\cdot t^{\sum_{j=l+1}^{d}s_{j}/3}\cdot \prod_{j=l+1}^{d}(s_j+1)\Big]\sum_{\rho_{1}=0}^{\infty}\dots \sum_{\rho_{l}=0}^{\infty} t^{\sum_{j=1}^{l}\frac{\rho_{j}}{3}} \prod_{j=1}^l\Big(\frac{\rho_j+r_j-k_j}{(r_1-k_1+\dots +r_l-k_l)}\Big)\Big\}\nonumber \\
& \leq &C_d\cdot  (t^{-1/3}\frac{t^{\frac{r-1}{3}}}{t^{\frac{k-1}{3}}})^2 \cdot( \frac{1}{(r_1-k_1+\dots +r_l-k_l)^{x_d-d}})^2\label{inters-bisbisbis} 
 \end{eqnarray}
where in the step from (\ref{inters-bisbis}) to (\ref{inters-bisbisbis})  we have used that  $x_d\geq d+1$ and that all the following quantities are bounded from above by a  $d$-dependent constant:\\
\begin{itemize}
\item
$$ (r_1-k_1+\dots +r_l-k_l)^{x_d-d}\cdot \,\sum_{w=r-k}^{r}\,w^{d-1}\frac{1}{w^{x_d}}$$
\item
$$\sum_{s_{l+1}=0}^{r}\dots \sum_{s_{d}=0}^{r}  \Big(\prod_{j=l+1}^{d}s_j\Big)\cdot t^{\sum_{j=l+1}^{d}s_{j}/3}\cdot \prod_{j=l+1}^{d}(s_j+1)$$
\item
$$\sum_{\rho_{1}=0}^{\infty}\dots \sum_{\rho_{l}=0}^{\infty} t^{\sum_{j=1}^{l}\frac{\rho_{j}}{3}} \prod_{j=1}^l\Big(\frac{\rho_j+r_j-k_j}{(r_1-k_1+\dots +r_l-k_l)}\Big)\,.$$
\end{itemize}

\noindent
\emph{Leading terms in  $(\ref{rest-2})$: Contribution proportional to  (\ref{2.158})}
\\

\noindent
Finally, we estimate (\ref{2.158}) by exploiting the inequality
 \begin{eqnarray}
 & &\Big\|\frac{1}{G_{J_{\bold{k},\bold{q}}}-E_{J_{\bold{k},\bold{q}}}}( \sum_{\bold{j}\in J_{\bold{k},\bold{q}}} P^{\perp}_{\Omega_{\bold{j}}}+1) \frac{1}{\sum_{\bold{j}\in J_{\bold{k},\bold{q}}} P^{\perp}_{\Omega_{\bold{j}}}+1}P^{(+)}_{J_{\bold{k},\bold{q}}}\,V^{(\bold{k},\bold{q})_{-1}}_{J_{\bold{k},\bold{q}}}\,P^{(-)}_{J_{\bold{k},\bold{q}}}\Big\| \label{bound-in}\\
 &\leq & \Big\|\frac{1}{G_{J_{\bold{k},\bold{q}}}-E_{J_{\bold{k},\bold{q}}}}(\sum_{\bold{j}\in J_{\bold{k},\bold{q}}} P^{\perp}_{\Omega_{\bold{j}}}+1)\Big\|\,\Big\| \frac{1}{\sum_{\bold{j}\in J_{\bold{k},\bold{q}}} P^{\perp}_{\Omega_{\bold{j}}}+1}P^{(+)}_{J_{\bold{k},\bold{q}}}\,V^{(\bold{k},\bold{q})_{-1}}_{J_{\bold{k},\bold{q}}}\,P^{(-)}_{J_{\bold{k},\bold{q}}} \Big\| \\
& \leq &3\cdot \frac{t^{\frac{k-1}{3}}}{k^{x_d+2d}},\label{bound-fin}
 \end{eqnarray}
where the first factor can be estimated to be less than $3$ provided $t_d$ is small enough, by using the  bound in  (\ref{specG})  (see Lemma  \ref{bound-lemma-gap})
  that holds due to S2) in the previous step; for the second factor we invoke the  inductive hypothesis in (\ref{R3-1}).
\\

Hence we conclude that at fixed $\bold{k}$, and with $l$ components different from the corresponding components of $\bold{r}$,
\begin{equation}\label{contributo-small-leading}
\|(\ref{lead-2.115})\| \leq C_d \cdot t^{2/3}\cdot \frac{ t^{\frac{r-1}{3}}}{ (r_1-k_1+\dots +r_l-k_l)^{x_d-d} \cdot k^{x_d+2d}}\,.
\end{equation}
\\

\noindent
\emph{Higher order terms in  $(\ref{rest-2})$}

\noindent
In order to show the bound in (\ref{R3-1}), with regard to $(\ref{rest-2})$ we have still to estimate:
\begin{itemize}
\item remainder  (\ref{rem-2.115})  (coming from the study of $(\ref{rest-2-first})_{small}$) and those corresponding to  $(\ref{rest-2-first})_{large}$, i.e., proportional to terms with $J_{\bold{k}',\bold{q}'}$ such that  $(\bold{k}',\bold{q}')\succ \bold{k},\bold{q})$;
\item  the contribution due to (\ref{rest-2-second}).
\end{itemize}
We observe that:

\noindent
i) in all these terms there are either two factors $S_{\bold{k},\bold{q}}$  or two factors $\|(V^{({\bold{k},\bold{q})_{-1}}}_{J_{\bold{k},\bold{q}}})_1\|$ (see (\ref{two-factors})) or $J_{\bold{k}',\bold{q}'}$ is large such that $(\bold{k}',\bold{q}')\succ (\bold{k},\bold{q})$;  thus we get at least  an extra factor $\mathcal{O}(t^{\frac{r-2 r^{1/4}-1}{3}})$;

\noindent
ii) the bound from above, $ \mathcal{O}(r^{2d-1})$,  of the number of the elements of $\mathcal{G}^{(\bold{k},\bold{q})}_{J_{\bold{r},\bold{i}}}$ (see Remark \ref{shapes}).

\noindent
Hence, just using the  inductive hypotheses (\ref{R3-2}) and (\ref{R3-3}) we can estimate
\begin{eqnarray}
& &\|(\ref{rem-2.115}) \|+\|(\ref{rest-2-first})_{large}\|+\|(\ref{rest-2-second})\| \label{sum-rem}\\
&\leq&C_d\cdot t\cdot r^{4d-1}\cdot t^{\frac{r-2\cdot r^{1/4}-1}{3}} \cdot \frac{ t^{\frac{r-1}{3}}}{ (r-k)^{x_d} \cdot k^{x_d}}\,.
\end{eqnarray}
At fixed $k$ there are at most $\mathcal{O}(r^d\cdot k^{d-1})$ contributions of type (\ref{sum-rem}).
\\

\noindent
\emph{Complete estimate of  (\ref{R3-1})}

\noindent
Finally, by the re-expansion outlined above and due to the estimates of (\ref{rest-1}), (\ref{lead-2.115}), (\ref{rem-2.115}), $(\ref{rest-2-first})_{large}$, and $(\ref{rest-2-second})$ that have been derived (see (\ref{contributo-small-leading}), (\ref{first-nonleading}), and (\ref{sum-rem})), we can conclude that 
\begin{eqnarray}
& &\|\frac{1}{\sum_{\bold{j}\in J_{\bold{r},\bold{i}}} P^{\perp}_{\Omega_{\bold{j}}}+1}P^{(+)}_{J_{\bold{r},\bold{i}}}\,V^{(\bold{k},\bold{q})}_{J_{\bold{r},\bold{i}}}\,P^{(-)}_{J_{\bold{r},\bold{i}}}\|\\
&\leq &\|\frac{1}{\sum_{\bold{j}\in J_{\bold{r},\bold{i}}} P^{\perp}_{\Omega_{\bold{j}}}+1}P^{(+)}_{J_{\bold{r},\bold{i}}}\,V^{(\bold{k},\bold{q})_{**}}_{J_{\bold{r},\bold{i}}}\,P^{(-)}_{J_{\bold{r},\bold{i}}}\| \label{first-summand}\\
& &+C_d\cdot t\cdot \sum_{l=1}^{d}\binom{d}{l}\cdot \sum_{k_1=0}^{r_1-1}\dots \sum_{k_l=0}^{r_{l}-1}\,\Theta(k-r+\lfloor r^\frac{1}{4}\rfloor )\,\times \label{th-in-0}\\
& &\quad\quad\quad\quad\quad\quad \times \frac{ t^{\frac{r-1}{3}}}{ (r_1-k_1+\dots +r_l-k_l)^{x_d-d} \cdot k^{x_d+2d}}\label{th-in}
  \\
& &+\sum_{k=r-\lfloor r^\frac{1}{4}\rfloor }^{k= r-1 }\,\Big\{  c_d\cdot r^d\cdot k^{d-1} \cdot t \cdot \frac{t^{\frac{k-1}{3}}\cdot t^{\frac{r-1}{3}}}{k^{x_d}\cdot r^{x_d}}\label{control-rest-1}\\
& &\quad\quad\quad\quad +C_d\cdot t\cdot r^{5d-1}\cdot k^{d-1}\cdot t^{\frac{r-2r^\frac{1}{4}-1}{3}} \cdot \frac{ t^{\frac{r-1}{3}}}{ (r-k)^{x_d} \cdot k^{x_d}}\Big\} \label{th-fin}
\end{eqnarray}
where $\Theta$ is the characteristic function of $\mathbb{R}^+$, indeed  $k\geq r- \lfloor r^\frac{1}{4}\rfloor$ in regime $\frak{R}3$.
In addition to summand (\ref{first-summand}) that  is smaller than $2\cdot \frac{t^\frac{r-1}{3}}{r^{x_d+2d}}$ by the inductive hypothesis (\ref{inductive-reg-2}),  on the r-h-s of the estimate above  we have three summands that we shall discuss in detail. Prior to this discussion, we explain why the final estimate in (\ref{PVP-1}) works.
\begin{rem}\label{whyregime3}
We point out that:
\begin{itemize}
\item[i)] regarding the expression in (\ref{th-in-0})-(\ref{th-in}), the factor $\frac{1}{k^{x_d+2d}}$ (coming from the inductive hypothesis used to estimate (\ref{2.158})) provides the expected behaviour since $k\geq r-\lfloor r^{\frac{1}{4}}\rfloor$ in regime $\mathfrak{R}3$, and the rest can be made less than $\frac{t^{\frac{r-1}{3}}}{3}$ due to the definition of $x_d$, as we explain below;
\item[ii)] regarding the expressions in (\ref{control-rest-1}) and (\ref{th-fin}), we exploit the extra powers $t^{\frac{k-1}{3}}$ and $t^{\frac{r-2r^\frac{1}{4}-1}{3}} $, respectively,  in order to control the sum over $k$ and provide the desired behavior.
\end{itemize}
\end{rem}

As for (\ref{th-in-0})-(\ref{th-in}),
we first observe that we have
\begin{eqnarray}
&&\sum_{k_1=0}^{r_1-1}\dots \sum_{k_l=0}^{r_{l}-1}\,\Theta(k-r+r^\frac{1}{4})\,\times\frac{ t^{\frac{r-1}{3}}}{ (r_1-k_1+\dots +r_l-k_l)^{x_d-d} \cdot k^{x_d+2d}}\\
&\leq &C_d\cdot  \sum_{s_1=1}^{r_1}\dots \sum_{s_l=1}^{r_{l}}\frac{ t^{\frac{r-1}{3}}}{ (s_1+s_2+\ldots +s_l)^{x_d-d} \cdot r^{x_d+2d}}\\
&\leq &C_d\cdot \frac{t^\frac{r-1}{3}}{r^{x_d+2d}}\sum_{s_1=1}^{\infty}\dots \sum_{s_l=1}^{\infty}\frac{ 1}{ (s_1+s_2+\ldots +s_l)^{x_d-d}}\\
&\leq & C_d\cdot \frac{t^\frac{r-1}{3}}{r^{x_d+2d}}
\end{eqnarray}
where, since $x_d-d>l$, $ \sum_{s_1=1}^{\infty}\dots \sum_{s_l=1}^{\infty}\frac{ 1}{ (s_1+s_2+\ldots +s_l)^{x_d-d}}$ is bounded by a $d$-dependent constant. 
Therefore, since $$x_d>d+d-1=2d-1\,,$$ we see that the overall quantity can be made less than
$\frac{1}{3}\cdot \frac{t^\frac{r-1}{3}}{r^{x_d+2d}}$ provided $t\geq 0$ is small enough.\\
As for  (\ref{control-rest-1}), we have
\begin{eqnarray}
&&\sum_{k=r-\lfloor r^\frac{1}{4} \rfloor }^{k= r-1 }\, c_d\cdot r^d\cdot k^{d-1} \cdot t \cdot \frac{t^{\frac{k-1}{3}}\cdot t^{\frac{r-1}{3}}}{k^{x_d}\cdot r^{x_d}}\\
&\leq & r^\frac{1}{4}\cdot  c_d\cdot r^d \cdot r^{d-1}\cdot t^{\frac{r-r^\frac{1}{4}-1}{3}}\cdot t^\frac{r-1}{3}\cdot \frac{1}{(r-r^\frac{1}{4})^{x_d}\cdot r^{x_d}}\\
&\leq&2^{x_d} \cdot c_d \cdot  t^{\frac{r-r^\frac{1}{4}-1}{3}}\cdot \frac{t^{\frac{r-1}{3}}}{r^{2x_d-2d+\frac{3}{4}}}\\
&\leq& \frac{1}{3} \cdot \frac{t^\frac{r-1}{3}}{r^{x_d+2d}}
\end{eqnarray}
since $x_d\geq 4d -\frac{3}{4}$ and  $t\geq 0$ is small enough.\\
As for (\ref{th-fin}), this quantity can be estimated in the following way:
\begin{eqnarray}
&&\sum_{k=r-\lfloor r^\frac{1}{4}\rfloor}^{r-1}C_d\cdot t\cdot r^{5d-1}\cdot k^{d-1}\cdot t^{\frac{r-2r^\frac{1}{4}-1}{3}} \cdot \frac{ t^{\frac{r-1}{3}}}{ (r-k)^{x_d} \cdot k^{x_d}}\\\
&\leq &
r^\frac{1}{4}\cdot 2^{x_d} \cdot C_d \cdot t \cdot t^{\frac{r-2r^\frac{1}{4}-1}{3}}\frac{t^\frac{r-1}{3}}{r^{x_d-6d+2}}\\
&=&2^{x_d} \cdot  C_d \cdot t\cdot  t^{\frac{r-2r^\frac{1}{4}-1}{3}}\cdot \frac{t^\frac{r-1}{3}}{r^{x_d-6d+\frac{7}{4}}}\\
&\leq &\frac{1}{3}\cdot \frac{t^\frac{r-1}{3}}{r^{x_d+2d}}
\end{eqnarray}
where the last inequality holds provided $t\geq 0$ is so small as to fulfill the inequality
$$2^{x_d} \cdot C_d \cdot t\cdot t^\frac{r-2r^\frac{1}{4}-1}{3}\leq \frac{1}{3\cdot r^{8d-\frac{7}{4}}}$$ uniformly in $r$. 

\noindent
Finally, for $t\geq 0$ small enough, we find
\begin{eqnarray}\label{PVP-1}
& &\|\frac{1}{\sum_{\bold{j}\in J_{\bold{r},\bold{i}}} P^{\perp}_{\Omega_{\bold{j}}}+1}P^{(+)}_{J_{\bold{r},\bold{i}}}\,V^{(\bold{k},\bold{q})}_{J_{\bold{r},\bold{i}}}\,P^{(-)}_{J_{\bold{r},\bold{i}}}\|\\
&\leq &2\cdot\frac{t^{\frac{r-1}{3}}}{r^{x_d+2d}} +3\cdot\frac{1}{3}\cdot \frac{t^{\frac{r-1}{3}}}{r^{x_d+2d}} \label{secondo}  \\
&=&3\cdot \frac{t^{\frac{r-1}{3}}}{r^{x_d+2d}}
\end{eqnarray}
as claimed.
\\

The control of 
\begin{eqnarray}
& &\frac{1}{\sum_{\bold{j}\in J_{\bold{r},\bold{i}}} P^{\perp}_{\Omega_{\bold{j}}}+1}\, P^{(+)}_{J_{\bold{r},\bold{i}}}\,V^{(\bold{k},\bold{q})}_{J_{\bold{r},\bold{i}}}\,P^{(+)}_{J_{\bold{r},\bold{i}}}\,\frac{1}{\sum_{\bold{j}\in J_{\bold{r},\bold{i}}} P^{\perp}_{\Omega_{\bold{j}}}+1}
\end{eqnarray}
is analogous. The study of
\begin{eqnarray}
& &\frac{1}{\sum_{\bold{j}\in J_{\bold{r},\bold{i}}} P^{\perp}_{\Omega_{\bold{j}}}+1}\, P^{(-)}_{J_{\bold{r},\bold{i}}}\,V^{(\bold{k},\bold{q})}_{J_{\bold{r},\bold{i}}}\,P^{(-)}_{J_{\bold{r},\bold{i}}}\,\frac{1}{\sum_{\bold{j}\in J_{\bold{r},\bold{i}}} P^{\perp}_{\Omega_{\bold{j}}}+1}
\end{eqnarray}
is actually simpler, since in the analogous re-expansion  the terms proportional to  \emph{small} $J_{\bold{k}',\bold{q}'}$ are identically  zero.
\\

\noindent
\underline{\emph{Proof of  (\ref{R3-2}) and (\ref{R3-3})}}
\\

\noindent
Concerning (\ref{R3-2}), we observe that
\begin{eqnarray}
& &\|P^{(\#)}_{J_{\bold{r},\bold{i}}}\,V^{(\bold{k},\bold{q})}_{J_{\bold{r},\bold{i}}}\,P^{(\hat{\#})}_{J_{\bold{r},\bold{i}}}\|\\
&\leq &\|\sum_{\bold{j}\in J_{\bold{r},\bold{i}}} P^{\perp}_{\Omega_{\bold{j}}}+1\|^2\cdot \|\frac{1}{\sum_{\bold{j}\in J_{\bold{r},\bold{i}}} P^{\perp}_{\Omega_{\bold{j}}}+1}\, P^{(\#)}_{J_{\bold{r},\bold{i}}}\,V^{(\bold{k},\bold{q})}_{J_{\bold{r},\bold{i}}}\,P^{(\hat{\#})}_{J_{\bold{r},\bold{i}}}\,\frac{1}{\sum_{\bold{j}\in J_{\bold{r},\bold{i}}} P^{\perp}_{\Omega_{\bold{j}}}+1}\|
\end{eqnarray}
 where $\#,\hat{\#}=\pm $, and $\|\sum_{\bold{j}\in J_{\bold{r},\bold{i}}} P^{\perp}_{\Omega_{\bold{j}}}+1\|^2\leq (r^d+1)^2\leq 4 \cdot r^{2d}$, then we use the estimates in (\ref{R3-1}) proven above.

\noindent
In order to prove (\ref{R3-3}), it is enough to exploit inequality (\ref{bound-V}) and  (\ref{R3-2}).\\

\medskip
\emph{Inductive step to prove S2)}

\noindent
Since we have already proven S1), the bound in (\ref{ass-2-multi}) is fulfilled for $t\geq 0$ sufficiently small, and we can use Lemma (\ref{bound-lemma-gap}) and Corollary (\ref{cor-gap}). Hence, S2) holds for $t\geq 0$ sufficiently small but independent of $N$, $\bold{k}$, and $\bold{q}$. \qed

Analogously to the treatment of the one-dimensional systems in \cite{FP}, we can now derive the main result of the paper.

\begin{thm}\label{main-res}
Under the assumption that (\ref{gaps}) and (\ref{potential}) hold, the Hamiltonian $K_{N}$ defined in (\ref{Hamiltonian}) has the following properties: There exists some $t_d > 0$ such that, for any $t\in \mathbb{R}$ with 
$\vert t \vert < t_d$, and for all $N < \infty$,
\begin{enumerate}
\item[(i)]{ $K_{N}\equiv K_{N}(t)$ has a unique ground-state; and}
\item[(ii)]{ the energy spectrum of $K_{N}$ has a strictly positive gap, $\Delta_{N}(t) \geq \frac{1}{2}$, above the ground-state energy.}
\end{enumerate}
\end{thm}

\noindent
\emph{Proof.}
The final transformed Hamiltonian is $K_{N}^{(\bold{N-1},\bold{1})} \equiv G_{J_{\bold{N-1},\bold{1}}}+tV^{(\bold{N-1},\bold{1})}_{J_{\bold{N-1},\bold{1}}}$. Hence the composition of the unitary conjugations associated with each block-diagonalization step yields the unitary operator 
$\text{exp}(S_{N}(t))$,  (see (\ref{conjug})), such that  the operator
$$e^{S_{N}(t)}K_{N}(t)e^{-S_{N}(t)}=G_{J_{\bold{N-1},\bold{1}}}+tV^{(\bold{N-1},\bold{1})}_{J_{\bold{N-1},\bold{1}}}=: \widetilde{K}_{N}(t),$$  
enjoys the properties in (\ref{block-diag}) and (\ref{gapss}), which follow from Theorem \ref{th-norms} and from (\ref{gap-in})-(\ref{gap-fin}), for $(\bold{k},\bold{q})=(\bold{N-1},\bold{1})$, where we also include the block-diagonalized potential $V^{(\bold{N-1},\bold{1})}_{J_{\bold{N-1},\bold{1}}}$. \qed


\setcounter{equation}{0}
\begin{appendix}
\section{Appendix}\label{appendix}
\begin{lem}\label{op-ineq-1} 
For any $J_{\bold{l},\bold{i}}$, recalling that
\begin{equation}\label{pro-plus-multi-bis-bis}
P^{(+)}_{J_{\bold{l},\bold{i}}}:=  \Big(\bigotimes_{\bold{j}\in J_{\bold{l},\bold{i}}}P_{\Omega_{\bold{j}}}\Big)^{\perp}\,,
\end{equation}
we have that the inequality
\begin{equation}\label{gen-lemmaA-1-bis}
 \sum_{\bold{j}\in J_{\bold{l},\bold{i}}} P^{\perp}_{\Omega_{\bold{j}}}\geq P^{(+)}_{J_{\bold{l},\bold{i}}} \,
\end{equation}
holds true where $P^{\perp}_{\Omega_{\bold{j}}}:=\charf_{\bold{j}}-P_{\Omega_{\bold{j}}}$.
\end{lem}

\noindent
\emph{Proof}

\noindent
Let us define the self-adjoint operator $A$ acting on $\mathcal{H}^{(N)}$ as

$$A:=\sum_{\bold{j}\in J_{\bold{l},\bold{i}}}  P^{\perp}_{\Omega_{\bold{j}}}+  \Big(\bigotimes_{\bold{j}\in J_{\bold{l},\bold{i}}}P_{\Omega_{\bold{j}}}\Big)\,.$$
Since $A$ is the sum of $(l_1+1)(l_2+1)\ldots(l_d+1)+1$ orthogonal projections that commute with one another, its spectrum 
must be contained in the set $$\{0, 1,2,\ldots, (l_1+1)(l_2+1)\ldots(l_d+1)+1  \}\,.$$
Next we intend to prove that 
$A$ is invertible, so the inequality $A\geq \charf_{\mathcal{H}^{(N^d)}}$ will follow, which is exactly
the sought inequality since by definition $$\charf_{\mathcal{H}^{(N^d)}}-\bigotimes _{\bold{j}\in J_{\bold{l},\bold{i}}}P_{\Omega_{\bold{j}}}=P^{(+)}_{J_{\bold{l},\bold{i}}}.$$ 
Proving the invertibility of   $A$ is equivalent to showing its injectivity.
Decomposing the Hilbert space $\mathcal{H}^{(N^d)}$ as $\mathcal{H}_1\otimes \mathcal{H}_2$, where
$\mathcal{H}_1$ and $\mathcal{H}_2$ are given by $\bigotimes_{\bold{j}\in \Lambda_N^d \setminus J_{\bold{l},\bold{i}}}\mathcal{H}_\bold{j}$ and $\bigotimes_{\bold{j}\in J_{\bold{l}, \bold{i}}}\mathcal{H}_\bold{j}$, respectively, yields a factorization of $A$ as $A_1\otimes A_2$, with $A_1$ and $A_2$ acting on $\mathcal{H}_1$ and $\mathcal{H}_2$, respectively. Therefore,
the injectivity of $A$ is equivalent to the injectivity of both $A_1$ and $A_2$. But only $A_2$ needs to be dealt with, for
$A_1$ is  just a multiple of the identity.
Thanks to the definition of $A$, $A_2$  is seen to coincide with
$$\sum_{\bold{j}\in J_{\bold{l}, \bold{i}}}P^\perp_{\Omega_\bold{j}}+\prod_{\bold{j}\in J_{\bold{l},\bold{i}}}P_{\Omega_{\bold{j}}}\,,$$ 
thus  also $A_2$ is the sum of projections that
commute with one another.

\noindent
Let $\Psi$ be a vector in $\mathcal{H}_2$ such that
$A_2\Psi=0$. From the equality $(\Psi, A_2\Psi)=0$, we see that 
\begin{equation}\label{proj-equals}
(\Psi, P^\perp_{\Omega_\bold{j}}\Psi)=0 \,\,\, \forall \bold{j}\in J_{\bold{l}, \bold{i}}\quad \text{and}\quad \prod_{\bold{j}\in J_{\bold{l},\bold{i}}}P_{\Omega_\bold{j}}\Psi=0\,.
\end{equation}
The first equalities in (\ref{proj-equals}) imply $\Psi=P_{\Omega_\bold{j}}\Psi$ for every $\bold{j}\in J_{\bold{l}, \bold{i}}$.
But then the second equality reads $\Psi=\prod_{\bold{j}\in J_{\bold{l},\bold{i}}}P_{\Omega_\bold{j}}\Psi=0$, which is what we wanted to prove.
\qed

From Lemma \ref{op-ineq-1} we derive:
\begin{cor}\label{op-ineq-2}
For any $J_{\bold{k},\bold{q}}$ the following inequality holds
\begin{equation}\label{gen-lemmaA-2-bis}
\sum_{\bold{i}\,:\,J_{\bold{l},\bold{i}}\subset  J_{\bold{k},\bold{q}} }P^{(+)}_{J_{\bold{l},\bold{i}}}\leq (l+1)^d \sum_{\bold{j}\in  J_{\bold{k},\bold{q}}} P^{\perp}_{\Omega_{\bold{j}}}\,
\end{equation}
where $l=|\bold{l}|$.
\end{cor}

\noindent
\emph{Proof}

\noindent
For fixed $\bold{l}$, we sum the l-h-s of  the inequality (see Lemma \ref{op-ineq-1})
\begin{equation}\label{gen-lemmaA-1-bis-bis}
P^{(+)}_{J_{\bold{l},\bold{i}}}\leq \sum_{\bold{j}\in  J_{\bold{k},\bold{q}}} P^{\perp}_{\Omega_{\bold{j}}}\,
\end{equation}
 over all $J_{\bold{l},\bold{i}}$ contained in  $J_{\bold{k},\bold{q}}$. Then   for each site $\bold{j}\in J_{\bold{k},\bold{q}}$  we get at most $$(l_1+1)(l_2+1)\dots(l_d+1)$$ terms of the type $P^{\perp}_{\Omega_{\bold{j}}}$.
Thus the inequality in (\ref{gen-lemmaA-2-bis}) is proven.
\qed

\begin{lem}\label{control-LS}
Assume $t>0$ sufficiently small,  $\|V^{(\bold{k},\bold{q})_{-1}}_{J_{\bold{k},\bold{q}}}\| \leq 48  \cdot \frac{t^{\frac{r-1}{3}}}{r^{x_d}}$ with $x_d=20d$, and $\Delta_{J_{k,q}}\geq \frac{1}{2}$. Then for any $N$ and $(\bold{k},\bold{q})$  the inequalities
\begin{equation}\label{bound-V}
\|V^{(\bold{k},\bold{q})}_{J_{\bold{k},\bold{q}}}\|\leq 2\|V^{(\bold{k},\bold{q})_{-1}}_{J_{\bold{k},\bold{q}}}\|\,,
\end{equation}
\begin{equation}\label{bound-S}
\|S_{J_{\bold{k}, \bold{q}}}\|\leq C\cdot t \cdot \|V^{(\bold{k},\bold{q})_{-1}}_{J_{\bold{k},\bold{q}}}\|\,,
\end{equation}
and
\begin{equation}\label{Vsquare}
\|\sum_{j=2}^{\infty}t^{j}(S_{J_{\bold{k},\bold{q}}})_j\|\leq C\cdot t^2 \cdot\|V^{(\bold{k},\bold{q})_{-1}}_{J_{\bold{k},\bold{q}}}\|^2
\end{equation}
hold true for a universal constant $C$.
\end{lem}

\noindent
\emph{Proof}

\noindent
We recall that 
\begin{equation}
V^{(\bold{k},\bold{q})}_{J_{\bold{k},\bold{q}}}:= \sum_{j=1}^{\infty}t^{j-1}(V^{(\bold{k},\bold{q})_{-1}}_{J_{\bold{k},\bold{q}}})^{diag}_j \,
\end{equation}
and
\begin{equation}
S_{J_{\bold{k},\bold{q}}}:=\sum_{j=1}^{\infty}t^j(S_{J_{\bold{k},\bold{q}}})_j\
\end{equation}
with
\begin{eqnarray}
& &(V^{(\bold{k},\bold{q})_{-1}}_{J_{\bold{k},\bold{q}}})^{diag}_j \,\label{def-Vj}\\
&:= &\sum_{p\geq 2, v_1\geq 1 \dots, v_p\geq 1\, ; \, v_1+\dots+v_p=j}\frac{1}{p!}\text{ad}\,(S_{J_{\bold{k},\bold{q}}})_{v_1}\Big(\text{ad}\,(S_{J_{\bold{k},\bold{q}}})_{v_2}\dots (\text{ad}\,(S_{I_{\bold{k},\bold{q}}})_{v_p}(G_{J_{\bold{k},\bold{q}}}))\dots \Big)\\
& &+\sum_{p\geq 1, v_1\geq 1 \dots, v_p\geq 1\, ; \, v_1+\dots+v_p=j-1}\frac{1}{p!}\text{ad}\,(S_{J_{\bold{k},\bold{q}}})_{v_1}\Big(\text{ad}\,(S_{J_{\bold{k},\bold{q}}})_{v_2}\dots (\text{ad}\,(S_{J_{\bold{k},\bold{q}}})_{v_p}(V^{(\bold{k},\bold{q})_{-1}}_{J_{\bold{k},\bold{q}}}))\dots \Big)\quad\quad\quad\quad\,.
\end{eqnarray}
and
\begin{equation}\label{formula-Sj-bis}
(S_{J_{\bold{k},\bold{q}}})_j:=ad^{-1}\,G_{J_{\bold{k},\bold{q}}}\,((V^{(\bold{k},\bold{q})_{-1}}_{J_{\bold{k},\bold{q}}})^{od}_j):=\frac{1}{G_{J_{\bold{k},\bold{q}}}-E_{J_{\bold{k},\bold{q}}}}P^{(+)}_{J_{\bold{k},\bold{q}}}\,(V^{(\bold{k},\bold{q})_{-1}}_{J_{\bold{k},\bold{q}}})_j\,P^{(-)}_{J_{\bold{k},\bold{q}}}-h.c.\,.
\end{equation}
From (\ref{formula-Sj-bis}) we get
\begin{eqnarray}
& &\text{ad}\,(S_{J_{\bold{k},\bold{q}}})_{r_p}(G_{J_{\bold{k},\bold{q}}})\\
&=&\text{ad}\,(S_{J_{\bold{k},\bold{q}}})_{r_p}(G_{J_{\bold{k},\bold{q}}}-E_{J_{\bold{k},\bold{q}}})\\
&=&\,[\frac{1}{G_{J_{\bold{k},\bold{q}}}-E_{J_{\bold{k},\bold{q}}}}P^{(+)}_{J_{\bold{k},\bold{q}}}\,(V^{(\bold{k},\bold{q})_{-1}}_{J_{\bold{k},\bold{q}}})_{r_p}\,P^{(-)}_{J_{\bold{k},\bold{q}}} \,,\,G_{J_{\bold{k},\bold{q}}}-E_{J_{\bold{k},\bold{q}}}]+h.c.\\
&=&-P^{(+)}_{J_{\bold{k},\bold{q}}}\,(V^{(\bold{k},\bold{q})_{-1}}_{J_{\bold{k},\bold{q}}})_{r_p}\,P^{(-)}_{J_{\bold{k},\bold{q}}}-P^{(-)}_{J_{\bold{k},\bold{q}}}\,(V^{(\bold{k},\bold{q})_{-1}}_{J_{\bold{k},\bold{q}}})_{r_p}\,P^{(+)}_{J_{\bold{k},\bold{q}}}\,
\end{eqnarray}
and 
\begin{equation}\label{ineq-S-V}
\|(S_{J_{\bold{k},\bold{q}}})_j\|\leq 2\frac{\|(V^{(\bold{k},\bold{q})_{-1}}_{J_{\bold{k},\bold{q}}})_j\|}{\Delta_{J_{\bold{k},\bold{q}}}}\leq 4\|(V^{(\bold{k},\bold{q})_{-1}}_{J_{\bold{k},\bold{q}}})_j\|\,
\end{equation}
since we have assumed $\Delta_{J_{\bold{k},\bold{q}}}\geq \frac{1}{2}$.
Next, using the definition in (\ref{def-Vj}), we can estimate
\begin{eqnarray}
& &\|(V^{(\bold{k},\bold{q})_{-1}}_{J_{\bold{k},\bold{q}}})_j\|\label{norm-vj}\\
&\leq  &\sum_{p=2}^{j}\,\frac{8^p}{p!}\,\sum_{ v_1\geq 1 \dots, v_p\geq 1\, ; \, v_1+\dots+v_p=j}\,\|\,(V^{(\bold{k},\bold{q})_{-1}}_{J_{\bold{k},\bold{q}}})_{v_1}\|\|\,(V^{(\bold{k},\bold{q})_{-1}}_{J_{\bold{k},\bold{q}}})_{v_2}\|\dots \|\,(V^{(\bold{k},\bold{q})_{-1}}_{J_{\bold{k},\bold{q}}})_{v_p}\|\\
& &+2\|V^{(\bold{k},\bold{q})_{-1}}_{J_{\bold{k},\bold{q}}}\|\, \sum_{p=2}^{j-1}\,\frac{8^p}{p!}\,\sum_{p\geq 1, v_1\geq 1 \dots, v_p\geq 1\, ; \, v_1+\dots+v_p=j-1}\,\|\,(V^{(\bold{k},\bold{q})_{-1}}_{J_{\bold{k},\bold{q}}})_{v_1}\|\|\,(V^{(\bold{k},\bold{q})_{-1}}_{J_{\bold{k},\bold{q}}})_{v_2}\|\dots \|\,(V^{(\bold{k},\bold{q})_{-1}}_{J_{\bold{k},\bold{q}}})_{v_p}\|\,.\quad\quad\quad\quad\,.
\end{eqnarray}
In order to estimate (\ref{norm-vj}) we refer to Theorem 3.2 in \cite{DFFR}, hence we consider the numbers $B_j$, $j\geq 1$, recursively defined by
\begin{eqnarray}
B_1&:= &\|V^{(\bold{k},\bold{q})_{-1}}_{J_{\bold{k},\bold{q}}}\| \\
B_j&:=&\frac{1}{a}\sum_{l=1}^{j-1}B_{j-l}B_l\,,\quad j\geq 2\,, \label{def-Bj}
\end{eqnarray}
with $a$ such that
\begin{equation}\label{a-eq}
\frac{e^{8a}-8a-1}{a}+e^{8a}-1=1\,.
\end{equation}
Following  \cite{DFFR}), by induction we get
\begin{equation}\label{bound-V-B}
\|(V^{(\bold{k},\bold{q})_{-1}}_{J_{\bold{k},\bold{q}}})_j\|\leq B_j\,\Big(\frac{e^{8a}-8a-1}{a}\Big)+2\|V^{(\bold{k},\bold{q})_{-1}}_{J_{\bold{k},\bold{q}}}\|\,B_{j-1}\Big(\frac{e^{8a}-1}{a}\Big)
\end{equation}
and
\begin{equation}\label{bound-b}
  B_j\geq \frac{2B_{j-1}\|\,V^{(\bold{k},\bold{q})_{-1}}_{J_{\bold{k},\bold{q}}}\,\|}{a}
\end{equation} 
that combined with (\ref{a-eq}) yields
\begin{equation}\label{bound-b-bis}
B_j\geq \|\,(V^{(\bold{k},\bold{q})_{-1}}_{J_{\bold{k},\bold{q}}})_j\|.
\end{equation} The numbers $B_j$ are seen to be the Taylor's coefficients of
\begin{equation}\label{taylor}
f(u):=\frac{a}{2}\cdot \Big(\,1-\sqrt{1-\frac{4}{a}\cdot \|V^{(\bold{k},\bold{q})_{-1}}_{J_{\bold{k},\bold{q}}}\|u }\,\Big)\,;
\end{equation}
see  \cite{DFFR}. Therefore, if we consider the norms $\|(V^{(\bold{k},\bold{q})_{-1}}_{J_{\bold{k},\bold{q}}})^{diag}_j \|$ as $u$-independent, the radius of analyticity, $t_0$,  of 
\begin{equation}
\sum_{j=1}^{\infty}u^{j-1}\|(V^{(\bold{k},\bold{q})_{-1}}_{J_{\bold{k},\bold{q}}})^{diag}_j \|=\frac{1}{u}\Big(\sum_{j=1}^{\infty}u^{j}\|(V^{(\bold{k},\bold{q})_{-1}}_{J_{\bold{k},\bold{q}}})^{diag}_j \|\Big)
\end{equation}
is bounded below by that of $\sum_{j=1}u^jB_j$, hence
\begin{equation}\label{radius}
t_0\geq \frac{a}{4\|V^{(\bold{k},\bold{q})_{-1}}_{J_{\bold{k},\bold{q}}}\|}\geq \frac{a}{192}\,,
\end{equation}
where in the last inequality we use the assumption on $\|V^{(\bold{k},\bold{q})_{-1}}_{J_{\bold{k},\bold{q}}}\|$.
The same  bound holds for the radius of convergence of the series $S_{J_{\bold{k},\bold{q}}}:=\sum_{j=1}^{\infty}t^j(S_{J_{\bold{k},\bold{q}}})_j\,$ as a consequence of  the inequality in (\ref{ineq-S-V}).

\noindent
For $0<t<1$ and in the interval $(0,\frac{1}{2}\cdot \frac{a}{ 192})$, due to  (\ref{bound-b-bis}) and (\ref{taylor}) the following holds true
\begin{eqnarray}
\sum_{j=1}^{\infty}t^{j-1}\|(V^{(\bold{k},\bold{q})_{-1}}_{I_{k,q}})^{diag}_j \|&\leq &\frac{1}{t}\sum_{j=1}^{\infty}t^jB_j\\
&=&\frac{1}{t}\cdot \frac{a}{2}\cdot \left(\,1-\sqrt{1- (\frac{4}{a}\cdot \|V^{(\bold{k},\bold{q})_{-1}}_{I_{k,q}}\|) \,t }\,\right)\\
&\leq &(1+C_a \cdot t )\,\|V^{(\bold{k},\bold{q})_{-1}}_{I_{k,q}}\|\label{A.36}
\end{eqnarray}
for some $a$-dependent constant $C_a>0$.
This implies the inequality in (\ref{bound-V}), by assuming $t>0$ sufficiently small but independent of $N$, $k$, and $q$. 
Likewise we derive (\ref{bound-S}).

\noindent
As for (\ref{Vsquare}), we start from 
\begin{equation}
\|\sum_{j=2}^{\infty}t^{j}(S_{J_{\bold{k},\bold{q}}})_j\|\leq\sum_{j=2}^\infty t^j\|(S_{J_{\bold{k},\bold{q}}})_j\|
\leq 4\sum_{j=2}^\infty t^j \|(V^{(\bold{k},\bold{q})_{-1}}_{J_{\bold{k},\bold{q}}})_j\|\leq 4\sum_{j=2}^{\infty}t^jB_j\\
\end{equation}
then, using $B_1\equiv \|V^{(\bold{k},\bold{q})_{-1}}_{J_{\bold{k},\bold{q}}}\|$ and a Taylor expansion, for $t$ in the interval  where (\ref{A.36}) holds, we estimate
\begin{eqnarray}
\sum_{j=2}^\infty t^jB_{j}&=&\frac{a}{2}\cdot \Big(\,1-\sqrt{1-\frac{4}{a}\cdot \|V^{(\bold{k},\bold{q})_{-1}}_{J_{\bold{k},\bold{q}}}\|t }\,\Big)\,-t\cdot \|V^{(\bold{k},\bold{q})_{-1}}_{J_{\bold{k},\bold{q}}}\|\\
&\leq&D_a\cdot t^2\cdot \|V^{(\bold{k},\bold{q})_{-1}}_{J_{\bold{k},\bold{q}}}\|^2
\end{eqnarray}
where $D_a$ depends only on $a$.
\qed

\begin{lem}\label{conn-rect}
Let $\{J_{\bold{s}^{(i)}, \bold{u}^{(i)}} \,;\, |\bold{s}^{(i)}|=k, i\in\{1\cdots n\}\}$ for some $k\in\mathbb{N}$ be such that $\cup_i J_{\bold{s}^{(i)}, \bold{u}^{(i)}} $ is connected. Then there is a closed path  (see Definition \ref{pathsdef}) $\gamma_k$ with supp$(\gamma_k)=\{J_{\bold{s}^{(i)}, \bold{u}^{(i)}} \,;\, |\bold{s}^{(i)}|=k, i\in\{1\cdots n\}\}$ and length $l_{\gamma_k}=2n-2$.
\end{lem}


\noindent
\emph{Proof}

\noindent
We proceed by  induction on the number, $n$, of elements of 
a collection of rectangles as in the statement. 
The statement is clearly true for $n=2$ rectangles. We  assume that it holds for collections of $n\geq 2$ rectangles and 
prove it continues to hold for collections of $n+1$ rectangles as well. Given any such collection, without loss of generality we can
suppose that the union of the first $n$ rectangles is still a connected set
of $\mathbb{R}^d$. Clearly the $n+1$-th rectangle
must intersect at least one of the previous $n$ rectangles. We can then pick one of those rectangles, say $J_{\bold{k}^{(i_{*})}, \bold{q}^{(i_{*})}}$, from the set, and consider the step $$(J_{\bold{k}^{(i_{*})}, \bold{q}^{(i_{*})} }, J_{\bold{k}^{(n+1)}, \bold{q}^{(n+1)}})$$ and the step back $(J_{\bold{k}^{(n+1)}, \bold{q}^{(n+1)}},J_{\bold{k}^{(i_{*})}, \bold{q}^{(i_{*})} })$ . Hence we get a path with two more steps and enjoying the required properties. 

\noindent
\qed

\begin{lem}\label{conn-rect-2}
For $\mathfrak{b}\in  \mathcal{B}_{V^{(\bold{k},\bold{q})}_{J_{\bold{r},\bold{i}}}}$, assume that
$$ \cup_{i\in\{1,\cdots,|\mathcal{R}_{\mathfrak{b}}|\}}J_{\bold{k}^{(i)},\bold{q}^{(i)}}=\cup_{\rho=k_0}^{k} \cup_{j=1}^{j_{\rho}}\mathcal{Z}^{(j)}_{\rho}$$  where $\{\mathcal{Z}^{(j)}_{\rho}, \quad j=1,\dots,j_{\rho}\}$  are distinct connected components of (unions of) rectangles of same size $\rho$. Then there is a path, $\Gamma_{\mathfrak{b}}$, of length $l_{\Gamma_\mathfrak{b}}$ such that  $$l_{\Gamma_\mathfrak{b}} \leq 2(n_{k_0}+\sum_{j=1}^{j_2}n_{k_0+1}^{(j)}+\dots +\sum_{j=1}^{j_k}n_k^{(j)})-2$$ with the following properties:
\begin{enumerate}
\item[A)]   the support of $\Gamma_\mathfrak{b}$ is  $\mathcal{R}_\mathfrak{\mathfrak{b}}$;
\item[B)] for each component $\mathcal{Z}^{(j)}_{\rho}$ consisting of the union of $n_{\rho}^{(j)}$ rectangles, at most $2n_{\rho}^{(j)}-2$ steps are implemented (i.e., there are at most $2n_{\rho}^{(j)}-2$ steps $\sigma\in\mathcal{S}_{\Gamma_\mathfrak{b}}$ for which $\sigma\in \text{supp}(\mathcal{Z}^{(j)}_{\rho})\times \text{supp}(\mathcal{Z}^{(j)}_{\rho})$);
\item[C)]  there are at most two steps connecting rectangles in $\text{supp}(\mathcal{Z}_\rho^{(j)})$ with rectangles
of lower size: more precisely, for every connected component $\mathcal{Z}_\rho^{(j)}$ there is at most one $J_{\bold{s},\bold{u}}$ in $\text{supp}(\mathcal{Z}_\rho^{(j)})$  such that $(J_{\bold{s}',\bold{u}'},J_{\bold{s},\bold{u}})\in\mathcal{S}_{\Gamma_\mathfrak{b}}$ with $s'<s$, and one $J_{\bold{s},\bold{u}}$ such that $(J_{\bold{s},\bold{u}},J_{\bold{s}',\bold{u}'})\in\mathcal{S}_{\Gamma_\mathfrak{b}}$ with $s<s'$.
\end{enumerate}

\end{lem}

\noindent
\emph{Proof}


\noindent
The construction is by induction in the size, $k$, of the rectangles.  
We call $\gamma^{(j)}_{\rho}$ the closed path that visits the rectangles of the component $\mathcal{Z}^{(j)}_{\rho}$ and constructed according to Lemma \ref{conn-rect}. Hence, for $k=k_0$ we just refer to Lemma \ref{conn-rect}. Notice that property c) does not apply for $k=k_0$. 

\noindent
Next we assume that we have constructed the path, say $\Gamma_b^{(k'-1)}$, with $k_0\leq k'\leq k$, fulfilling A), B), and C) for the set $\cup_{\rho=k_0}^{k'-1} \cup_{j=1}^{j_{\rho}}\mathcal{Z}^{(j)}_{\rho}$, which is connected by Property P-i).  Then from this path we derive a new one,  that we call $\Gamma_b^{(k')}$, with the desired properties
for the set $\cup_{\rho=k_0}^{k'} \cup_{j=1}^{j_{\rho}}\mathcal{Z}^{(j)}_{\rho}$. The path is constructed following the prescriptions below:

\begin{itemize}

\item
we follow $\Gamma_{\mathfrak{b}}^{(k'-1)}$ until it reaches a rectangle that has an overlap with a rectangle of one of the components $\mathcal{Z}^{(j)}_{k'}$, then we  implement a ``turning step" that means we stop proceeding along $\Gamma_{\mathfrak{b}}^{(k'-1)}$ and start to follow the closed path $\gamma^{(j)}_{k'}$ along the component $\mathcal{Z}^{(j)}_{k'}$, by starting and ending at the rectangle of the turning step;
\item
we proceed in the same way along the remaining part of the path $\Gamma_{\mathfrak{b}}^{(k'-1)}$, that means we implement a turning step as soon as the rectangle that has been reached has an overlap with another component, say $\mathcal{Z}^{(j')}_{k'}$,  not visited yet;
\item
we iterate this procedure until  all the components $\mathcal{Z}^{(j)}_{k'}$ have been visited and the path $\Gamma_{\mathfrak{b}}^{(k'-1)}$ has been completed to the initial rectangle. 
\end{itemize}

\noindent
\qed

\end{appendix}

\end{document}